\documentclass[12pt,preprint]{aastex}

\usepackage{amsmath}          
\usepackage{natbib}           
\usepackage{graphicx}         
\usepackage{rotating}         

\bibpunct{(}{)}{;}{a}{}{,}    

\newcommand{\mrm}[1]{\ensuremath{\mathrm{#1}}}

\defcitealias{Korm-pc08-psBG}{K08}

\shorttitle{Bulgeless galaxies}
\shortauthors{Bizzocchi et~al.}

\begin{document}


\title{Bulgeless Galaxies at Intermediate Redshift: Sample Selection, Colour Properties,
       and the Existence of Powerful AGN}


\author{Luca Bizzocchi\altaffilmark{1},
        Mercedes E.~Filho\altaffilmark{3},
        Elvira Leonardo\altaffilmark{1},
        Marco Grossi\altaffilmark{1,2},
        Roger L. Griffith\altaffilmark{4},
        Jos\'e Afonso\altaffilmark{1,2},
        Cristina Fernandes\altaffilmark{1},
        Jo\~ao Retr\^e\altaffilmark{1},
        Sonia Anton\altaffilmark{5,6},
        Eric F. Bell\altaffilmark{7},
        Jarle Brinchmann\altaffilmark{8},
        Bruno Henriques\altaffilmark{9},
        Catarina Lobo\altaffilmark{3,10},
        Hugo Messias\altaffilmark{11}}


\altaffiltext{1}{Centro de Astronomia e Astrof\'isica da Universidade de Lisboa,
                 Observat\'orio Astron\'omico de Lisboa, Tapada da Ajuda,
                 1349-018 Lisboa (Portugal)}
\altaffiltext{1}{Centro de Astronomia e Astrof\'{\i}sica da Universidade de Lisboa,  
                 Observat\'{o}rio Astron\'{o}mico de Lisboa, Tapada da Ajuda, 
                 1349-018 Lisbon, Portugal}
\altaffiltext{2}{Department of Physics, Faculty of Sciences, University of Lisbon, 
                 Campo Grande, 1749-016 Lisbon, Portugal}
\altaffiltext{3}{Centro de Astrof\'isica da Universidade do Porto,
                 Rua das Estrelas, 4150-762 Porto (Portugal)}
\altaffiltext{4}{Infrared Processing and Analysis Center, California Institute of Technology,
                 Pasadena, CA~91125 (USA)}
\altaffiltext{5}{Centro de Investiga\c{c}\~ao em Ci\^encias Geo-Espaciais, Faculdade
                 de Ci\^encias da Universidade do Porto, Porto (Portugal)}
\altaffiltext{6}{SIM, Faculdade de Ci\^encias da Universidade de Lisboa, Lisboa (Portugal)}
\altaffiltext{7}{Department of Astronomy, University of Michigan, 500~Church St., Ann Arbor 
                 MI~48109 (USA)}
\altaffiltext{8}{Leiden Observatory, Leiden University, P.O. Box 9513, NL--2300 RA Leiden,
                 (Netherlands)}
\altaffiltext{9}{Max-Planck-Institut f\"ur Astrophysik, Karl Schwarzschild Stra\ss e 1,
                 D-85748 Garching bei M\"unchen  (Germany)}
\altaffiltext{10}{Departamento de F\'isica e Astronomia, Faculdade de C\^encias da Universidade
                 do Porto, Rua do Campo Alegre, 687, 4169-007, Porto (Portugal)}
\altaffiltext{11}{Departamento de astronom\'ia,
                 Av.\ Esteban Iturra 6to piso, Facultad de Ciencias F\'isicas y Matem\'aticas,
                 Universidad de Concepci\'on (Chile)}


\begin{abstract}
\noindent
We present a catalogue of bulgeless galaxies, which includes 19\,225 objects selected in 
four of the deepest, largest multi-wavelength datasets available -- COSMOS, AEGIS, GEMS and 
GOODS -- at intermediate redshift ($0.4 \leq z \leq 1.0$).
The morphological classification was provided by the Advanced Camera for Surveys General 
Catalogue (ACS-GC), which used publicly available data obtained with the ACS instrument on 
the \textit{Hubble Space Telescope}.
Rest-frame photometric quantities were derived using \texttt{kcorrect}.
We analyse the properties of the sample and the evolution of pure-disc systems with redshift.
Very massive [$\log (M_\star/M_{\odot}) > 10.5$] bulgeless galaxies contribute to $\sim 30$\% 
of the total galaxy population number density at $z \geq 0.7$, but their number density drops 
substantially with decreasing redshift.
We show that only a negligible fraction of pure discs appear to be quiescent systems, and 
red sequence bulgeless galaxies show indications of dust-obscured star formation.
X-ray catalogues were used to search for X-ray emission within our sample. 
After visual inspection and detailed parametric morphological fitting we identify 30~AGN 
that reside in galaxies without a classical bulge. 
The finding of such peculiar objects at intermediate redshift shows that while AGN growth 
in merger-free systems is a rare event (0.2\% AGN hosts in this sample of bulgeless 
galaxies), it can indeed happen relatively early in the Universe history.
\end{abstract}


\section{Introduction} \label{sec:intro}
One of the cornerstones of modern cosmology is understanding the formation and evolution 
of galaxies and the relative importance of merger processes.
According to Cold Dark Matter ($\Lambda$CDM) structure formation models, galaxies form 
hierarchically: dark matter collapses into halos in the early Universe, which virialise 
and cluster hierarchically to form large-scale structures.
Rotating discs start forming in slowly growing dark matter halos \citep{Fall-MNRAS80-Bulg} 
and become the building blocks of galaxies \citep{Cole-MNRAS00-Hierarc}.
In this framework of a merger-dominated Universe, galaxies without a bulge are difficult 
to produce: the discs will thicken or warp due to interactions between low mass halos 
\citep{Benson-MNRAS04-discs} or be completely disrupted by major mergers 
\citep{Koda-ApJ09-mergers}, leaving behind a classical bulge.
Halos that somehow escape or at least avoid major mergers since $z\sim$3 could, in
principle, form bulgeless galaxies, especially if some form of heuristic prescription for
stellar feedback is included in the simulations
\citep{dOnghia-ApJL04-BulgL,Robertson-ApJ04-disk}, but the simulated disks end up being 
smaller, denser and having lower angular momentum than the observed ones.
However, more recent hydrodynamical simulations of dwarf galaxies show that outflows can
prevent the formation of classical merger-driven bulges in low mass systems by expelling
most of the low angular momentum gas generated in such events 
\citep{Brook-MNRAS12-GF,Dutton-MNRAS09-FB}.

The observed correlations between black hole (BH) mass and host galaxy properties (such as, 
stellar velocity dispersion, circum-nuclear star formation, mass, and luminosity; see e.g., 
\citealt{Kormendy-ARAA95-BHs,Magorrian-AJ98-BHs,Tremaine-ApJ02-BHsigma,%
         Haring-ApJL04-BHM,Koda-ApJ09-mergers,%
         Schaw-ApJ10-SMBH}) 
are rarely observed in galaxies with no bulges \defcitealias{Kormendy-Nat11-SMBHbar}{K11}
\citep[hereafter referred as K11]{Kormendy-Nat11-SMBHbar}, 
thus suggesting that the formation and evolution of galaxy bulges are linked to the 
formation and evolution of the host BHs, albeit perhaps differently in early- and late-type 
objects \citep{Schaw-ApJ10-SMBH}.

The growth of a BH in a bulgeless galaxy remains difficult to understand in the 
absence of merger processes.
On the other hand, several observations of BHs and AGN in nearby bulgeless galaxies 
currently indicate that a classical bulge is not a requirement for an (active) nuclear BH: 
NGC~4395 \citep{Filipp-ApJL03-NGC4395,Peterson-ApJ05-NGC4395} is a bulgeless Sm galaxy with 
a Seyfert~1 spectrum, Pox~52 \citep{Thornton-ApJ08-POX52} is a dwarf Seyfert~1 galaxy, 
Henize~2-10 is the first ever (irregular) dwarf discovered to have a supermassive BH 
\citep{Reines-NAt11-H210}, whereas NGC~3621, NGC~4178, NGC~3367, NGC~4536, and NGC~4561 
\citep{Satyapal-ApJL07-NGC3621,Barth-ApJ09-NGC6321,Satyapal-ApJ09-NGC4178,%
       Secrest-ApJ12-NGC4178,McAlpine-ApJ11-NGC3367,Araya-ApJ12-NGC4561}, are late-type Sd 
       galaxies with AGN activity from low mass black holes.
Also, \citet{Desroches-ApJ09-AGN} found a high incidence of AGN activity in a sample of
late-type spirals observed with \emph{Chandra}, indicating that BHs can evolve in
galaxies with little or no evidence for bulges.

\citeauthor{Kormendy-Nat11-SMBHbar} \citepalias{Kormendy-Nat11-SMBHbar} found no correlation 
between BH and disc masses in a sample of 11~galaxies. 
\citet{Mathur-ApJ12-SMBHs} analysed ten pseudo-bulge Narrow Line Seyfert~1 (NLS~1) galaxies 
and compared the BH mass to the pseudo-bulge luminosity, finding that the BHs were 
undermassive compared to the host galaxies. 
This supports the hypothesis that BHs in pseudo-bulges/disc galaxies are in the growth mode 
at the present epoch and that this growth is triggered by secular processes.
In this context, BHs/AGN found in bulgeless galaxies are thought to be the smallest BHs grown
by local processes and are the seeds of the larger merger-grown SMBHs
found in massive galaxies \citepalias{Kormendy-Nat11-SMBHbar}.

The Sloan Digital Sky Survey (SDSS) has been used to establish a benchmark for the local 
Universe \citep{Kautsch-AA06-Eon} and to investigate the significance of AGN feedback in 
bulgeless galaxies, including quenching of star formation 
\citep{Bell-ApJ08-BLess,Coelho-MNRAS13-BLess}. 
Recently, a sample of 13~AGN in bulgeless galaxies with BH masses in the range 
$10^6-10^7\,M_\odot$ was selected from the SDSS using visual classification from the Galaxy 
Zoo \citep{Simmons-MNRAS12-Gzoo}.
The authors showed that significant BH growth can take place in the absence of mergers or 
violent disc instabilities.

Given the importance of bulgeless galaxies to understand galaxy evolution, a study of these 
objects at higher redshifts is fundamental. 
The existence of a set of very deep and high-resolution observations with the 
\textit{Hubble Space Telescope} (\textit{HST}) allows to extend morphological local studies, 
as those performed with SDSS, to much higher redshifts, and start mapping the evolution of 
bulgeless galaxies. 
Here we present a catalogue and a first analysis of such systems at intermediate redshift 
($0.4 \leq z \leq 1.0$), using publicly available data from four of the largest, deepest 
multi-wavelength surveys: 
the Cosmological Evolutionary Survey\footnote{http://cosmos.astro.caltech.edu} 
\citep[COSMOS;][]{Scov-ApJS07-COSMOS}, the All-wavelength Extended Groth Strip (EGS) 
International Survey\footnote{http://aegis.ucolick.org} \citep[AEGIS;][]{Davis-ApJ07-AEGIS}, 
the Galaxy Evolution from Morphology and SEDs (GEMS) 
survey\footnote{http://www.mpia.de/GEMS/gems.htm} \citep{Cald-ApJS08-GEMS}, and the Great
Observatories Origins Deep Survey\footnote{http://www.stsci.edu/science/goods/}
\citep[GOODS;][]{Dick-pc03-GOODS,Giava-ApJ04-GOODS}. 
The latter survey targets two separate fields, the Hubble Deep Field North (HDF-N, hereafter
referred to as GOODS-N) and the Chandra Deep Field South (CDF-S, hereafter referred to as 
GOODS-S). 
The \textit{HST} data for all these fields have been recently
assembled by \citet{Griff-ApJS12-ACSGC} into a single homogeneously analysed data-set.

Besides high-resolution optical imaging, these fields also possess very comprehensive
multi-wavelength coverage, from X-rays to radio, as well as vast spectroscopic information.
The synergy of such a panchromatic data-set offers us the opportunity to investigate the
significance of bulgeless galaxies and bulgeless AGN hosts when the Universe was less than
one-half of its current age (more than 7\,Gyr in the past), an epoch much more active than 
the current one.
This can provide valuable insight into the understanding of the evolutionary path of 
bulgeless systems within the overall galaxy formation and evolutionary scenario.

The paper is organised as follows:
in \S~\ref{sec:sample} and \S~\ref{sec:anc} we describe the data-set and present the
derivation of the relevant rest-frame quantities; in \S~\ref{sec:BGcat} we present the final 
magnitude-limited ($m_\mrm{AB}<24$) catalogue containing $19\,225$ bulgeless galaxies at 
intermediate redshift ($0.4 \leq z \leq 1.0$), and investigate the reliability of the adopted 
morphological classification, as well as the accuracy of the mass estimates. In
\S~\ref{sec:BGprop} an overview of the sample properties, focusing on bulgeless galaxy 
evolution and on peculiar subsamples, such as red and massive bulgeless galaxies, is provided.
In \S~\ref{sec:AGN} we present the first systematic search for bulgeless AGN at intermediate 
redshift, based on X-ray emission, and presenting an accurate analysis of these objects' 
morphology.
Finally, a summary of this work is given in \S~\ref{sec:summ}.

Throughout the article we use H$_{\rm 0}$=71 km s$^{-1}$ Mpc$^{-1}$, $\Omega_{\rm M}$=0.27
and $\Omega_{\Lambda}$=0.73 \citep[WMAP7;][]{Larson-ApJS11-WMAP7}.
All magnitudes are (unless otherwise noted) expressed in the AB system.

\section{Sample selection} \label{sec:sample}
The present study is based on the Advanced Camera for Surveys (ACS) \textit{HST} public
morphology catalogue, which has been recently assembled by \citet{Griff-ApJS12-ACSGC}, 
merging the largest \textit{HST} imaging surveys.
The catalogue contains a total of 469\,501 sources in five sky fields: COSMOS (304\,688 
sources), AEGIS (70\,142 sources), GEMS (59\,447 sources), GOODS-S (10\,999 sources), and 
GOODS-N (24\,225 sources).
This large ACS data-set has been homogeneously analysed using the code \textsc{Galapagos}
\citep{Barden-MNRAS12-GALAP}, an application that combines object detection
with \textsc{SExtractor} \citep{Bertin-AAS96-SEx} and light profile modelling with
\textsc{Galfit} \citep{Peng-AJ02-GALFIT}.
These data provide the basis to extract precise morphological parameters for the galaxies.
A thorough description of the ACS general catalogue (henceforth referred to as the ACS-GC), 
its contents, additional data products, and details of its construction from HST/ACS
images, are given in \citet{Griff-ApJS12-ACSGC}.
A brief summary of the ACS-GC imaging data is also reported here in Table~\ref{tab:surveys},
giving central coordinates for the surveys, survey size, filters and pixel scales.
In the following, we outline those details which are pertinent to the
context of the present investigation.

\subsection{Imaging and morphology data} \label{sec:morphology}
The morphological classification was derived from the analysis of the surface brightness
profiles, obtained at the reddest filter image available for each field, namely $I_{F814W}$
for the COSMOS and EGS fields, $z_{F850LP}$ for the GEMS + GOODS-S fields, and $i_{F775W}$
for the GOODS-N field.
The light distribution of the galaxies was analysed by fitting a single S\'ersic profile
\citep{Sersic-68}:
\begin{equation}
 \Sigma(r) \propto \exp\{-\kappa(r/r_e)^{-1/n}-1\} \,,
\end{equation}
\noindent where the S\'ersic index $n$ describes the shape of the light profile, $r_e$ is 
the effective radius of the galaxy, and $\kappa$ is a positive parameter that is coupled to
$n$, such that half of the total flux is always within $r_e$.
An index $n = 1$ corresponds to an exponential profile of a typical pure-disc galaxy, whereas 
$n = 4$ corresponds to the de~Vancouleurs' profile of elliptical or spheroidal galaxies.

In order to improve the robustness of the morphology selection, we rejected the faintest 
galaxies by imposing a magnitude cut, $m_\mrm{AB} \le 24$, in the filter considered for 
each field (see above). 
To avoid sources with very little or no morphological information, we removed point sources 
from the sample in the following way. 
Pixel sizes in the adopted surveys range from 0.05\,arcsec for COSMOS (ACS/Wide Field Channel), 
to 0.03\,arcsec (ACS/High Resolution Channel) for AEGIS, GEMS and GOODS.
For an average ACS PSF of~0.075\,arcsec, point-like sources occupy just over one to two 
pixels.
We thus discarded catalogue entries with a half-light radius smaller than 0.15\,arcsec
(two times the PSF). 
We further discarded all sources with a $b/a$ axis ratio less than 0.5 (equivalent to 
inclinations larger than 60\,deg), in order to minimise the effects of dust extinction on 
galaxy colours.
At this stage the galaxy sample contains 78\,830 sources.

It should be considered that galaxy appearance may depend on the rest-frame wavelength at 
which it is observed, thus comparison of the morphology of galaxies observed at different 
redshift or through different filters should be treated with additional care.
Given the redshift interval under study ($0.4 \leq z \leq 1.0$) and the filters adopted, 
we are sampling a rest-frame range of $\lambda$ 3900--6100\,\AA, where the morphological 
$K$-correction is found to be not significant for most galaxies.
This has been pointed out in the detailed COSMOS morphological study of 
\citet{Scarlata-ApJS07-ZEST} (see also, detailed discussions in 
\citealt{Lotz-AJ04-morph,Cassata-MNRAS05-GOODS}).

\subsection{Redshift data} \label{sec:redshift}
The ACS-GC contains spectroscopic and photometric redshifts from different sources,
as summarised in the following \citep[see][for details]{Griff-ApJS12-ACSGC}.
\begin{itemize}
%
\item
COSMOS: $\sim10\,300$ spectroscopic redshifts are provided from the VIMOS/VLT
zCOSMOS survey\footnote%
{http://archive.eso.org/cms/eso-data/data-packages/zcosmos-data-release-dr2/}
\citep{Lilly-ApJS07-zCOSMOS}.
This data release provides a classification for the quality of the spectroscopic measurements:
in the present work we considered classes 3 and 4 (very secure redshifts, $P \geq 99.5\%$),
class 2 (probable redshift, $P >92\%$), and class 9 (one identified line, high photo-$z$
consistency) as reliable, whereas other redshift classes were discarded.
Photometric redshifts ($\sim 252\,000$) are from the COSMOS public catalogue
\citep{Ilbert-ApJ09-COSMOS}, and are based on 30-band photometry, which spans the wavelength
range from the ultraviolet to the mid-infrared.
%
\item
AEGIS: the ACS catalogue provides $\sim5\,700$ spectroscopic redshifts from
the Deep Extragalactic Evolutionary probe, DEEP2\footnote%
{http://deep.berkeley.edu/DR3/dr3.primer.html},
obtained with DEIMOS on Keck \citep{Davis-SPIE03-DEEP2,Davis-ApJ07-AEGIS}.
Following DEEP2 team's recommendations, only the entries with redshift quality ($z_q$)
greater than or equal to 3 ($P >90\%$) were treated as having a reliable redshift 
measurement.
Sources having lower redshift quality were discarded.
The catalogue also contains $\sim 43\,800$ photometric redshifts, obtained from the 
CFHT/MEGACAM deep multi-colour data comprising 11 bands \citep{Ilbert-AA06-EGSphz}.
%
\item
GEMS + GOODS-S: $\sim7\,000$ spectroscopic redshifts are provided from various
sources, mostly obtained with ESO facilities \cite[see Table 1 of][]{Griff-ApJS12-ACSGC}.
We retained redshift entries labelled with the quality flags 4 and 3 (high and good quality,
respectively).
The photometric redshifts ($\sim44\,200$) for GEMS are from the COMBO-17 survey\footnote{%
http://www.mpia-hd.mpg.de/COMBO/combo\_CDFSpublic.html} \citep{Wolf-AA04-COMBO,Wolf-AA08-COMBO},
whereas for the GOODS-S field we merged the relevant section of the COMBO-17 catalogue with 
that of the latest redshift release of \citet{Dahlen-ApJ10-GOODSS}, which comprises 
$\sim32\,500$ photometric redshifts derived from the 12-band HST/ACS, VLT/VIMOS, VLT/ISAAC, 
and \emph{Spitzer} photometry data.
%
\item
GOODS-N: the ACS catalogue contains $\sim2\,800$ spectroscopic redshifts from
\citet{Barger-ApJ08-GOODSN} and from the GOODS-N-ALL spectroscopic survey%
\footnote{http://tkserver.keck.hawaii.edu/tksurvey/data\_products/goods\_desc.html},
which includes the Team Keck Treasury Redshift Survey \citep[TKRS,][]{Wirth-AJ04-TKRS},
and the \citet{Cowie-AJ04-GOODSN} survey.
As spectroscopic redshift quality metrics are not available for these data,
they are all considered as reliable.
As for the photometric redshifts, we merged the $\sim6\,300$ catalogue entries from
\citet{Bundy-ApJ09-MOIRCS} (6 bands) with the $\sim 9\,700$ source redshifts of
\citet{Kaji-PASJ11-MOIRCS} (11 bands), selected from the Subaru/MOIRCS near-infrared data.
\end{itemize}

\begin{figure}[tbh]
  \centering
  \includegraphics[bb=9 0 518 299,angle=0,width=16cm]{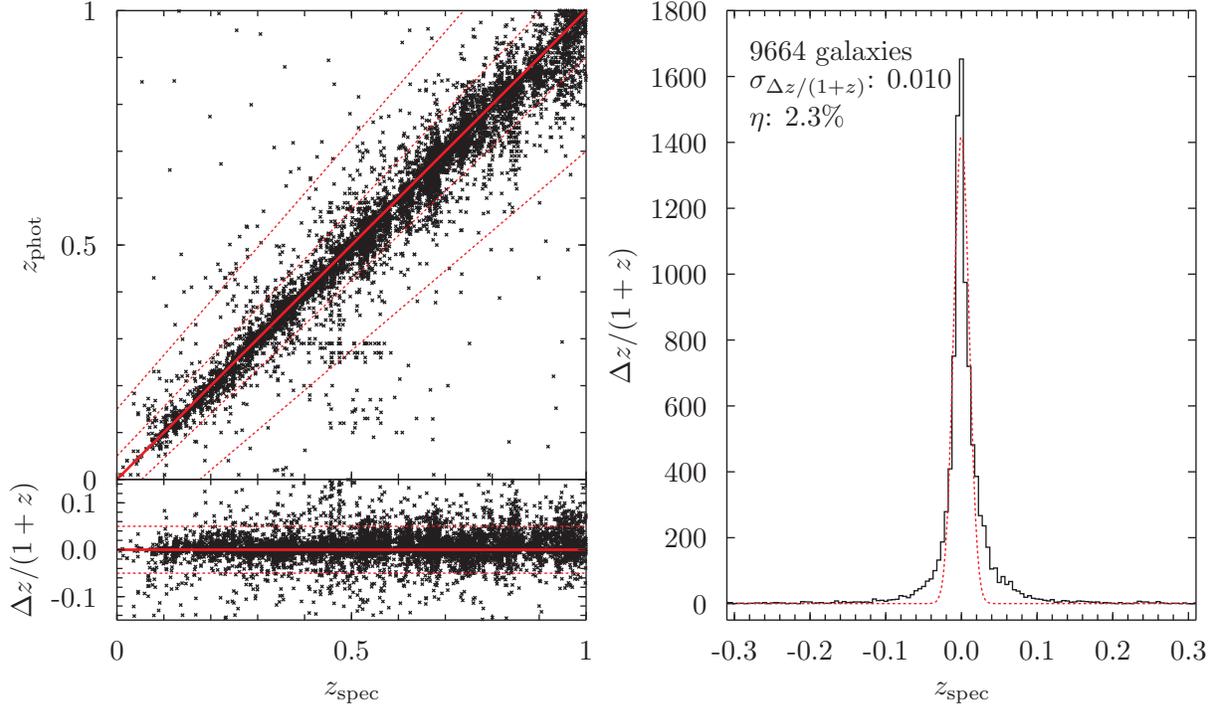}
  \caption{Left panel: comparison between $z_\mrm{phot}$ and $z_\mrm{spec}$ for our
           final galaxy sample (all fields) with \texttt{MAG\_BEST\_HI} $ < 24$.
           The dotted and dashed lines are for
           $z_\mrm{phot} = z_\mrm{spec} \pm 0.15(1 + z_\mrm{spec})$ and
           $z_\mrm{phot} = z_\mrm{spec} \pm 0.05(1 + z_\mrm{spec})$, respectively.
           The $1\sigma$ dispersion and the fraction of catastrophic failures $\eta$
           are listed in the top-left corner of the right panel.
           Right panel: $\Delta z/(1 + z_\mrm{spec})$ distribution.
           The dashed line is a Gaussian distribution with $\sigma = 0.010$.}
  \label{fig:zs-vs-zp}
\end{figure}

Given that we assembled redshift data from various sources, it is useful to review the
overall photo-$z$ quality of the galaxies entering our sample through comparison with the
available spec-$z$.
If $\Delta z = z_\mrm{spec} - z_\mrm{phot}$, we can estimate the redshift accuracy from
$\sigma_{\Delta z/(1 + z_\mrm{spec})}$, using the normalized median absolute deviation
\citep[NMAD;][]{Hoaglin-83}, defined as
$1.48\times\mrm{median}\{|z_\mrm{spec} - z_\mrm{phot}|/(1 + z_\mrm{spec})\}$.
The NMAD is directly comparable to other works, which quote the quantity
$\mrm{rms}/(1 + z)$.
This dispersion estimate is robust with respect to ``catastrophic errors''
(i.e., objects with $|z_\mrm{spec} - z_\mrm{phot}|/(1 + z_\mrm{spec}) > 0.15$), whose
percentage is denoted by $\eta$.
Figure~\ref{fig:zs-vs-zp} (left panel) shows the comparison between $z_\mrm{spec}$ and
$z_\mrm{phot}$ for the 9\,664~galaxies (all fields) with apparent magnitude $m_\mrm{AB} < 24$ 
and $z_\mrm{spec} \leq 1$.
We obtain an accuracy of $\sigma_{\Delta z/(1 + z_\mrm{spec})} = 0.010$ for the photometric 
redshifts, with a percentage of ``catastrophic'' failures of 2.3\%.

For each field, all the available redshift data-sets were merged together: whenever available,
a ``reliable'' spectroscopic redshift value was used, while for the remainder of the
sample, photometric redshifts were instead adopted.

\section{Ancillary data-sets} \label{sec:anc}
In order to allow for a detailed characterisation of the selected galaxies, and in particular 
to obtain galaxy mass estimates from optical-to-NIR spectral energy distribution (SED) fits 
to the rest-frame $K$-corrected magnitudes, several multi-wavelength data-sets were also 
assembled, as detailed below.

\subsection{Optical, UV and Near-infrared surveys} \label{sec:OptNIR}

\subsubsection{COSMOS} \label{sec:kc-COSMOS}
The COSMOS photometric data were taken from the publicly available COSMOS Intermediate and 
Broad-Band Photometric catalogue\footnote%
{http://irsa.ipac.caltech.edu/data/COSMOS/gator\_docs/cosmos\_ib\_colDescriptions.html}, 
the UltraVISTA Survey Data Release~1 \citep{McCrack-AA12-UVista}\footnote{%
http://www.eso.org/sci/observing/phase3/data\_releases/ultravista\_dr1.html} and 
\emph{Spitzer} IRAC bands 1--2 (3.6 and 4.5\,$\mu$m) data from the S-COSMOS public 
catalogue\footnote%
{http://irsa.ipac.caltech.edu/data/COSMOS/gator\_docs/scosmos\_irac\_colDescriptions.html}
\citep{Sanders-ApJS07-SCOSMOS}.

The COSMOS catalogue comprises: ultraviolet data from GALEX and CFHT/$u^\ast$-band imaging; 
in the optical, the ACS/$I_{F814W}$-band data and the COSMOS-21 survey on Subaru
\citep{Tanig-ApJS07-COSMOS}, including six broad-band ($B_J$ , $V_J$ , $g^+$, $r^+$, $i^+$,
$z^+$), 12 medium-band (IA427, IA464, IA484, IA505, IA527, IA574, IA624, IA679, IA709,
IA738, IA767, IA827) and two narrow-band (NB711, NB816) filters.
Total magnitudes were obtained from the corresponding aperture values by applying the
recommended aperture corrections.
Only reliable photometric measurements were considered, i.e., data with bad pixel band mask
flags different than zero were discarded.
Recommended photometric offsets \citep{Ilbert-ApJ09-COSMOS} and galactic extinction
corrections \citep{Capak-ApJS07-COSMOS} were also applied.

In the near-infrared, we used the VIRCAM/$Y$-, $H$-, $J$-, and $K_s$-bands imaging data from
the UltraVISTA survey and \emph{Spitzer} IRAC bands 1--2
(3.6 and 4.5\,$\mu$m) data from the S-COSMOS.
\textsc{SExtractor} ``auto'' (or Kron-like) magnitudes were used for near-infrared UltraVISTA 
data, whereas IRAC channels 1--2 total magnitudes were calculated from the aperture-corrected
fluxes of the S-COSMOS data release.
Unreliable flux entries (e.g., close to bright stars or close to the image margins) were
discarded.

\subsubsection{AEGIS} \label{sec:kc-EGS}
For the AEGIS field, the following data were considered:
GALEX far-ultraviolet and near-ultraviolet data from the EGS multi-wavelength data-set\footnote%
{http://www.galex.caltech.edu/researcher/techdoc-ch5a.html} \citep{Davis-ApJ07-AEGIS},
HST/ACS $V_{F606W}$- and $I_{F814W}$-band photometry\footnote%
{http://aegis.ucolick.org/acs\_datasets.html} \citep{Lotz-APJ08-EGS},
CFH12K $B$-, $R$- and $I$-band photometry\footnote%
{http://deep.berkeley.edu/DR1/dr1.primer.html}
\citep{Coil-ApJ04-EGS}, CFHT Megacam $u$-, $g$-, $r$-, $i$- and $z$-band photometry%
\footnote{http://www3.cadc-ccda.hia-iha.nrc-cnrc.gc.ca/community/CFHTLS-SG/docs/cfhtls.html}
\citep{Gwyn-PASP08-MegaCam}, Palomar/WIRC $J$- and $K_s$-band data%
\footnote{http://www.astro.caltech.edu/AEGIS/} \citep{Bundy-ApJ06-EGS} and \emph{Spitzer}
IRAC bands 1--2 (3.6 and 4.5\,$\mu$m) data\footnote{http://www.cfa.harvard.edu/irac/egs/}
\citep{Barmby-ApJS08-EGS}. Together, these datasets provide up to 15 bands for 
spectral energy distribution analysis.

\subsubsection{GEMS} \label{sec:kc-GEMS}
GEMS sources were analysed using the optical multi-band photometry
from the COMBO-17 survey \citep{Wolf-AA04-COMBO,Wolf-AA08-COMBO}, which provides observations 
in 17 optical filters in the $365-914$\,nm wavelength range, obtained with the Wide Field Imager
at the MPG/ESO 2.2-m telescope\footnote%
{See {http://www.mpia.de/COMBO/combo\_filters.html} for a comprehensive description of the
filter set}. Aperture-corrected photon fluxes were considered in the analysis. 

\subsubsection{GOODS-S} \label{sec:kc-GOODSS}
For the GOODS-S sources we used a 12-band selection of the high
quality multi-wavelength data of GOODS-MUSIC\footnote%
{http://cdsarc.u-strasbg.fr/viz-bin/qcat?J/A+A/449/951}
\citep[Multi-wavelength Southern Infrared catalogue;][]{Grazian-AA06-MUSIC}.
This includes $U$-band data from the 2.2\,ESO and VLT/VIMOS, HST/ACS $B_{435}$, $V_{606}$,
$i_{775}$, and $z_{850}$ filters, the $JHK_s$ VLT data, as well as the 3.6 and 4.5\,$\mu$m
\emph{Spitzer} data.
MUSIC provides \textsc{SExtractor} ``auto'' (or Kron) magnitudes for the ACS images, whereas
for ground-based and \emph{Spitzer} data, reliable estimates of the total magnitudes were
retrieved through PSF-matching.

\subsubsection{GOODS-N} \label{sec:kc-GOODSN}
Optical photometric data for the GOODS-N field \citep{Giava-ApJ04-GOODS} were taken from the GOODS 
\texttt{r2.0z}\footnote{http://archive.stsci.edu/pub/hlsp/goods/v2/h\_goods\_v2.0\_rdm.html} 
ACS multi-band source catalogue  and, at longer wavelengths, from 
the WIRCam ultra-deep catalogue of GOODS-N \citep{Wang-ApJS10-GOODSNKs}.
We thus considered a 7-band catalogue, which comprises the $B_{F435W}$-, $V_{F606W}$-, 
$i_{F775W}$-, and $z_{F850LP}$-filter data, plus $K_s$- and IRAC\,1--2 bands 
(3.6 and 4.5\,$\mu$m).
\textsc{SExtractor} ``best'' and  ``auto'' magnitudes were chosen for the visible bands and
$K_s$ data, respectively.
The 3.6\,$\mu$m and 4.5\,$\mu$m fluxes derived by \citet{Wang-ApJS10-GOODSNKs} were instead
obtained by a deconvolution procedure based on a $K_s$ image prior, which is expected to
deliver highly accurate near-infrared and mid-infrared colours.

\subsection{Mid-IR surveys} \label{sec:MIR-data}
\emph{Spitzer} Space Telescope mid-infrared data were used to evaluate star-forming rates 
(SFRs). 
MIPS 24\,$\mu$m public catalogues are available for COSMOS (S-COSMOS GO2 and GO3 
releases\footnote{%
http://irsa.ipac.caltech.edu/data/COSMOS/gator\_docs/scosmos\_mips\_24\_go3\_colDescriptions.html}
\citealt{Frayer-AJ09-SCOSMOS}) and for AEGIS~\citep{Barro-APJS11-SFR}, while for 
GOODS-S/GEMS fields we used a catalogue of sources extracted from FIDEL~24 image (v0.5) of the
Extended~CDFS (6\,arcsec PSF matched photometry, X.Z.~Zheng, in preparation).

\subsection{X-ray catalogues} \label{sec:Xray-data}
X-ray surveys have targeted areas within the fields presented in our study.
Given the relevance of those data for the detection of AGN, we have also assembled the 
available public data-sets existing in these regions, namely:
\begin{itemize}
 \item the COSMOS-\emph{Chandra} catalogue \citep[C-COSMOS,][]{Elvis-ApJS09-CCosm}, and the
       AEGIS catalogue in the \emph{Chandra} field \citep[AEGIS-X,][]{Laird-ApJS09-AEGISX},
       with 1\,761 and~1\,325 X-ray point sources detected, respectively.
 \item The XMM-\emph{Newton}-COSMOS catalogue
       \citep[XMM-COSMOS,][]{Capp-AA09-XMM,Brusa-ApJ10-XMM}, which detected
       1\,887~unique sources over an area of 2\,deg$^2$.
 \item The catalogues of the 2\,Ms \emph{Chandra} Deep Field-North
       \citep[CDF-N,][]{Alex-AJ03-CDFN}, with a total of 503 detections, the 4\,Ms 
       \emph{Chandra} Deep Field-South \citep[CDF-S,][]{Xue-ApJS11-CDFS}, which contains 
       776~sources, and also the wider coverage 1\,Ms Extended \emph{Chandra} Deep Field-South
       \citep[E-CDF-S,][]{Lehmer-ApJS05-ECDFS}, with 795 sources.
\end{itemize}

\section {The final catalogue of bulgeless galaxies} \label{sec:BGcat}
After having assembled the various data-sets, we have obtained a homogeneous sample of 
rest-frame ultraviolet to NIR magnitudes for all galaxies by using the publicly  available 
\texttt{kcorrect} code (version v4\_2)\footnote{%
  available at http://howdy.physics.nyu.edu/index.php/Kcorrect.} 
described in \citet{Blanton-AJ07-Kcorr}.
\texttt{kcorrect} is designed to extract the most physically realisable SED by using linear 
combinations of five SEDs that are characteristic of physical states of galaxies, from 
intense starburst to quiescent objects.
Also, the implementation of the code makes it particularly suitable for modelling optical
and near-infrared observations in the redshift range $0 < z < 1.0$.
The galaxy templates are based on the \citet{Bruzual-MNRAS03-SES} stellar evolution synthesis
models and, together with rest-frame magnitudes, this tool also provides estimates of the
stellar mass-to-light ratio (see \S~\ref{sec:masses}).
\texttt{kcorrect} seeks the best fit to each source photometry using a linear combination
of five spectral templates $F(\lambda)$, which are, in turn,
derived by non-negative matrix factorisation (NMF) of 485 basis templates spanning grids
of metallicities ($Z = 0.0001 \sim 0.05$) and ages (1\,Myr $\sim$ 13.75\,Gyr), 
with three choices of dust models (no extinction, Milky Way-, and Small Magellanic Cloud-type 
extinction).
Thirty-five templates of line emission from the ionised gas
\citep[MAPPING-III,][]{Kewley-ApJ01-SBmodel} are also included in the model.
$K$-corrections were calculated for all entries of our sample (irrespective of $n$),
which have both $z \leq 1.0$ and at least five photometric detections.

We thus obtained rest-frame absolute magnitudes for all objects in our sample. 
For the objects lacking one or more photometric data-points, rest-frame absolute quantities
have been derived directly from the fitted SEDs.
We have then generated a homogeneous 11-filter data-set comprising: GALEX far-ultraviolet, 
near-ultraviolet, Johnson's $U$, $B$, $V$, $I$, SDSS $g$, $r$, $z$, Vircam\footnote{%
  VIRCAM is a wide field imager for survey mounted on the VISTA telescope. 
  See http://www.eso.org/sci/facilities/paranal/instruments/vircam/inst/ for more 
  information on the filter set.}
$J$ and $K_s$ bands.
The reduced $\chi^2$ of each SED fit is retained in the final catalogue.

\begin{sidewaysfigure}[tbh]
  \centering
  \includegraphics[angle=0,width=20cm]{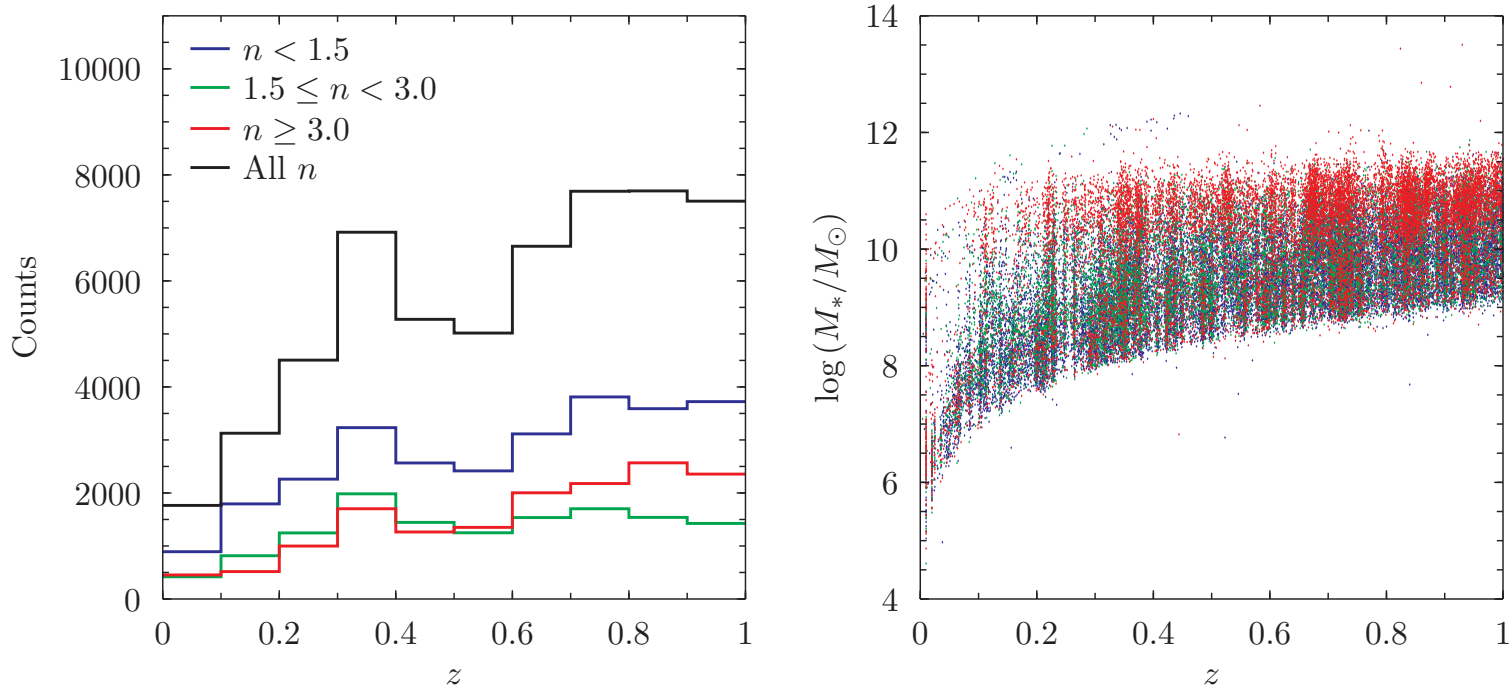}
  \caption{Left panel: Redshift distribution for the 39\,852 galaxies belonging to the 
           morphological catalogue defined in the present work.
           Different S\'ersic index $n$ ranges are indicated by colours: blue $n < 1.5$, 
           green $1.5\leq n < 3.0$, and red $n \geq 3.0$\@.
           Right panel: Distribution of galaxy stellar masses derived using \texttt{kcorrect} 
           (see \S~\ref{sec:BGcat}) as a function of redshift. 
           Blue, green, and red symbols label the three morphological types as in the 
           left panel.}
  \label{fig:z-dist}
\end{sidewaysfigure}

A definition of $n \leq 1.5$ was applied to select galaxies with little or no contribution
of the bulge to the light profile \citep[bulgeless galaxies;][]{Gadotti-MNRAS09-SDSS}.
As control samples we also selected galaxies with different morphological properties, adopting
the following criteria: $1.5 < n \leq 3.0$ corresponding to disc galaxies with an increasingly
prominent bulge component, and $n > 3$ for bulge-dominated galaxies.
These samples were used to test the morphological evolutionary scenarios discussed in recent
literature \citep[e.g.,][]{Panella-ApJ09-COSMOS,Oesch-ApJ10-COSMOS}.
A summary of the final sample composition is reported in Table~\ref{tab:n-summ}.
The redshift distribution of the galaxy sample including all the morphological types 
is shown in the left panel of Figure~\ref{fig:z-dist}. In the right panel we show the 
distribution of galaxy stellar masses as a function of redshift.

In summary, we have created a magnitude-limited ($m_\mrm{AB} < 24$, in the reddest available 
HST band, $\sim$7750 to 8500\,\AA), low-inclination, point-source free, $K$-corrected galaxy 
catalogue for five fields (COSMOS, AEGIS, GOODS-S, GEMS, and GOODS-N) with robust intermediate 
($0.4 \leq z \leq 1.0$) redshift measurements (photometric and spectroscopic) and reliable 
morphological data, containing a total of $19\,225$ bulgeless galaxies out of $39\,852$ 
sources.
A conservative estimate for the mass completeness of this catalogue, considering
the mass of the model SED used by \texttt{kcorrect} with the highest mass-to-light ratio, 
results in a minimum measurable stellar mass of $1.0\times10^{9}\,M_\odot$ at $z = 0.4$, 
and $4.3\times10^{10}\,M_\odot$ at $z = 1$.

\begin{figure}[tbh]
  \centering
  \includegraphics[angle=0,width=16cm]{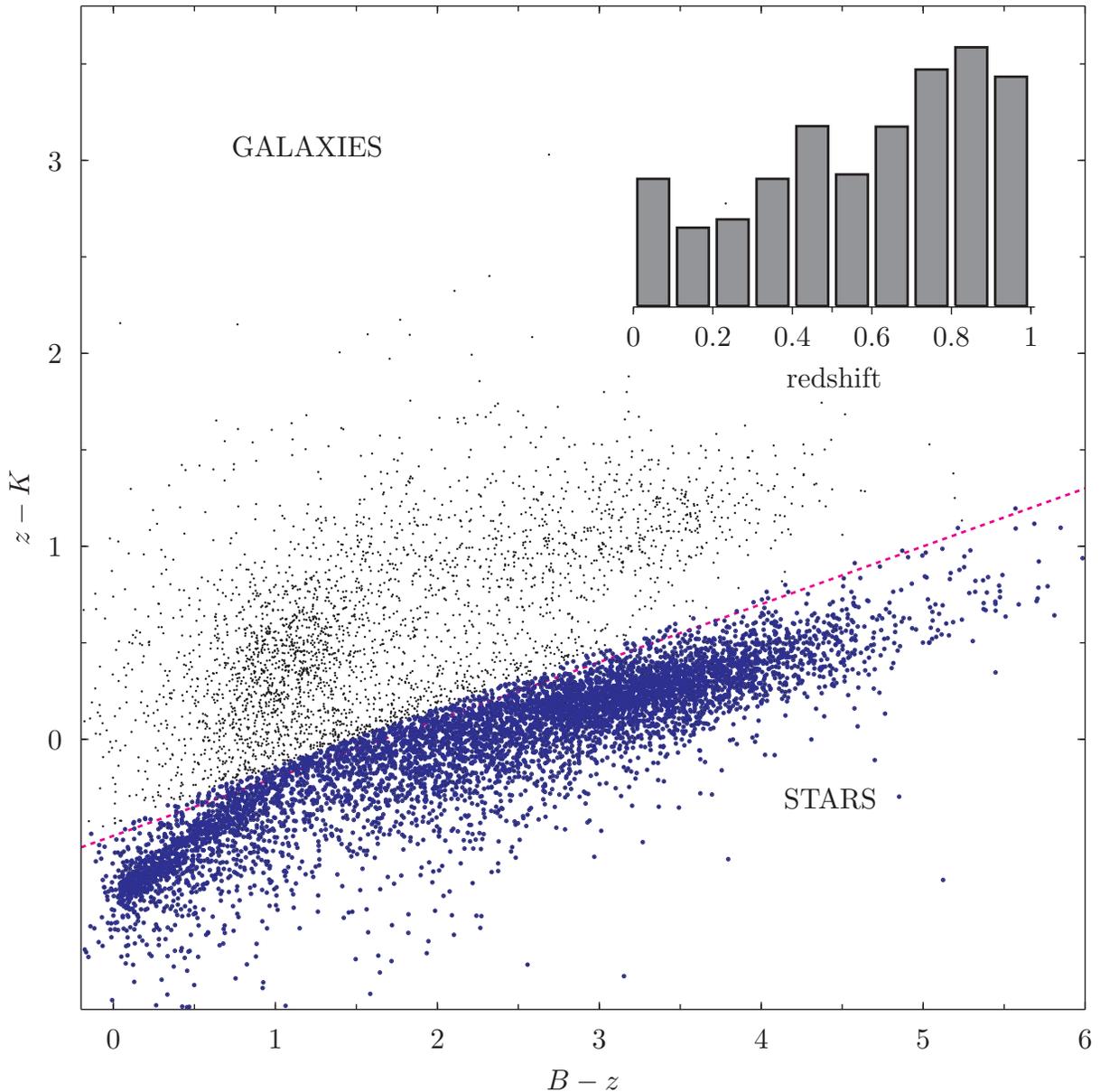}
  \caption{($z - K$) \emph{vs} ($B - z$) colour--colour diagram for the compact objects 
           selected as described in \S~\ref{sec:morphology} and removed from the final
           morphological catalogue.
           The diagonal dotted line defines the region $(z - K) < 0.3(B - z) - 0.5$, which 
           is preferentially occupied by stars \citep{Daddi-ApJ04-galcolor}.
           Stars selected using this criterion are labelled with blue asterisks.
           Black dots located above the dotted line identify the point-like galaxies having 
           little or no morphological information and that are rejected from the final sample.
           The inset shows the redshift distribution of this compact galaxy sample.}
  \label{fig:point-sou}
\end{figure}

As mentioned in \S~\ref{sec:morphology}, we adopted a filtering criterion to exclude point
sources from the final galaxy sample in order to reduce the contamination produced by 
spurious morphological classification.
This led us to reject 9\,049 sources in the redshift interval $0.4\leq z \leq 1.0$, 
corresponding to 23\% of the sample size.
Due to the lack of reliable morphology, these missing objects might potentially alter the 
content of the catalogue in a not well predictable way.
We thus inspected the point-source sample to asses its nature and to estimate the magnitude 
of the potential biases produced in the galaxy-type class analysis.
Figure~\ref{fig:point-sou} shows the $B -z$ versus $z - K$ colour--colour plot of 9\,979 
compact source at $z\leq 1.0$.
In the $BzK$ plot, stars are segregated in the region defined by the relation
$(z - K) < 0.3(B - z) - 0.5$ \cite[e.g.,][]{Daddi-ApJ04-galcolor} and can be efficiently 
separated from galaxies.
Blue symbols in Figure~\ref{fig:point-sou} label the 7\,015 stars that can be identified using
this method, whereas the black dots located in the upper part of the $BzK$ plane (2\,964 
objects, $\sim$30\%) are likely to be small size galaxies that we ultimately discarded from 
our morphological catalogue.
This test indicates that the adopted point-source rejection criterion led at most to a 7\% 
decrement of the magnitude-limited galaxy sample in the interval $0.4\leq z \leq 1.0$\@.
The neglected population shows only a mild redshift dependence (see the inset of
Figure~\ref{fig:point-sou}), thus we are thus confident that the main results of this work 
are not significantly affected.

\begin{figure}
  \includegraphics[bb=90 50 770 690, width=\textwidth]{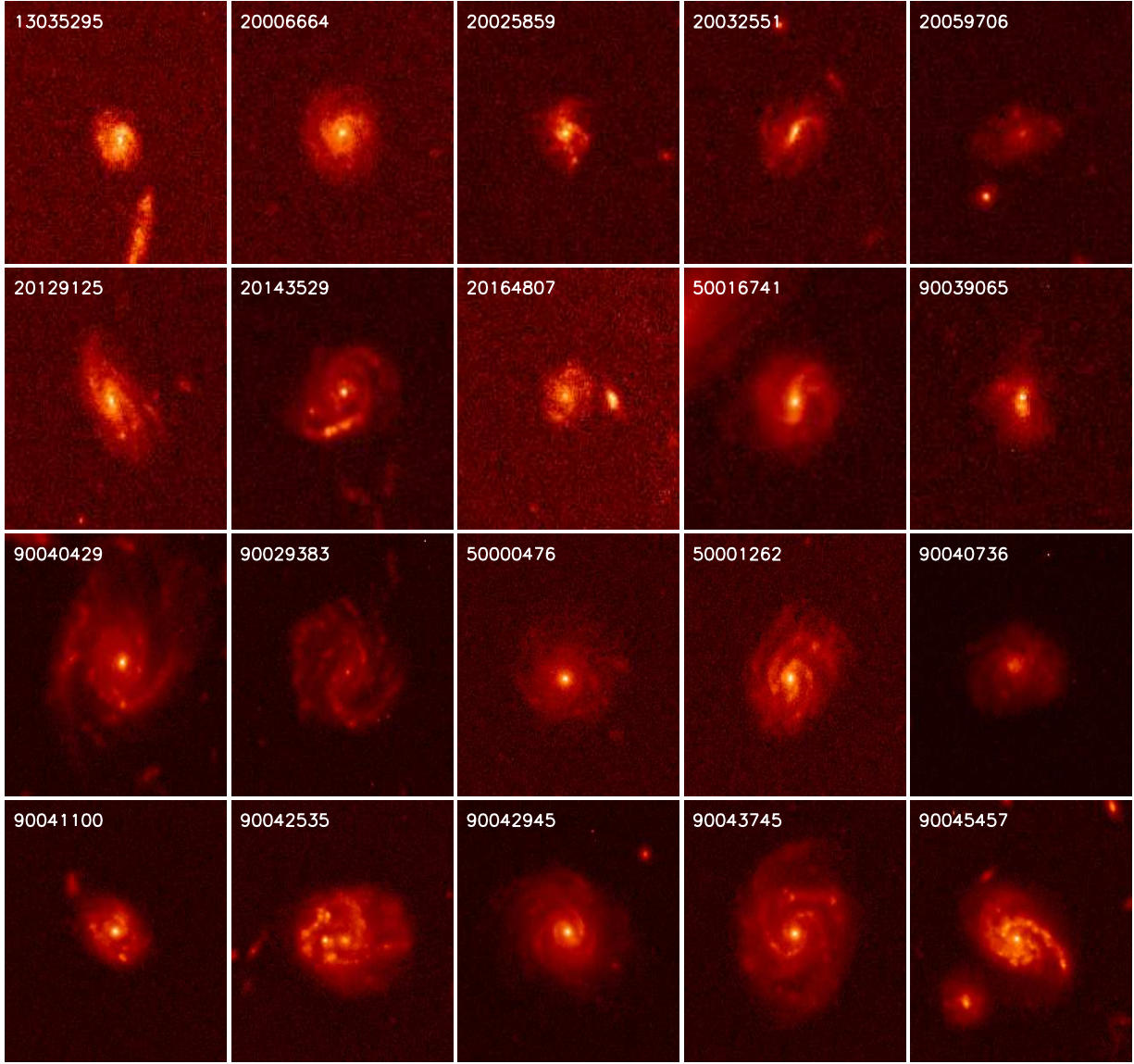}
  \caption{HST/ACS sample images of twenty selected bulgeless galaxies spanning the
           $0.4 \leq z \leq 1.0$ redshift range and $6\times 10^9 < M/M_\odot < 2\times10^{11}$ 
           stellar mass interval.
           The first two upper rows show $160\times160$ pixel cutouts, while the third and 
           fourth rows show $256\times256$ pixel cutouts.
          }
  \label{fig:postage-stamps}
\end{figure}

As an example of the catalogue content, we show in Figure~\ref{fig:postage-stamps} sample
cutouts of selected bulgeless galaxies spanning the full redshift range and the
$6\times 10^9 < M/M_\odot < 2\times 10^{11}$ stellar mass interval.
The catalogue of the absolute magnitudes, including morphological information and stellar
mass estimates analysed in this work is made publicly accessible at the following URL:  
http://www.oal.ul.pt/$\sim$jafonso/Bulgeless

\subsection{Reliability of the morphological classification} \label{sec:morph-rel}
One of the goals of this work is the selection of a reliable sample of intermediate redshift
galaxies having very little or no contribution from a bulge to their light profile.
The sample was homogeneously assembled from the imaging data of four different
\textit{HST}/ACS fields, and the morphological classification relied entirely on an automated
single-component parametric-fit method.
Due to the large content of the sample ($\sim$40\,000 objects), no extensive visual 
confirmation of the estimated morphology is feasible.
Nevertheless, it is important to assess the reliability of the bulgeless classification 
of our sample through comparison with some independent analysis
and a careful consideration of the different sources of uncertainty.

\begin{figure}[tbh]
  \centering
  \includegraphics[angle=0,width=14cm]{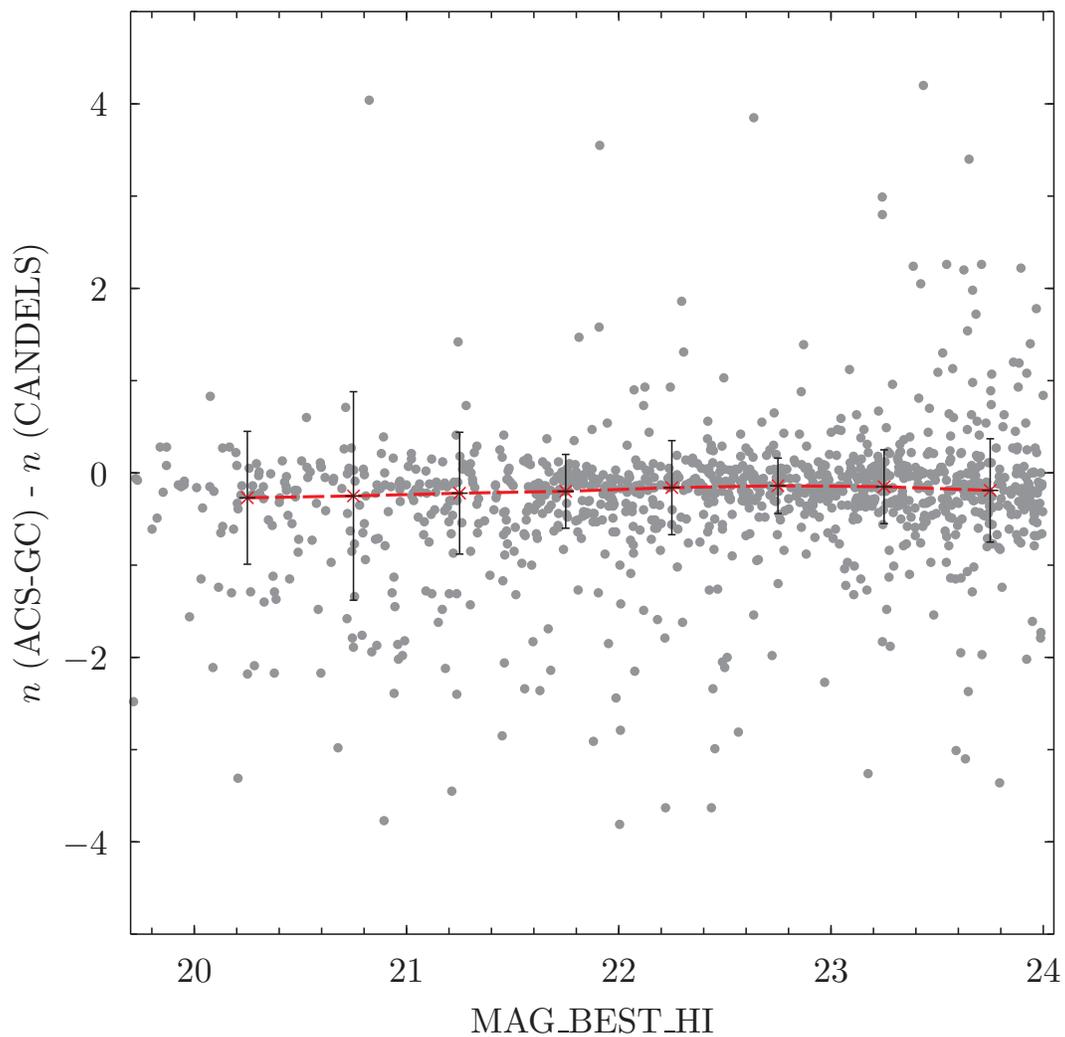}
  \caption{Comparison between \textsc{GALFIT} generated S\'ersic index $n$ in GOODS-S 
           sources using ACS-GC $z_{F850LP}$ and CANDELS $H_{160}$ bands as a function of 
           the apparent $z$ magnitude (all morphological types).
           Grey dots indicate $\Delta n$ deviations. 
           The red line illustrates the trend of the median taken over 0.5\,Mag bins. 
           Error bars mark the central 50\textit{th} percentile.}
  \label{fig:n-check}
\end{figure}

First of all, to assess the wavelength dependence of galaxy morphology, we cross-checked our 
ACS-GC \textsc{Galfit} classification with the results of a similar structural analysis 
performed by \citet{Wuyts-ApJ11-CANDELS} on GOODS-S sources spanning 
$0.1\lesssim z \lesssim 2.5$ using CANDELS $H_{160}$ data.
The results of comparison are shown in Figure~\ref{fig:n-check}.
In spite of the large rest-frame wavelength difference probed by the two surveys, the 
comparison between the two sets of S\'ersic $n$ index is fair, and the systematic 
discrepancies are limited to a few percent over the entire magnitude interval covered by 
our sample.

Secondly, we evaluated the overall reliability of our \textsc{Galfit}-based morphologies
through the comparison to a method based on different classification criteria.
In \citet{Scarlata-ApJS07-ZEST}, a sample of 56\,000 galaxies in COSMOS was the 
subject of a detailed morphological study, adopting the Zurich Estimator of Structural Types 
(ZEST), a sofisticated algorithm that uses a combination of non-parametric and parametric 
quantification of galaxy structure. 
Its classification scheme comprises type~1 (spheroids with no visible disc), type~2 
(disc galaxies), and type 3~(irregular) galaxies. 
A ``bulgeness'' index is also provided for type~2 disc galaxies, which coarsely correlates 
with bulge-to-disc ratio: it ranges from pure disk galaxies (type~2.3) to bulge-dominated 
discs (type~2.0)\@.
Since their analysis was also performed on the ACS $F814W$ COSMOS images, the comparison 
with our S\'ersic-index based morphological classification presented here is straightforward. 

We cross-matched the publicly available Zurich morphology catalogue\footnote%
{http://irsa.ipac.caltech.edu/data/COSMOS/datasets.html}
with our COSMOS sub-sample, obtaining 30\,104 galaxies with both a \textsc{Galfit}-derived
S\'ersic index and a ZEST classification.
The numerical results of this comparison split by ZEST classes are shown in 
Table~\ref{tab:zurich-comp}. 
We found that 13\,073 out of 13\,680 galaxies (96\%) of our bulgeless ($n \leq 1.5$) sample 
are classified as having little or no bulge contribution in ZEST (types 2.2, 2.3, and 3.0), 
570 (4\%) are classified as intermediate disc galaxies (type 2.1),
whereas a mere 37 objects (0.2\%) are considered as bulge-dominated in ZEST (type 1.0 and 2.0).
This comparison indicates that our bulgeless sample might have a 5\% contamination by 
misclassified bulge-dominated objects.
This error should be summed in quadrature to the uncertainty due to the point-source 
rejection ($\sim$ 7\% see \S~\ref{sec:BGcat}) to yield a conservative overall uncertainty of 
8\% in the morphological classification of the bulgeless sample.

Not surprisingly, the agreement between our ``bulgy'' defined sample and the ZEST 
spheroidal-type classification is only moderate.
ZEST type~1.0 and~2.0 collect 4\,562 out of 9\,494 objects corresponding to 48\%\@, while 
2\,760 further galaxies (29\%) are classifies as intermediate type 2.1 (disc with prominent 
bulges).
It should be noted that the ZEST classes~1.0 and~2.0 mostly include classical spheroidal 
objects whose surface brightness is best reproduced by De Vaucouleurs' profile with $n = 4$\@.
On the other hand, the $n \geq 3$ criterion we adopted to define the ``bulgy'' sample is 
broader and it is expected to include more objects having a significant disc component.
Given that this study is focused on bulgeless object, this comparison provides a confirmation
of the methodology used, i.e., that the S\'ersic index criterion has a good reliability in 
selecting galaxies with little bulge contribution.

\subsection{Accuracy of mass estimates} \label{sec:masses}
\begin{sidewaysfigure}
  \centering
  \includegraphics[angle=0,width=20cm]{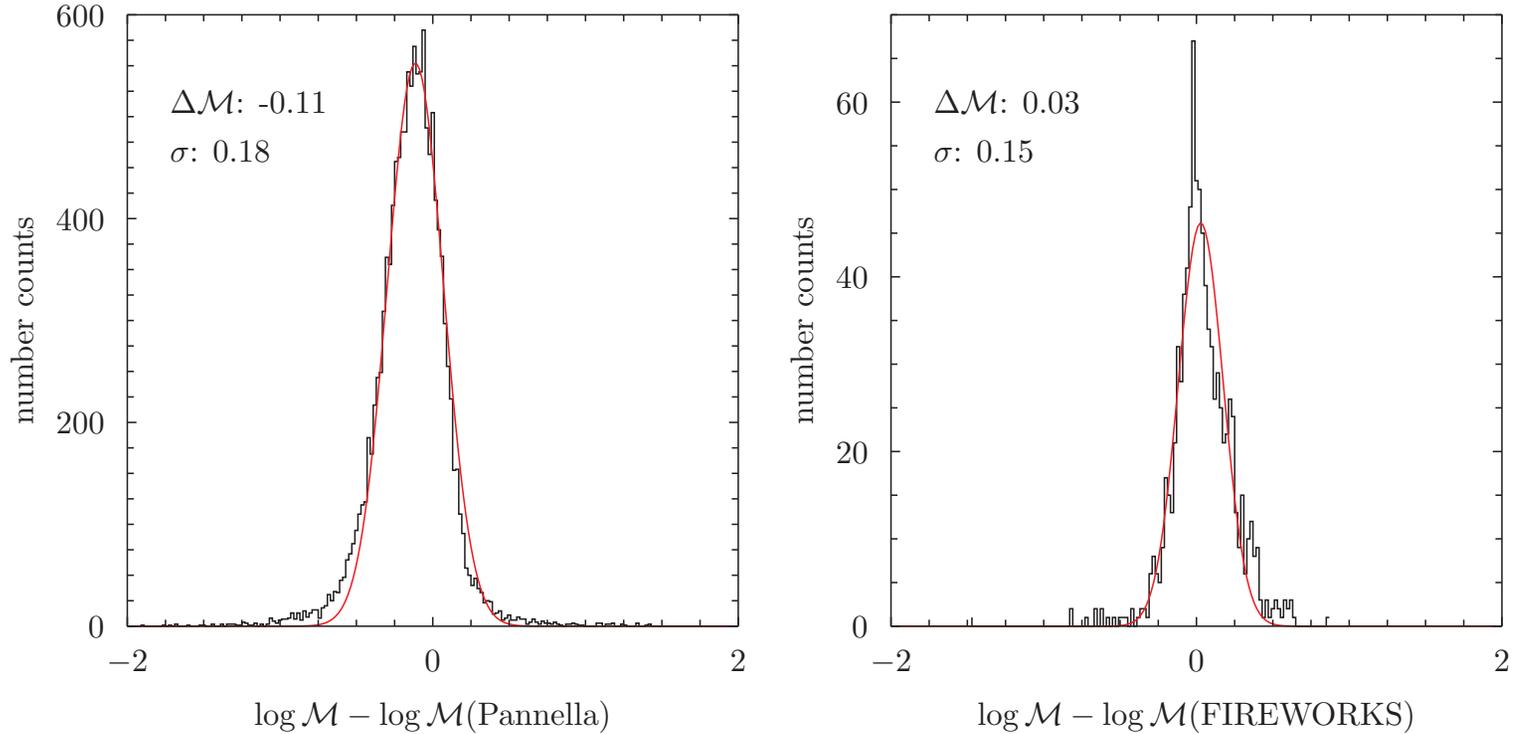}
  \caption{Black line solid histograms show the difference between the stellar masses
           computed using \texttt{kcorrect} on our ACS-GC derived sample and those
           calculated by \citet{Panella-ApJ09-COSMOS} on COSMOS galaxies (left panel),
           or by the FIREWORKS teams \citep{March-ApJ09-SMF} on GOODS-S sources
           (right panel).
           The red lines are Gaussian distributions with $\sigma = 0.18$ (left panel)
           and $\sigma = 0.15$ (right panel)}.
  \label{fig:mass-comp}
\end{sidewaysfigure}

The overall view in the literature
\citep[e.g.,][]{Bell-ApJ01-MLR,Drory-ApJ04-galmass,Muzz-ApJ09-NIR-Kgal,Hain-ApJ11-mass-SMGs} 
is that uncertainties in the various assumptions (i.e., redshift, initial mass function, 
stellar population synthesis model (SPS), star-formation history, dust extinction etc.) 
imply an overall conservative uncertainty of up to a $\sim$2--3 factor ($\sim$0.2--0.5\,dex) 
for each individual mass estimate.
Therefore, we compared our \texttt{kcorrect} results with those independently presented by
\citet{Panella-ApJ09-COSMOS} for COSMOS and by the FIREWORKS team \citep{March-ApJ09-SMF} 
for GOODS-S.
The comparison of the stellar masses is shown in Figure~\ref{fig:mass-comp}.
All three methods employ the \citet{Bruzual-MNRAS03-SES} SPS model.
The FIREWORKS mass determinations closely match our \texttt{kcorrect}-derived 
values, whereas those of \citet{Panella-ApJ09-COSMOS} are 0.11\,dex larger, on average, 
which is within the range of uncertainties referred by these authors.
In both cases, the scatter is less than 0.2\,dex, still within the overall uncertainty
referred above.

\section{Bulgeless galaxy properties} \label{sec:BGprop}

\subsection{Colour evolution at intermediate redshift} \label{sec:colours}
\begin{figure}[tbh]
  \centering
  \includegraphics[angle=0,width=16cm]{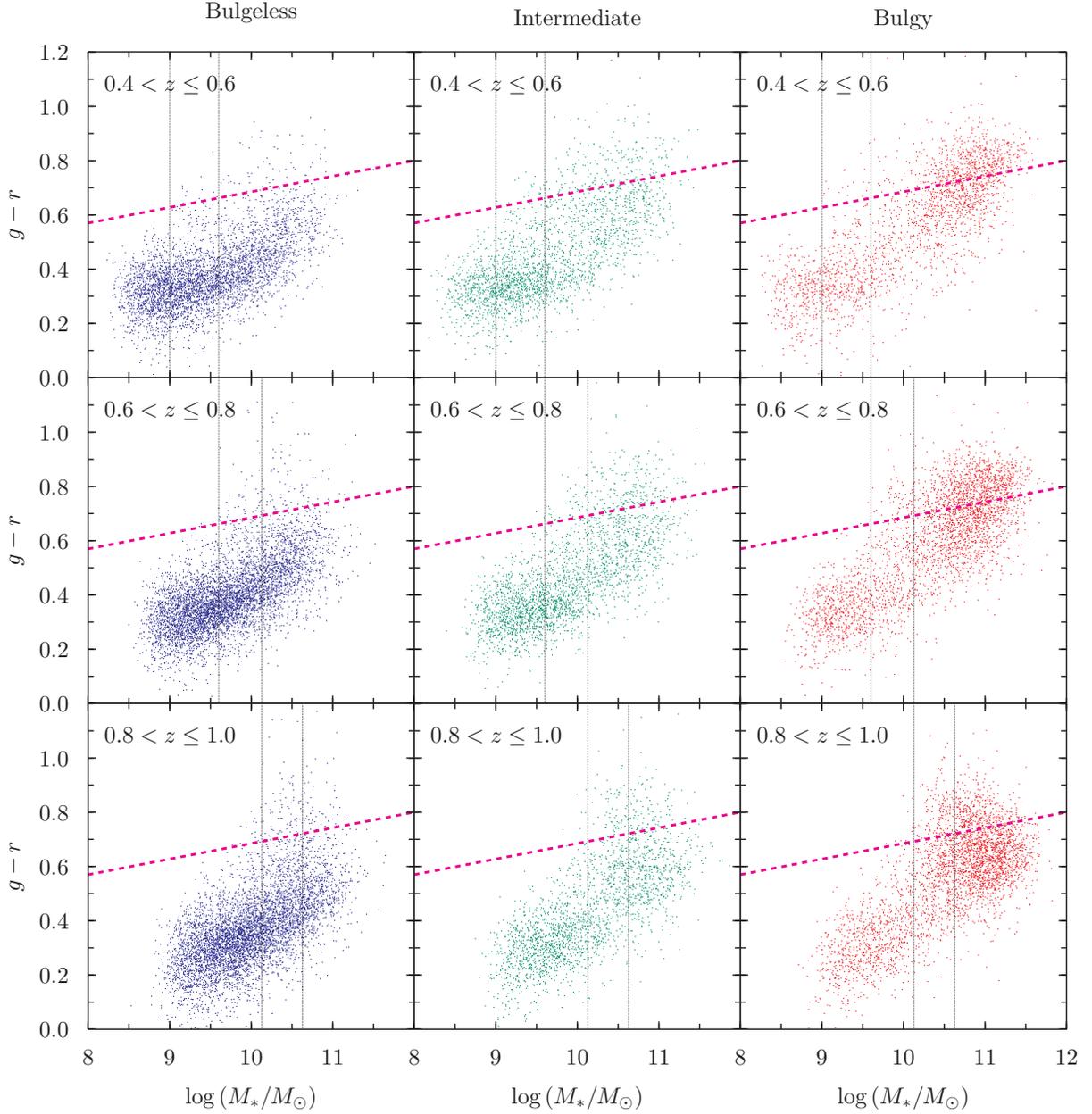}
  \caption{Rest-frame $g-r$ colour--mass distribution for the final, low-inclination,
           magnitude-limited catalogue.
           The galaxies are separated into three sub-samples: those lacking significant
           bulge (S\'ersic index $n < 1.5$; \emph{left}), those with an intermediate 
           disc-to-bulge ratio ($1.5\leq n < 3.0$; \emph{central}), and those which are 
           bulge-dominated ($n \geq 3.0$; \emph{right}.
           Also, each sub-sample has been divided in three redshift bins: $0.4\leq z < 0.6$
           (\emph{top}), $0.6\leq z < 0.8$ (\emph{middle}), and $0.8\leq z <1.0$ (\emph{bottom}).
           The dashed purple lines indicates the red sequence locus (see text), the two vertical
           dotted lines show the mass completeness limit at the upper and lower edges of each
           redshift bin.}
  \label{fig:g-r_vs_mass}
\end{figure}

Figure~\ref{fig:g-r_vs_mass} shows the rest-frame $g-r$ colour as a function of mass, 
redshift, and morphological classification.
Three redshift ranges are considered (row-wise): $0.4\leq z \leq 0.6$, $0.6\leq z \leq 0.8$, 
and $0.8\leq z \leq 1.0$.
The oblique dashed line represents the red sequence, defined as by \citet{Bell-ApJ08-BLess}
in the SDSS sample: $(g - r) > 0.57 + 0.0575\log(M_\ast/10^8 M_\odot)$\@.

The \textit{left-hand} panels show the colour-mass properties of the bulgeless sample, which 
is the goal of the present investigation.
The bulgeless population at all redshifts exhibits a striking segregation at bluer colours
compared to the other morphological types, with few objects extending beyond the red 
sequence line.
The properties of the bulgeless sample are discussed in more details in \S~\ref{sec:red-seq} 
and \S~\ref{sec:massive}.

Bulge-dominated galaxies ($n \geq 3.0$, \textit{right panels} of Figure~\ref{fig:g-r_vs_mass}),
are mostly located in two regions in the colour--mass diagram, with a clear segregation in mass.
Massive objects ($M_\ast > 5\times 10^{10}\,M_\odot$) are located on the red sequence
at all redshifts. 
They do not show an appreciable evolution in mass (the fraction of galaxies
with $\log(M_\star/M_\odot) > 10.5$ remains approximately constant at $\sim 0.9$ irrespective 
of redshift bin) and move slightly towards redder colours with cosmic time, an effect of the 
ageing of the stellar population.
On the lower mass end of the distribution ($M_\ast < 10^{10}\,M_\odot$), there exists a less
numerous, but still appreciable, population of bulge-dominated galaxies, reaching the blue
cloud.
These could be low-mass bulge-dominated galaxies hosted in low-density groups, where they are 
able to sustain appreciable star formation, possibly via ``wet merger'' events.
Indeed, early-type galaxies with young stellar populations at low-$z$ are predominantly 
low-velocity dispersion systems and tend to live in lower density regions 
\citep{Suh-ApJS10-SDSS,Thomas-MNRAS10-SelReg}.
Moreover, some Blue Compact Dwarfs (BCDs), particularly those classified as nE types in 
\citet{Loose-MitAG86-Morph}, also show surface brightness profiles fitted by a 
de~Vaucouleurs $R^{\frac{1}{4}}$ profile 
\citep{Loose-ApJ86-SBs,Kunth-AA88-BluGal,Cairos-ApJS01-BluGal}.
Assessing the proper nature of these objects is beyond the scope of this study, however we 
refer to \citet{Panella-ApJ09-COSMOS} for a more detailed analysis of the environment of 
blue early-type galaxies at intermediate redshift.

The intermediate--type population (with $1.5\leq n < 3.0$, \textit{central panels} of 
Figure~\ref{fig:g-r_vs_mass}) is mainly composed of disc or spiral galaxies having a 
prominent bulge.
In the high-redshift end of the diagram they extend from the red sequence to the 
blue cloud region, whereas at lower-$z$ they are more concentrated in the blue/low-mass 
region. This agrees with the conclusions of previous studies on galaxy 
evolution \citep{Panella-ApJ09-COSMOS,Oesch-ApJ10-COSMOS}, although these authors have used 
a different morphological classification scheme.

\begin{figure}[tbh]
  \centering
  \includegraphics[angle=0,width=15cm]{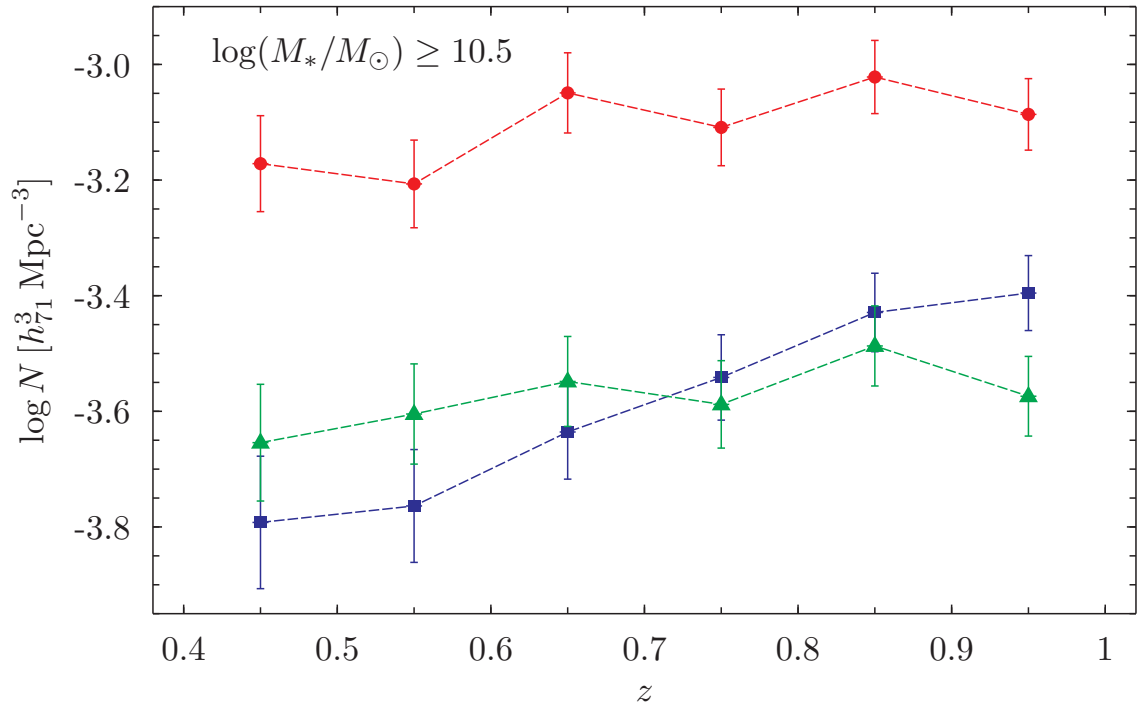}
  \caption{Number densities as a function of redshift and morphology for galaxy masses
           above the completeness limit.
           Symbols are coloured according to the morphological classification (blue: bulgeless,
           green, intermediate, red: bulgy).
           The error bars were calculated by adding in quadrature the following contributions: 
           ($i$) cosmic variance \citep{Moster-ApJ11-CosVar}, ($ii$) the square root of the 
           number of galaxies per redshift bin, and ($iii$) the 0.5\,dex error estimated in 
           the stellar mass determination.
          }
  \label{fig:n-dens}
\end{figure}

Concerning the bulgeless galaxies, if we focus on the high-mass end 
(e.g., for $\log(M_\star/M_\odot) > 10.5$, where our sample is complete at all $z$), one
can see a decreasing number of objects from $z\sim 1$ to the lowest redshift bin.
To take properly into account the survey volume effects, we evaluated the number density 
evolution of each morphological type for stellar masses $\log(M_\star/M_\odot) > 10.5$. 
The result is illustrated in Figure~\ref{fig:n-dens}, where it is clear that for the 
high-mass range considered, bulge-dominated systems constitute the majority of the galaxy 
population in the redshift range $0.4\leq z \leq 1.0$.

Looking at the evolution of the different morphological types, one can indeed see that, 
contrary to the number densities of massive bulgy and intermediate galaxies, which do not 
evolve significantly between $z\sim 0.4$ and $z\sim 1$, the bulgeless galaxy population 
shows a significant decrease in their number density as cosmic time increases.
The fact that early-type objects, in the selected mass range, show little or no evolution 
is in agreement with the most massive E/S0 galaxies being already in place at $z \sim 1$ and 
evolve only weakly since then \citep{Collins-Nat09-Massive}.
On the other hand, the observed disappearance of massive bulgeless galaxies at later cosmic 
times necessarily implies a process that accompanies the growth of their stellar masses and 
tends to displace them to an early-type morphological bin through the formation of a 
classical bulges.

\citet{Oesch-ApJ10-COSMOS} attempted to quantify this observed trend assuming the
morphological transformation of galaxies as due exclusively to merger events.
Their simulations were unable to reasonably fit the observed data for all morphological 
types simultaneously, even including small accretion events in the model.
Most notably, all the models over-predict the mass fraction of bulgeless disc galaxies as 
cosmic time proceeds. Thus, it is likely that these objects undergo a bulge-building process 
driven by disc dynamical instability, possibly triggered by secular accretion activity 
\citep{KeK-ARAA04-bulges}.

As a final remark, we remind that the possible bias introduced by the point-source rejection 
is estimated to be weak and scarcely redshift dependent (see discussion in \S~\ref{sec:BGcat}),
so that its effects are expected to be within the error bars shown in Figure~\ref{fig:n-dens}.

\subsection{Red bulgeless galaxies: quiescent or dusty?} \label{sec:red-seq}
Recent studies \citep{Bell-ApJ08-BLess,Bell-ApJ12-CANDELS} demonstrated that the S\'ersic 
index, among other galaxy quantities, correlates best with the lack of star-formation 
activity.
At $z < 0.05$, large Sers\'ic indices correlate extremely well with quiescence, as shown
by \citet{Bell-ApJ08-BLess} using a SDSS sample. It is argued that ``genuine'' bulgeless 
quiescent galaxies could be either ($i$) 
satellite galaxies in high-mass halos, whose gas is stripped in a deep potential well, or 
($ii$) possibly a result of incorrect morphological classification, given that many of them 
show a hint of a bulge once visual examination is performed.
Quenched bulgeless galaxies are rather rare up to high redshift 
\citep[$z\sim 2.2$,][]{Bell-ApJ12-CANDELS}, and only a few pure disc, quiescent 
systems are reported in the literature at $z\gtrsim 1$ 
\citep[e.g.,][]{McGrath-ApJ08-Morph,vdW-ApJ11-diskgal,Bundy-ApJ10-diskgal}.

\begin{sidewaysfigure}
  \centering
  \includegraphics[angle=0,width=20cm]{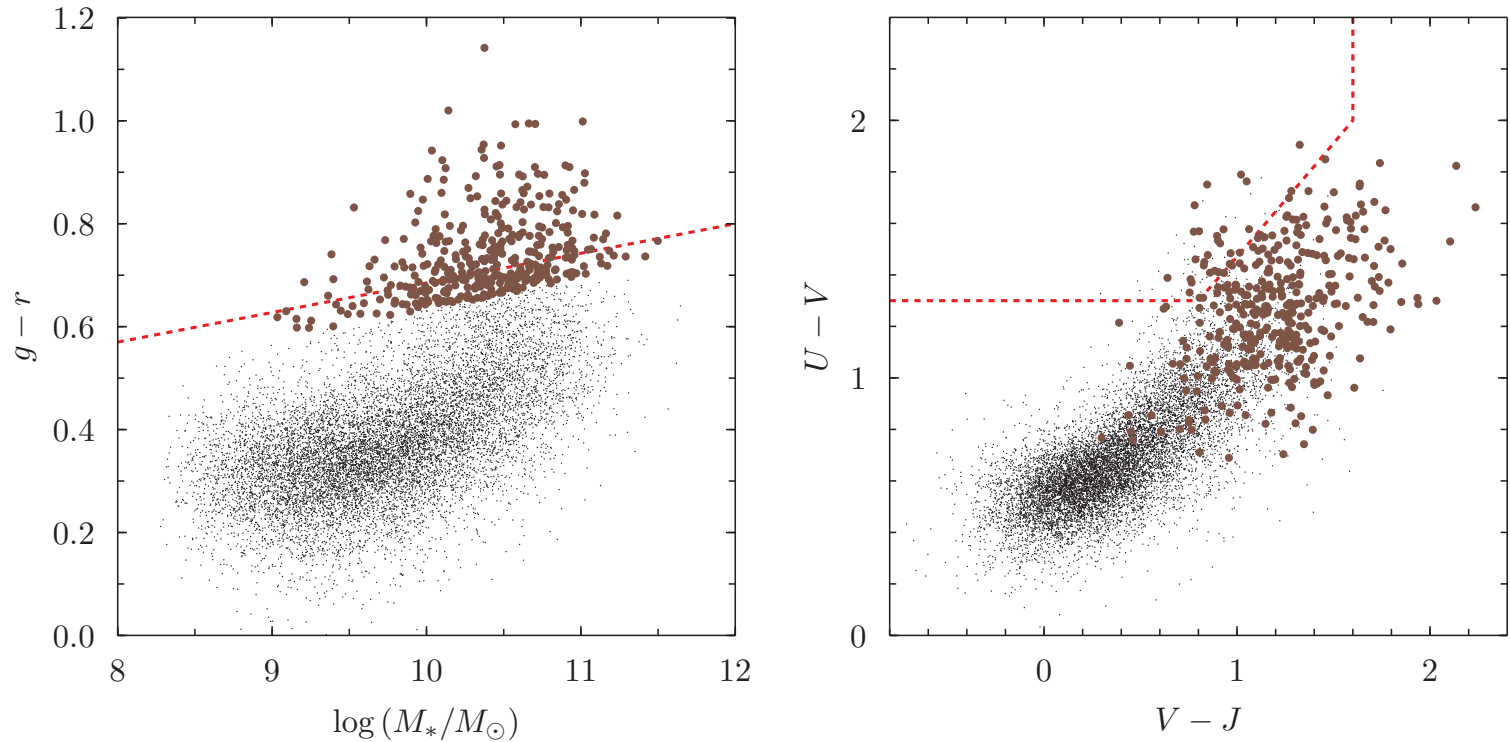}
  \caption{Left panel: Selection of the red sequence bulgeless galaxies (orange dots) in
           the $g - r$ colour--mass plane ($0.4 \leq z \leq 1.0$).
           The cut has been applied 0.05\,mag below the red sequence definition
           line described in the text (dashed line).
           Right panel: Rest-frame $U-V$ as a function of $V-J$ colour for the same
           bulgeless galaxy sample.
           The rest-frame colours cuts defined in \citet{Williams-ApJ09-quiegal} as a separation 
           between dusty star-forming and red quiescent galaxies are indicated as dashed lines.
           According to this plot, the vast majority of red bulgeless galaxies are red because of 
           the dust and not because of an older stellar population.}
  \label{fig:red-seq}
\end{sidewaysfigure}

As discussed in \S~\ref{sec:colours}, it is known that the red sequence is mostly 
composed of galaxies with elliptical morphology and prominent bulges 
\citep[e.g.,][]{Cassata-ApJS07-COSMOS}. 
Nonetheless, it also hosts a small fraction of dusty, late-type, reddened star-forming 
objects \citep{Williams-ApJ09-quiegal} and Spi/Irr galaxies, with quenched star-formation
\citep[see e.g.,][and references therein]{Bell-ApJ12-CANDELS}.
Indeed, figure~\ref{fig:red-seq} (\textit{left panel}) shows that a significant fraction of 
the bulgeless population is located in the red sequence region of the colour--mass diagram.
Finding red bulgeless galaxies gives evidence for an intriguing population of late-type 
objects -- their optically red nature may result from substantial dust obscuration but it 
could also indicate that they have ceased forming stars, thus revealing a mechanism that 
induces quiescence without significantly altering the morphology.

A way to break the degeneracy between dusty star-forming and red quiescent galaxies
is the use of rest-frame near-infrared--optical colours $U-V/V-J$
\citep{Wuyts-ApJ07-IRAC,Williams-ApJ09-quiegal}.
Essentially, the $U-V$ colour provides a proxy for unobscured star-formation activity,
whereas $V-J$ helps in highlighting dust-free quiescent galaxies that exhibit a bluer
colour than dusty objects, and thus occupy a different \emph{locus} in the $UVJ$ plane.
The relevant plot is shown in the \textit{right panel} of Figure~\ref{fig:red-seq}.
The superimposed dashed lines are the redshift-dependent selection criteria for
quiescent galaxies, described by \citet{Williams-ApJ09-quiegal}.
To discriminate between quiescency and dust obscuration, we highlighted in the $UVJ$ plane 
the sub-sample of red bulgeless lying down to 0.05\,mag below the red sequence definition 
line indicated in \S~\ref{sec:colours}.
The plot indicates that most of the bulgeless galaxies that are red in the $g-r$/mass 
diagram owe their optical colour to dust instead of an old stellar population, with only a 
few objects located in the region of the diagram corresponding to quiescent galaxies.

\subsection{Massive bulgeless galaxies} \label{sec:massive}
Figures \ref{fig:g-r_vs_mass} and \ref{fig:n-dens} show that our sample includes a 
non-negligible fraction of massive [$\log(M/M_\odot) > 10.5$] bulgeless galaxies which 
contribute to $\sim$ 30\% of the total galaxy population number density at $z\gtrsim 0.7$.
Overall, we find~2\,339 out of~19\,233 bulgeless galaxies (12\%) 
with $\log(M/M_\odot) > 10.5$, and~1.5\% of the sample consists of very massive systems 
with $\log(M/M_\odot) > 11$\@.

The existence of such massive bulgeless galaxies challenges the current picture of galaxy 
formation, because, in principle, they require a hierarchical growth to reach such large 
masses. 
It is not entirely clear how they can increase their masses through mergers without 
destroying their stellar discs and forming a classical bulge \citep{Toth-ApJ92-disks}.
However, it was shown that gas-rich mergers can produce a large disc instead of a spheroidal 
system \citep{Spring-ApJ05-spirals,Robert-ApJ06-mergers}. 
Discs can survive or rapidly regrow after a merging event provided that the gas fraction 
in the discs of the progenitors is high 
\citep{Robert-ApJ06-mergers,Robert-ApJ08-disks,Hopk-ApJ09-discs}.
At $z\sim 1$ massive bulgeless galaxies are less numerous compared to early-type galaxies, 
but are more abundant than intermediate-type at similar mass range (Figure \ref{fig:n-dens}). 
Their number density decreases with redshift, reducing by a factor of 2 from $z \sim 1$ 
to~0.4, and they are rare in the local Universe \citep{Silk-RAA12-galaxies}.
This trend is also found in similar studies of galactic morphological evolution within
the same redshift range \citep{Panella-ApJ09-COSMOS,Oesch-ApJ10-COSMOS}.
The gradual disappearance of very massive bulgeless galaxies at recent times might imply
a process that accompanies the growth of their stellar masses and tends to transform them
into an earlier type morphology through the formation of a classical bulge.

To explore this possibility we have used \emph{Spitzer} and \emph{Galex} data to derive both 
ultraviolet and infrared star forming rates (SFRs), to investigate the star-forming properties 
of the most massive galaxies in our sample.
As mentioned in section \S~\ref{sec:MIR-data}, \emph{Spitzer} Space Telescope observations 
are available for the COSMOS, AEGIS and GEMS/GOODS-S fields. Overall, 5\,574 sources (29\%) 
out of the entire bulgeless galaxy sample have a good photometric measurement at 24\,$\mu$m. 
Interestingly, $\sim$80\% (228/296) of the very massive ($M\ge10^{11}\,M_\odot$) bulgeless 
sample has a 24\,$\mu$m measurement.
The total infrared luminosity, $L_\mrm{IR}$ (8-1000\,$\mu$m), is determined following the 
procedure described in \citet{Caputi-ApJ08-MIPS}. 
Starting from the 24$\mu$m flux densities, we used the SED templates of NGC\,3351 to 
reproduce the corresponding rest-frame luminosity and to derive the integrated $L_\mrm{IR}$ 
of that template. 
Infrared and ultraviolet SFRs were determined from $L_\mrm{IR}$ and $L_\mrm{NUV}$ rest-frame 
luminosities, and added together to obtain an estimate of the total SFR for each galaxy.

\begin{sidewaysfigure}
  \centering
  \includegraphics[angle=0,width=20cm]{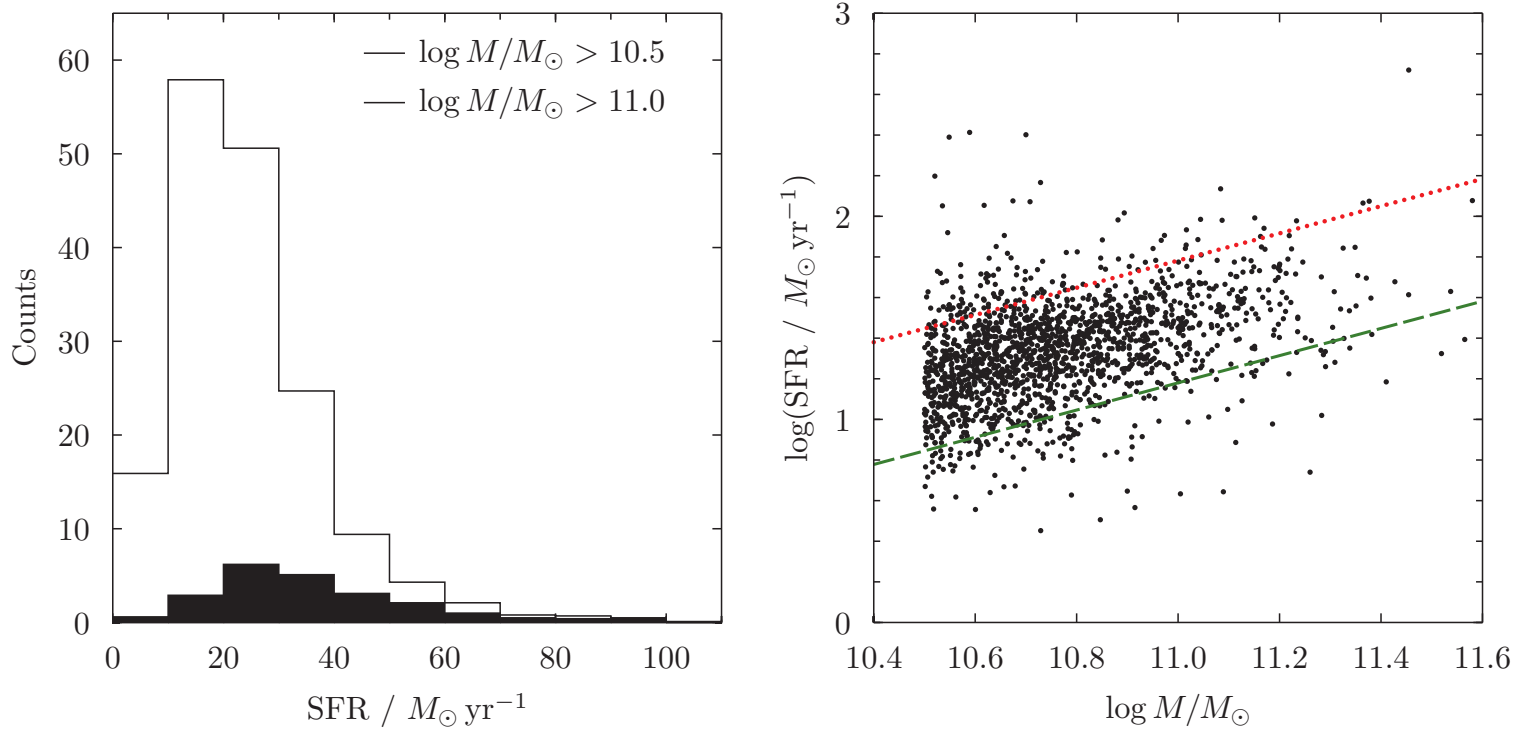}
  \caption{Left panel: Star forming rate (SFR) distributions for bulgeless galaxies with 
           $M_\ast > 10.5$ (open histogram), and $M_\ast > 11$ (filled histrogram).
           Right panel: SFR \emph{vs} $M_\ast$ for the massive ($M_\ast > 10.5$) bulgeless 
           galaxy sample.
           The green dashed line plots the relation found by \citet{Noeske-ApJ07-AEGIS} for 
           the same mass range and $z = 0.2 - 0.7$.
           The red dotted line shows the same relation for a four times higher SFR.}
  \label{fig:massive}
\end{sidewaysfigure}

The SFRs range between 10 and 100\,$M_\odot$\,yr$^{-1}$ and the distribution peaks at 
roughly 20-30\,$M_\odot$\,yr$^{-1}$ (Figure~\ref{fig:massive}; \emph{left panel}). 
For comparison, we plotted the relation found by \citet{Noeske-ApJ07-AEGIS} for galaxies 
with masses $10^{10}-10^{11}$\,$M_\odot$ at $z = 0.2 - 0.7$ (SF main sequence), and the 
corresponding line defining a four time higher SFR\@.
Despite the scatter it appears that massive bulgeless have a higher (roughly twice) specific 
SFR than same mass galaxies located along the SF main sequence, but they do not show evidence
of starburst activity which would imply SFR ten time higher than the trend defined by the 
SF main sequence.

The most massive galaxies in our sample ($\log(M/M_\odot) > 11$) have an average SFR of 
about 50\,$M_\odot$\,yr$^{-1}$, implying that they can nearly double their stellar mass 
due to their star-formation activity over $\sim2$\,Gyr. 
This is less than the look-back time spanned by our redshift range, corresponding  to 
$\sim3.5$\,Gyr\@. 
These galaxies are clearly different in nature to a fairly massive spiral like the Milky Way. 
Their SFR is more than 30~times higher than the Galactic one, while not reaching the extreme 
nature of Ultra-luminous Infrared Galaxies (ULIRGs).
If their intense star-formation activity, as suggested by their high far-infrared luminosities 
$(L_\mrm{FIR}\lesssim 10^{12} L_{\odot})$, is concentrated in the centre, it is plausible 
that a fraction of the additional mass formed from their star-formation activity may lead 
to the formation of a bulge, contributing to the morphological transformation necessary to 
evolve into a massive disc galaxy with a bulge. 
This might explain why massive bulgeless galaxies are relatively rare in the local Universe.
The presence of a bar may also be related to the morphological transformation of bulgeless 
galaxies.
In the disc-instability scenario, low angular momentum material in the centre of a disc is 
assumed to form a bar due to a global instability \citep{Shen-pc03-warps}.
Galactic bars induce inflow of interstellar gas towards the centre, and the gas accumulated 
at the galactic centre provides raw material for the bulge component \citep{Friedli-AA93-Bars}.

Finally, one cannot exclude that the disappearance of massive bulgeless galaxies is related 
to mergers producing bulge-dominated discs or spheroidals. Thus, it is important to analyse 
the local environment where these systems reside in order to investigate this possibility in 
more detail.
The study of the environmental properties of massive bulgeless galaxies, as well as a more 
detailed analysis of the multi-colour HST images of these systems to search for centrally 
enhanced star-formation activity or the presence of bars, will be the focus of future papers.

\section{Systematic search of bulgeless AGN hosts in X-ray surveys} \label{sec:AGN}
Having assembled a statistically significant sample of bulgeless galaxies at intermediate 
redshift, we carried out a sistematic search for Active Galactic Nuclei (AGN) candidates to
verify their existence in pure disc systems.
X-ray emission is generally recognised as a robust indicator of AGN activity since it does 
not suffer from heavy dust or gas extinction and contamination from star-formation in the 
host galaxy is relatively weak. 

X-rays from AGN are produced in the inner and hottest nuclear region of the galaxy, where 
accretion onto the black hole (BH) occurs, and their penetrating power (especially for hard 
X-rays, 2--10\,keV) allows them to carry information from the central engine without being 
substantially affected by absorption. 
Since detection in the soft band (0.5--2\,keV) only could miss obscured AGN 
\citep{Coma-ApJ04-EGS}, we assembled a dataset including objects detected in either hard or 
soft X-ray bands.
However, even the deepest of the current X-ray surveys are likely to miss the most heavily 
obscured ``Compton thick'' AGN \citep{Gilli-AA07-Xray}.

\subsection{Identification of AGN} \label{sec:AGNid}
In order to find X-ray emitting bulgeless, we matched the X-ray catalogues described in 
\S~\ref{sec:Xray-data} with our optical database --- including all the morphological
types --- using an adaptation of the likelihood-ratio method \citep{SuthSau-MNRAS92-likel}.
This method assigns to every matching pair a probability of being a \emph{non-false}
match; then we chose only the pairs for which the method gave high probability values 
($P > 90\%$), and selected only the bulgeless objects in the redshift range of interest 
($n \leq 1.5$ and $0.4 \leq z \leq 1.0$).

We also visually inspected the optical images to spot any obvious mis-identification due
to a failure of the matching algorithm (e.g., because of a nearby very bright source).
We excluded from our sample extended X-ray sources, which are usually related to galaxy 
clusters rather than AGN \citep{Finog-ApJS07-AGN-clusters}, and objects which have large 
cross-band positional offsets \citep[$>$ 2.5\,arcsec,][]{Alex-AJ03-CDFN} to obtain a 
reliable final sample.

The rest-frame X-ray luminosity was calculated using photometric (or spectroscopic, when 
available) redshifts using the formula:
\begin{equation} \label{eq:Xlum}
  L_\mrm{X} = 4\pi d_L^2 f_\mrm{X} (1 + z)^{(-2+\Gamma)} \,,
\end{equation}
where $d_L$ is the luminosity distance in cm, and $f_\mrm{X}$ is the observed full X-ray 
flux in units of \mbox{erg\,cm$^{-2}$\,s$^{-1}$}.
The photon index, $\Gamma$, is assumed to be equal to~1.8 for all sources, as the use of
individual indices results in only minor differences in the rest-frame luminosities
\citep{Barger-AJ02-CDFN,Barger-ApJ07-mJypop}.

Following \citet{Szok-ApJS04-XcritAGN} the criteria adopted to identify and classify an AGN 
are the following:
\begin{itemize} 
\item AGN-1: $10^{42}\leq L_\mrm{X} < 10^{44}$\,erg\,s$^{-1}$ and $\mrm{HR}\leq -0.2$,
      which show narrow and broad lines in the optical bands;
\item AGN-2: $10^{41}\leq L_\mrm{X} < 10^{44}$\,erg\,s$^{-1}$ and $\mrm{HR}\geq -0.2$,
      for which only narrow lines are observed in the optical bands.
\end{itemize}
where HR is the hardness ratio defined as $H-S/H+S$; $H$ is the number of photon 
counts in the hard band and $S$ is the number of counts in the soft band.
The HR can be used to roughly estimate the obscuration affecting the sources, as the most
efficient and reddening independent method to select obscured type~2 AGN is the presence
of luminous X-ray emission and hard X-ray colours.
Where a source has only upper limits in the soft or hard bands, the HR is assumed as~+1
and~-1, respectively.

Though widely used, this method suffers from some uncertainties. 
It is worth noting that some high redshift absorbed/type~2 sources can be identified as 
type~1 when adopting a X-ray classification based on the HR\@. 
In fact, increasing absorption makes the sources harder, while a high redshift makes them 
softer.
Nevertheless, there is evidence that the method is robust as it gives results that are in
good agreement with those obtained through optical indicators, such as BPT diagrams
\citep[see for instance,][]{ELBouc-MNRAS09-FIRST}.
This classification yielded~43 AGN candidates. 

Typically, hard X-ray detections with luminosity higher that 10$^{41}$\,erg\,s$^{-1}$ are 
evident tracers of AGN activity; nonetheless, for sources in which only soft X-rays were 
detected (21~in our sample), we tested whether the X-ray luminosities could be produced 
by star-formation activity instead of an AGN.
SFRs derived in \S~\ref{sec:massive} for most of the AGN candidates give values in the range 
4--50\,$M_{\odot}$\,yr$^{-1}$, with just one outlier showing 167\,$M_{\odot}$\,yr$^{-1}$.
Applying the relation \citep{Mineo-arXiv12-LxSFR}: 
\begin{equation} \label{eq:XSFR}
 L_\mrm{X}/\text{SFR} = (3.5 \pm 0.4)\times 10^{39} \text{(erg\,s$^{-1}$/$M_{\odot}$\,yr$^{-1}$)} \,
\end{equation}
\noindent that holds at $0.1\leq z \leq 1.3$, we can conclude that the X-ray luminosities of 
our sample are one to three orders of magnitude higher than the ones calculated 
with Eq.~\eqref{eq:XSFR}.

\subsection{Morphological analysis of AGN candidates} \label{sec:AGNmorph}
The morphological classification of the sample extracted from the ACS-GC, as described in
\citet{Griff-ApJS12-ACSGC} and in \S~\ref{sec:morphology}, relied on \textsc{galfit} results; 
the reliability of such a classification was assessed in \S~\ref{sec:morph-rel}, and it 
proved to be appropriate for statistical purposes.
However, since the AGN host candidates represent a very small fraction of our sample, we 
double checked the morphology of each object with an X-ray counterpart to exclude possible 
mis-classifications due to the automatic procedure. Given that bulgeless AGN are extremely 
rare, any morphological  mis-classification would likely be clearly revealed in this sample.

For each target we retrieved a 256$\times$256\ pixel cut-out from the HST/ACS archive.
We visually inspected the 43~cut-out images and identified 4~sources which we classified 
as mergers and rejected from the sample without attempting any further analysis.
One additional source showed an extended dust lane feature and was also discarded.
The morphology of the remaining sources was tested with a detailed analysis of the surface
brightness (SB) profiles.
We measured the surface photometry in IRAF, using the task \texttt{ellipse} in STSDAS.
We then determined the best fit S\'ersic profiles for each AGN host candidate galaxy.
If a single S\'ersic model did not provide an adequate fit to the data, we tried a 
combination of a S\'ersic plus an exponential disc profile.
According to \citet{KeK-ARAA04-bulges} late-type galaxies can host \emph{pseudo-bulges}
in the central regions which are described by low S\'ersic indices ($n<2$), and have central
light distribution similar to that of the outer disc.
From this analysis we rejected three candidates showing a bulge component with $n > 1.5$.
Finally, five additional objects were rejected because they had an ambiguous match between 
the optical and the X-ray source position.

The final sample is formed by thirty objects, eleven of them are classified as type~I, the 
remaining nineteen are type~II AGN.
Their X-ray properties are displayed in Table~\ref{tab:AGN-X}.
Their brightness surface profiles are described by either a single S\'ersic profile with 
$n < 1.5$ (22~out of~30; 74\%), or a combination of a central ``pseudo-bulge'' S\'ersic 
profile with $n < 1.5$ plus an extended disc (8~out of~30; 26\%).
The best-fit parameters are collected in Table~\ref{tab:morphfit}. 
Figures~\ref{fig:AGN-prof} and~\ref{fig:AGN-prof-dubious} show the HST images of the
30~AGN host galaxy candidates and the SB profiles with the corresponding best fits.
Substructures in the discs are visible in some cases affecting the goodness of the fit.

Visual inspection of the images shows that among the sample there are six targets which,
despite the low $n$ index, show a compact/spheroidal morphology 
(Figure~\ref{fig:AGN-prof-dubious}).
Colour-mass and colour-colour plots of these galaxies show that they are at the edge 
of the red sequence (red dots, Figure~\ref{fig:AGN-colours}, \textit{left panel}).
However, only two objects fall in the $UVJ$ region of red quiescent galaxies.
The remaining ones appear to be star-forming systems with a substantial amount of dust 
(red dots, Figure~\ref{fig:AGN-colours}, right panel).
Although these galaxies should be treated with additional care, we show that they do have 
low S\'ersic indices and most of them have evidence of star-formation activity, thus 
suggesting that they can be considered as reliable bulgeless AGN host candidates.

A more detailed analysis of the X-ray properties of the AGN candidates, including an 
estimate of the BH masses, will be presented in a forthcoming paper (Leonardo~et~al., 
2013, in preparation).

\section{Summary and conclusion} \label{sec:summ}
In this paper we presented a magnitude-limited ($m_\mrm{AB} < 24$), low-inclination,
point-source free, $K$-corrected sample of~19\,233 bulgeless galaxies at intermediate 
redshift ($0.4 \leq z \leq 1$) based on the ACS-GC public morphology catalogue.
It includes a photometric and morphological database derived from the COSMOS, AEGIS, 
GEMS, GOODS-N and GOODS-S surveys.
We assembled all the ancillary data available for these five fields and generated a 
homogeneous data-set, including photometric measurements from the far-ultraviolet to 
mid-infrared wavelength range. 
The catalogue is made publicly available to the scientific community and it contains 
morphological classifications, photometric and/or spectroscopic redshifts, rest-frame 
magnitudes and stellar masses which have been estimated by fitting the multi-colour 
photometry to a grid of composite stellar population models using the \texttt{kcorrect} 
code. 
Comparison with existing catalogues of COSMOS and GOODS-S fields 
\citep{Panella-ApJ09-COSMOS,March-ApJ09-SMF} shows that our results are in agreement with 
both the morphological classification and mass estimates already available in the literature.
The applied magnitude limit in the $I$ band corresponds to a minimum detectable mass of 
10$^9$\,$M_{\odot}$ and $4.3\times 10^{10}$\,$M_{\odot}$ at $z = 0.4$ and $z = 1$, 
respectively.

We analysed the properties of the sample and the evolution of pure-disc systems with 
redshift.
As expected, the bulk of bulgeless galaxies occupies the low-mass end region of the 
colour-mass diagram ($\log (M_\star/M_{\odot}) < 10$) and is predominantly blue, indicating 
recent star-formation activity, as opposed to the early-type objects, which dominate the 
red, high-mass end of the diagram.

Within the bulgeless sample, we found a non-negligible population of very massive galaxies 
with $\log (M_\star/M_{\odot}) > 10.5$, which contributes to $\sim 30$\% of the total galaxy 
population number density at $z \gtrsim 0.7$.
Analysis of the evolution of the number density of the different morphological types with 
redshift, above our completeness mass limit ($3\times 10^{10}$\,$M_{\odot}$), shows a 
decrease of the bulgeless number density with time compared to the ``bulgy''/early-type 
systems. 
This implies that internal bulge growth through either star-formation activity or mergers, 
and interactions with nearby companions make massive pure-disc systems evolve to earlier 
type morphologies through time.

The most massive galaxies among our sample are indeed characterized by SFRs ranging 
between~10 and~100\,$M_{\odot}$\,yr$^{-1}$, suggesting that a prolongued star-formation 
activity ($>10^8$\,yr) at such rate might contribute to the build-up of nuclear mass and 
formation of a classical bulge.
The disappearance of very massive bulgeless galaxies is confirmed by low redshift studies.
Pure disc/pseudo-bulge galaxies are the most common morphological types within the 
local volume ($< 11$\,Mpc) at stellar masses below 10$^{10}$\,$M_{\odot}$, while more massive 
pure-disc galaxies are rare \citep{Fisher-ApJ10-LocVol}.

Since bulges and AGN are thought to co-evolve in galaxies, we searched for possible AGN 
counterparts to bulgeless systems, in order to investigate the formation of massive nuclear 
BHs in galaxies with an apparently uneventful merger history.
Because X-ray emission provides one of the most effective and reliable tools to confirm 
the existence of an AGN compared to the radio continuum or mid-infrared observations, we 
combined the optical database with X-ray catalogues to search for X-ray sources within our 
sample.
We found only~30 X-ray detections with a reliable bulgeless optical counterpart, after having 
visually inspected the corresponding HST images and re-analysed the $I$-band surface 
brightness profiles.
These AGN host candidates are well-fit by either a single S\'ersic model with index 
$n < 1.5$ or by a combination of a pseudo-bulge (with $n < 1.5$) and a disc component. 

The 30~detections represent 0.2\% of the bulgeless sample; this fraction must be considered 
as a stringent lower limit.
Comparing with the other morphological classes, we found that this detection ratio is only
marginally smaller than the one of the intermediate-type population (0.3\%) and 6~times 
smaller than that of the bulge-dominated sample (1.2\%)\@.

\section*{Acknowledgments}
We thank the anonymous referee for the useful comments and suggestions.
This work is based on (GO-10134, GO-09822, GO-09425.01, GO-09583.01, GO-9500) program 
observations with the NASA/ESA Hubble Space Telescope, obtained at the Space Telescope 
Science Institute, which is operated by the Association of Universities for Research in 
Astronomy, Inc., under NASA contract NAS 5-26555. 
This research has made use of the NASA/ IPAC Infrared Science Archive, which is operated by 
the Jet Propulsion Laboratory, California Institute of Technology, under contract with the 
National Aeronautics and Space Administration.
This work is also based on zCOSMOS observations carried out using the Very Large Telescope 
at the ESO Paranal Observatory under Program ID: LP175.A-0839.
This study makes use of data from AEGIS, a multiwavelength sky survey conducted with the 
Chandra, GALEX, Hubble, Keck, CFHT, MMT, Subaru, Palomar, Spitzer, VLA, and other telescopes 
and supported in part by the NSF, NASA, and the STFC.
Funding for the DEEP2 survey has been provided by NSF grants AST95-09298, AST-0071048, 
AST-0071198, AST-0507428, and AST-0507483 as well as NASA LTSA grant NNG04GC89G.
Some of the data were obtained at the W. M. Keck Observatory, which is operated as a 
scientific partnership among the California Institute of Technology, the University of 
California and the National Aeronautics and Space Administration. 
The Observatory was made possible by the generous financial support of the W. M. Keck 
Foundation. 
Work on this paper is based on observations obtained with MegaPrime/MegaCam, a joint project 
of CFHT and CEA/DAPNIA, at the Canada-France-Hawaii Telescope (CFHT), which is operated by 
the National Research Council (NRC) of Canada, the Institut National des Science de l'Univers 
of the Centre National de la Recherche Scientifique (CNRS) of France, and the University 
of Hawaii.
We gratefully acknowledge the GOODS Team (http://cosmos.astro.caltech.edu) for providing 
all the imaging material available worldwide.
We are also indebted to the COMBO-17 \citep{Wolf-AA04-COMBO} and MUSIC teams 
\citep{Grazian-AA06-MUSIC} for the catalogues publicly supplied.
The authors gratefully acknowledge financial support from the Science and Technology 
Foundation (FCT, Portugal) through the research grants PTDC/CTE-AST/105287/2008, 
PTDC/FIS-AST/2194/2012, PEst-OE/FIS/UI2751/2011, and PEst-OE/FIS/UI2751/2014\@.
L.B., E.L. and M.E.F. gratefully acknowledge support from the Science and Technology 
Foundation (FCT, Portugal) through the Fellowships SFRH/BPD/62966/2009, SFRH/BPD 
/71278/2010 and SFRH/BPD/36141/2007\@.


\begin{thebibliography}{126}
\expandafter\ifx\csname natexlab\endcsname\relax\def\natexlab#1{#1}\fi

\bibitem[{{Alexander} {et~al.}(2003){Alexander}, {Bauer}, {Brandt},
  {Schneider}, {Hornschemeier}, {Vignali}, {Barger}, {Broos}, {Cowie},
  {Garmire}, {Townsley}, {Bautz}, {Chartas}, \& {Sargent}}]{Alex-AJ03-CDFN}
{Alexander}, D.~M., {Bauer}, F.~E., {Brandt}, W.~N., {et~al.} 2003, AJ, 126,
  539

\bibitem[{{Araya Salvo} {et~al.}(2012){Araya Salvo}, {Mathur}, {Ghosh},
  {Fiore}, \& {Ferrarese}}]{Araya-ApJ12-NGC4561}
{Araya Salvo}, C., {Mathur}, S., {Ghosh}, H., {Fiore}, F., \& {Ferrarese}, L.
  2012, ApJ, 757, 179

\bibitem[{{Barden} {et~al.}(2012){Barden}, {H{\"a}u{\ss}ler}, {Peng},
  {McIntosh}, \& {Guo}}]{Barden-MNRAS12-GALAP}
{Barden}, M., {H{\"a}u{\ss}ler}, B., {Peng}, C.~Y., {McIntosh}, D.~H., \&
  {Guo}, Y. 2012, MNRAS, 422, 449

\bibitem[{{Barger} {et~al.}(2002){Barger}, {Cowie}, {Brandt}, {Capak},
  {Garmire}, {Hornschemeier}, {Steffen}, \& {Wehner}}]{Barger-AJ02-CDFN}
{Barger}, A.~J., {Cowie}, L.~L., {Brandt}, W.~N., {et~al.} 2002, AJ, 124, 1839

\bibitem[{{Barger} {et~al.}(2007){Barger}, {Cowie}, \&
  {Wang}}]{Barger-ApJ07-mJypop}
{Barger}, A.~J., {Cowie}, L.~L., \& {Wang}, W.-H. 2007, ApJ, 654, 764

\bibitem[{{Barger} {et~al.}(2008){Barger}, {Cowie}, \&
  {Wang}}]{Barger-ApJ08-GOODSN}
{Barger}, A.~J., {Cowie}, L.~L., \& {Wang}, W.-H. 2008, ApJ, 689, 687

\bibitem[{{Barmby} {et~al.}(2008){Barmby}, {Huang}, {Ashby}, {Eisenhardt},
  {Fazio}, {Willner}, \& {Wright}}]{Barmby-ApJS08-EGS}
{Barmby}, P., {Huang}, J.-S., {Ashby}, M.~L.~N., {et~al.} 2008, ApJS, 177, 431

\bibitem[{{Barro} {et~al.}(2011){Barro}, {P{\'e}rez-Gonz{\'a}lez}, {Gallego},
  {Ashby}, {Kajisawa}, {Miyazaki}, {Villar}, {Yamada}, \&
  {Zamorano}}]{Barro-APJS11-SFR}
{Barro}, G., {P{\'e}rez-Gonz{\'a}lez}, P.~G., {Gallego}, J., {et~al.} 2011,
  ApJS, 193, 13

\bibitem[{{Barth} {et~al.}(2009){Barth}, {Strigari}, {Bentz}, {Greene}, \&
  {Ho}}]{Barth-ApJ09-NGC6321}
{Barth}, A.~J., {Strigari}, L.~E., {Bentz}, M.~C., {Greene}, J.~E., \& {Ho},
  L.~C. 2009, ApJ, 690, 1031

\bibitem[{{Bell}(2008)}]{Bell-ApJ08-BLess}
{Bell}, E.~F. 2008, ApJ, 682, 355

\bibitem[{{Bell} \& {de Jong}(2001)}]{Bell-ApJ01-MLR}
{Bell}, E.~F. \& {de Jong}, R.~S. 2001, ApJ, 550, 212

\bibitem[{{Bell} {et~al.}(2012){Bell}, {van der Wel}, {Papovich}, {Kocevski},
  {Lotz}, {McIntosh}, {Kartaltepe}, {Faber}, {Ferguson}, {Koekemoer}, {Grogin},
  {Wuyts}, {Cheung}, {Conselice}, {Dunlop}, {Giavalisco}, {Herrington}, {Koo},
  {McGrath}, {de Mello}, {Rix}, {Robaina}, \& {Williams}}]{Bell-ApJ12-CANDELS}
{Bell}, E.~F., {van der Wel}, A., {Papovich}, C., {et~al.} 2012, ApJ, 753, 167

\bibitem[{{Bertin} \& {Arnouts}(1996)}]{Bertin-AAS96-SEx}
{Bertin}, E. \& {Arnouts}, S. 1996, A\&AS, 117, 393

\bibitem[{{Blanton} \& {Roweis}(2007)}]{Blanton-AJ07-Kcorr}
{Blanton}, M.~R. \& {Roweis}, S. 2007, AJ, 133, 734

\bibitem[{{Brook} {et~al.}(2012){Brook}, {Stinson}, {Gibson}, {Ro{\v s}kar},
  {Wadsley}, \& {Quinn}}]{Brook-MNRAS12-GF}
{Brook}, C.~B., {Stinson}, G., {Gibson}, B.~K., {et~al.} 2012, MNRAS, 419, 771

\bibitem[{{Brusa} {et~al.}(2010){Brusa}, {Civano}, {Comastri}, {Miyaji},
  {Salvato}, {Zamorani}, {Cappelluti}, {Fiore}, {Hasinger}, {Mainieri},
  {Merloni}, {Bongiorno}, {Capak}, {Elvis}, {Gilli}, {Hao}, {Jahnke},
  {Koekemoer}, {Ilbert}, {Le Floc'h}, {Lusso}, {Mignoli}, {Schinnerer},
  {Silverman}, {Treister}, {Trump}, {Vignali}, {Zamojski}, {Aldcroft},
  {Aussel}, {Bardelli}, {Bolzonella}, {Cappi}, {Caputi}, {Contini},
  {Finoguenov}, {Fruscione}, {Garilli}, {Impey}, {Iovino}, {Iwasawa},
  {Kampczyk}, {Kartaltepe}, {Kneib}, {Knobel}, {Kovac}, {Lamareille},
  {Leborgne}, {Le Brun}, {Le Fevre}, {Lilly}, {Maier}, {McCracken}, {Pello},
  {Peng}, {Perez-Montero}, {de Ravel}, {Sanders}, {Scodeggio}, {Scoville},
  {Tanaka}, {Taniguchi}, {Tasca}, {de la Torre}, {Tresse}, {Vergani}, \&
  {Zucca}}]{Brusa-ApJ10-XMM}
{Brusa}, M., {Civano}, F., {Comastri}, A., {et~al.} 2010, ApJ, 716, 348

\bibitem[{{Bruzual} \& {Charlot}(2003)}]{Bruzual-MNRAS03-SES}
{Bruzual}, G. \& {Charlot}, S. 2003, MNRAS, 344, 1000

\bibitem[{{Bundy} {et~al.}(2006){Bundy}, {Ellis}, {Conselice}, {Taylor},
  {Cooper}, {Willmer}, {Weiner}, {Coil}, {Noeske}, \&
  {Eisenhardt}}]{Bundy-ApJ06-EGS}
{Bundy}, K., {Ellis}, R.~S., {Conselice}, C.~J., {et~al.} 2006, ApJ, 651, 120

\bibitem[{{Bundy} {et~al.}(2009){Bundy}, {Fukugita}, {Ellis}, {Targett},
  {Belli}, \& {Kodama}}]{Bundy-ApJ09-MOIRCS}
{Bundy}, K., {Fukugita}, M., {Ellis}, R.~S., {et~al.} 2009, ApJ, 697, 1369

\bibitem[{{Bundy} {et~al.}(2010){Bundy}, {Scarlata}, {Carollo}, {Ellis},
  {Drory}, {Hopkins}, {Salvato}, {Leauthaud}, {Koekemoer}, {Murray}, {Ilbert},
  {Oesch}, {Ma}, {Capak}, {Pozzetti}, \& {Scoville}}]{Bundy-ApJ10-diskgal}
{Bundy}, K., {Scarlata}, C., {Carollo}, C.~M., {et~al.} 2010, ApJ, 719, 1969

\bibitem[{{Cair{\'o}s} {et~al.}(2001){Cair{\'o}s}, {V{\'{\i}}lchez},
  {Gonz{\'a}lez P{\'e}rez}, {Iglesias-P{\'a}ramo}, \&
  {Caon}}]{Cairos-ApJS01-BluGal}
{Cair{\'o}s}, L.~M., {V{\'{\i}}lchez}, J.~M., {Gonz{\'a}lez P{\'e}rez}, J.~N.,
  {Iglesias-P{\'a}ramo}, J., \& {Caon}, N. 2001, ApJS, 133, 321

\bibitem[{{Caldwell} {et~al.}(2008){Caldwell}, {McIntosh}, {Rix}, {Barden},
  {Beckwith}, {Bell}, {Borch}, {Heymans}, {H{\"a}u{\ss}ler}, {Jahnke}, {Jogee},
  {Meisenheimer}, {Peng}, {S{\'a}nchez}, {Somerville}, {Wisotzki}, \&
  {Wolf}}]{Cald-ApJS08-GEMS}
{Caldwell}, J.~A.~R., {McIntosh}, D.~H., {Rix}, H.-W., {et~al.} 2008, ApJS,
  174, 136

\bibitem[{{Capak} {et~al.}(2007){Capak}, {Aussel}, {Ajiki}, {McCracken},
  {Mobasher}, {Scoville}, {Shopbell}, {Taniguchi}, {Thompson}, {Tribiano},
  {Sasaki}, {Blain}, {Brusa}, {Carilli}, {Comastri}, {Carollo}, {Cassata},
  {Colbert}, {Ellis}, {Elvis}, {Giavalisco}, {Green}, {Guzzo}, {Hasinger},
  {Ilbert}, {Impey}, {Jahnke}, {Kartaltepe}, {Kneib}, {Koda}, {Koekemoer},
  {Komiyama}, {Leauthaud}, {Le Fevre}, {Lilly}, {Liu}, {Massey}, {Miyazaki},
  {Murayama}, {Nagao}, {Peacock}, {Pickles}, {Porciani}, {Renzini}, {Rhodes},
  {Rich}, {Salvato}, {Sanders}, {Scarlata}, {Schiminovich}, {Schinnerer},
  {Scodeggio}, {Sheth}, {Shioya}, {Tasca}, {Taylor}, {Yan}, \&
  {Zamorani}}]{Capak-ApJS07-COSMOS}
{Capak}, P., {Aussel}, H., {Ajiki}, M., {et~al.} 2007, ApJS, 172, 99

\bibitem[{{Cappelluti} {et~al.}(2009){Cappelluti}, {Brusa}, {Hasinger},
  {Comastri}, {Zamorani}, {Finoguenov}, {Gilli}, {Puccetti}, {Miyaji},
  {Salvato}, {Vignali}, {Aldcroft}, {B{\"o}hringer}, {Brunner}, {Civano},
  {Elvis}, {Fiore}, {Fruscione}, {Griffiths}, {Guzzo}, {Iovino}, {Koekemoer},
  {Mainieri}, {Scoville}, {Shopbell}, {Silverman}, \& {Urry}}]{Capp-AA09-XMM}
{Cappelluti}, N., {Brusa}, M., {Hasinger}, G., {et~al.} 2009, A\&A, 497, 635

\bibitem[{{Caputi} {et~al.}(2008){Caputi}, {Lilly}, {Aussel}, {Sanders},
  {Frayer}, {Le F{\`e}vre}, {Renzini}, {Zamorani}, {Scodeggio}, {Contini},
  {Scoville}, {Carollo}, {Hasinger}, {Iovino}, {Le Brun}, {Le Floc'h}, {Maier},
  {Mainieri}, {Mignoli}, {Salvato}, {Schiminovich}, {Silverman}, {Surace},
  {Tasca}, {Abbas}, {Bardelli}, {Bolzonella}, {Bongiorno}, {Bottini}, {Capak},
  {Cappi}, {Cassata}, {Cimatti}, {Cucciati}, {de la Torre}, {de Ravel},
  {Franzetti}, {Fumana}, {Garilli}, {Halliday}, {Ilbert}, {Kampczyk},
  {Kartaltepe}, {Kneib}, {Knobel}, {Kovac}, {Lamareille}, {Leauthaud}, {Le
  Borgne}, {Maccagni}, {Marinoni}, {McCracken}, {Meneux}, {Oesch}, {Pell{\`o}},
  {P{\'e}rez-Montero}, {Porciani}, {Ricciardelli}, {Scaramella}, {Scarlata},
  {Tresse}, {Vergani}, {Walcher}, {Zamojski}, \& {Zucca}}]{Caputi-ApJ08-MIPS}
{Caputi}, K.~I., {Lilly}, S.~J., {Aussel}, H., {et~al.} 2008, ApJ, 680, 939

\bibitem[{{Cassata} {et~al.}(2005){Cassata}, {Cimatti}, {Franceschini},
  {Daddi}, {Pignatelli}, {Fasano}, {Rodighiero}, {Pozzetti}, {Mignoli}, \&
  {Renzini}}]{Cassata-MNRAS05-GOODS}
{Cassata}, P., {Cimatti}, A., {Franceschini}, A., {et~al.} 2005, MNRAS, 357,
  903

\bibitem[{{Cassata} {et~al.}(2007){Cassata}, {Guzzo}, {Franceschini},
  {Scoville}, {Capak}, {Ellis}, {Koekemoer}, {McCracken}, {Mobasher},
  {Renzini}, {Ricciardelli}, {Scodeggio}, {Taniguchi}, \&
  {Thompson}}]{Cassata-ApJS07-COSMOS}
{Cassata}, P., {Guzzo}, L., {Franceschini}, A., {et~al.} 2007, ApJS, 172, 270

\bibitem[{Chambers {et~al.}(2004)Chambers, Miley, van Breugel, \&
  Huang}]{Benson-MNRAS04-discs}
Chambers, K.~C., Miley, G.~K., van Breugel, W. J.~M., \& Huang, J.-S. 2004,
  MNRAS, 351, 1215

\bibitem[{{Coelho} {et~al.}(2013){Coelho}, {Ant{\'o}n}, {Lobo}, \&
  {Ribeiro}}]{Coelho-MNRAS13-BLess}
{Coelho}, B., {Ant{\'o}n}, S., {Lobo}, C., \& {Ribeiro}, B. 2013, MNRAS

\bibitem[{{Coil} {et~al.}(2004){Coil}, {Newman}, {Kaiser}, {Davis}, {Ma},
  {Kocevski}, \& {Koo}}]{Coil-ApJ04-EGS}
{Coil}, A.~L., {Newman}, J.~A., {Kaiser}, N., {et~al.} 2004, ApJ, 617, 765

\bibitem[{{Cole} {et~al.}(2000){Cole}, {Lacey}, {Baugh}, \&
  {Frenk}}]{Cole-MNRAS00-Hierarc}
{Cole}, S., {Lacey}, C.~G., {Baugh}, C.~M., \& {Frenk}, C.~S. 2000, MNRAS, 319,
  168

\bibitem[{{Collins} {et~al.}(2009){Collins}, {Stott}, {Hilton}, {Kay},
  {Stanford}, {Davidson}, {Hosmer}, {Hoyle}, {Liddle}, {Lloyd-Davies}, {Mann},
  {Mehrtens}, {Miller}, {Nichol}, {Romer}, {Sahl{\'e}n}, {Viana}, \&
  {West}}]{Collins-Nat09-Massive}
{Collins}, C.~A., {Stott}, J.~P., {Hilton}, M., {et~al.} 2009, Nature, 458, 603

\bibitem[{{Comastri} \& {Fiore}(2004)}]{Coma-ApJ04-EGS}
{Comastri}, A. \& {Fiore}, F. 2004, ApJS, 294, 63

\bibitem[{{Cowie} {et~al.}(2004){Cowie}, {Barger}, {Hu}, {Capak}, \&
  {Songaila}}]{Cowie-AJ04-GOODSN}
{Cowie}, L.~L., {Barger}, A.~J., {Hu}, E.~M., {Capak}, P., \& {Songaila}, A.
  2004, AJ, 127, 3137

\bibitem[{{Daddi} {et~al.}(2004){Daddi}, {Cimatti}, {Renzini}, {Fontana},
  {Mignoli}, {Pozzetti}, {Tozzi}, \& {Zamorani}}]{Daddi-ApJ04-galcolor}
{Daddi}, E., {Cimatti}, A., {Renzini}, A., {et~al.} 2004, ApJ, 617, 746

\bibitem[{{Dahlen} {et~al.}(2010){Dahlen}, {Mobasher}, {Dickinson}, {Ferguson},
  {Giavalisco}, {Grogin}, {Guo}, {Koekemoer}, {Lee}, {Lee}, {Nonino}, {Riess},
  \& {Salimbeni}}]{Dahlen-ApJ10-GOODSS}
{Dahlen}, T., {Mobasher}, B., {Dickinson}, M., {et~al.} 2010, ApJ, 724, 425

\bibitem[{{Davis} {et~al.}(2003){Davis}, {Faber}, {Newman}, {Phillips},
  {Ellis}, {Steidel}, {Conselice}, {Coil}, {Finkbeiner}, {Koo}, {Guhathakurta},
  {Weiner}, {Schiavon}, {Willmer}, {Kaiser}, {Luppino}, {Wirth}, {Connolly},
  {Eisenhardt}, {Cooper}, \& {Gerke}}]{Davis-SPIE03-DEEP2}
{Davis}, M., {Faber}, S.~M., {Newman}, J., {et~al.} 2003, in Society of
  Photo-Optical Instrumentation Engineers (SPIE) Conference Series, ed.
  {P.~Guhathakurta}, Vol. 4834, 161

\bibitem[{{Davis} {et~al.}(2007){Davis}, {Guhathakurta}, {Konidaris}, {Newman},
  {Ashby}, {Biggs}, {Barmby}, {Bundy}, {Chapman}, {Coil}, {Conselice},
  {Cooper}, {Croton}, {Eisenhardt}, {Ellis}, {Faber}, {Fang}, {Fazio},
  {Georgakakis}, {Gerke}, {Goss}, {Gwyn}, {Harker}, {Hopkins}, {Huang},
  {Ivison}, {Kassin}, {Kirby}, {Koekemoer}, {Koo}, {Laird}, {Le Floc'h}, {Lin},
  {Lotz}, {Marshall}, {Martin}, {Metevier}, {Moustakas}, {Nandra}, {Noeske},
  {Papovich}, {Phillips}, {Rich}, {Rieke}, {Rigopoulou}, {Salim},
  {Schiminovich}, {Simard}, {Smail}, {Small}, {Weiner}, {Willmer}, {Willner},
  {Wilson}, {Wright}, \& {Yan}}]{Davis-ApJ07-AEGIS}
{Davis}, M., {Guhathakurta}, P., {Konidaris}, N.~P., {et~al.} 2007, ApJ, 660,
  L1

\bibitem[{{Desroches} \& {Ho}(2009)}]{Desroches-ApJ09-AGN}
{Desroches}, L.~B. \& {Ho}, L.~C. 2009, ApJ, 690, 267

\bibitem[{{Dickinson} {et~al.}(2003){Dickinson}, {Giavalisco}, \& {GOODS
  Team}}]{Dick-pc03-GOODS}
{Dickinson}, M., {Giavalisco}, M., \& {GOODS Team}. 2003, in The Mass of
  Galaxies at Low and High Redshift, 324

\bibitem[{D'Onghia \& Burkert(2004)}]{dOnghia-ApJL04-BulgL}
D'Onghia, E. \& Burkert, A. 2004, ApJ, 612, L13

\bibitem[{{Drory} {et~al.}(2004){Drory}, {Bender}, \&
  {Hopp}}]{Drory-ApJ04-galmass}
{Drory}, N., {Bender}, R., \& {Hopp}, U. 2004, ApJ, 616, L103

\bibitem[{{Dutton} \& {van den Bosch}(2009)}]{Dutton-MNRAS09-FB}
{Dutton}, A.~A. \& {van den Bosch}, F.~C. 2009, MNRAS, 396, 141

\bibitem[{{El Bouchefry}(2009)}]{ELBouc-MNRAS09-FIRST}
{El Bouchefry}, K. 2009, MNRAS, 396, 2011

\bibitem[{{Elvis} {et~al.}(2009){Elvis}, {Civano}, {Vignali}, {Puccetti},
  {Fiore}, {Cappelluti}, {Aldcroft}, {Fruscione}, {Zamorani}, {Comastri},
  {Brusa}, {Gilli}, {Miyaji}, {Damiani}, {Koekemoer}, {Finoguenov}, {Brunner},
  {Urry}, {Silverman}, {Mainieri}, {Hasinger}, {Griffiths}, {Carollo}, {Hao},
  {Guzzo}, {Blain}, {Calzetti}, {Carilli}, {Capak}, {Ettori}, {Fabbiano},
  {Impey}, {Lilly}, {Mobasher}, {Rich}, {Salvato}, {Sanders}, {Schinnerer},
  {Scoville}, {Shopbell}, {Taylor}, {Taniguchi}, \&
  {Volonteri}}]{Elvis-ApJS09-CCosm}
{Elvis}, M., {Civano}, F., {Vignali}, C., {et~al.} 2009, ApJS, 184, 158

\bibitem[{Fall \& Efstathiou(1980)}]{Fall-MNRAS80-Bulg}
Fall, S.~M. \& Efstathiou, G. 1980, MNRAS, 193, 189

\bibitem[{{Filippenko} \& {Ho}(2003)}]{Filipp-ApJL03-NGC4395}
{Filippenko}, A.~V. \& {Ho}, L.~C. 2003, ApJ, 588, L13

\bibitem[{{Finoguenov} {et~al.}(2007){Finoguenov}, {Guzzo}, {Hasinger},
  {Scoville}, {Aussel}, {B{\"o}hringer}, {Brusa}, {Capak}, {Cappelluti},
  {Comastri}, {Giodini}, {Griffiths}, {Impey}, {Koekemoer}, {Kneib},
  {Leauthaud}, {Le F{\`e}vre}, {Lilly}, {Mainieri}, {Massey}, {McCracken},
  {Mobasher}, {Murayama}, {Peacock}, {Sakelliou}, {Schinnerer}, {Silverman},
  {Smol{\v c}i{\'c}}, {Taniguchi}, {Tasca}, {Taylor}, {Trump}, \&
  {Zamorani}}]{Finog-ApJS07-AGN-clusters}
{Finoguenov}, A., {Guzzo}, L., {Hasinger}, G., {et~al.} 2007, ApJS, 172, 182

\bibitem[{{Fisher} \& {Drory}(2010)}]{Fisher-ApJ10-LocVol}
{Fisher}, D.~B. \& {Drory}, N. 2010, ApJ, 716, 942

\bibitem[{{Frayer} {et~al.}(2009){Frayer}, {Sanders}, {Surace}, {Aussel},
  {Salvato}, {Le Floc'h}, {Huynh}, {Scoville}, {Afonso-Luis}, {Bhattacharya},
  {Capak}, {Fadda}, {Fu}, {Helou}, {Ilbert}, {Kartaltepe}, {Koekemoer}, {Lee},
  {Murphy}, {Sargent}, {Schinnerer}, {Sheth}, {Shopbell}, {Shupe}, \&
  {Yan}}]{Frayer-AJ09-SCOSMOS}
{Frayer}, D.~T., {Sanders}, D.~B., {Surace}, J.~A., {et~al.} 2009, AJ, 138,
  1261

\bibitem[{{Friedli} \& {Benz}(1993)}]{Friedli-AA93-Bars}
{Friedli}, D. \& {Benz}, W. 1993, A\&A, 268, 65

\bibitem[{{Gadotti}(2009)}]{Gadotti-MNRAS09-SDSS}
{Gadotti}, D.~A. 2009, MNRAS, 393, 1531

\bibitem[{{Giavalisco} {et~al.}(2004){Giavalisco}, {Ferguson}, {Koekemoer},
  {Dickinson}, {Alexander}, {Bauer}, {Bergeron}, {Biagetti}, {Brandt},
  {Casertano}, {Cesarsky}, {Chatzichristou}, {Conselice}, {Cristiani}, {Da
  Costa}, {Dahlen}, {de Mello}, {Eisenhardt}, {Erben}, {Fall}, {Fassnacht},
  {Fosbury}, {Fruchter}, {Gardner}, {Grogin}, {Hook}, {Hornschemeier}, {Idzi},
  {Jogee}, {Kretchmer}, {Laidler}, {Lee}, {Livio}, {Lucas}, {Madau},
  {Mobasher}, {Moustakas}, {Nonino}, {Padovani}, {Papovich}, {Park},
  {Ravindranath}, {Renzini}, {Richardson}, {Riess}, {Rosati}, {Schirmer},
  {Schreier}, {Somerville}, {Spinrad}, {Stern}, {Stiavelli}, {Strolger},
  {Urry}, {Vandame}, {Williams}, \& {Wolf}}]{Giava-ApJ04-GOODS}
{Giavalisco}, M., {Ferguson}, H.~C., {Koekemoer}, A.~M., {et~al.} 2004, ApJ,
  600, L93

\bibitem[{{Gilli} {et~al.}(2007){Gilli}, {Comastri}, \&
  {Hasinger}}]{Gilli-AA07-Xray}
{Gilli}, R., {Comastri}, A., \& {Hasinger}, G. 2007, A\&A, 463, 79

\bibitem[{{Grazian} {et~al.}(2006){Grazian}, {Fontana}, {de Santis}, {Nonino},
  {Salimbeni}, {Giallongo}, {Cristiani}, {Gallozzi}, \&
  {Vanzella}}]{Grazian-AA06-MUSIC}
{Grazian}, A., {Fontana}, A., {de Santis}, C., {et~al.} 2006, A\&A, 449, 951

\bibitem[{{Griffith} {et~al.}(2012){Griffith}, {Cooper}, {Newman}, {Moustakas},
  {Stern}, {Comerford}, {Davis}, {Lotz}, {Barden}, {Conselice}, {Capak},
  {Faber}, {Kirkpatrick}, {Koekemoer}, {Koo}, {Noeske}, {Scoville}, {Sheth},
  {Shopbell}, {Willmer}, \& {Weiner}}]{Griff-ApJS12-ACSGC}
{Griffith}, R.~L., {Cooper}, M.~C., {Newman}, J.~A., {et~al.} 2012, ApJS, 200,
  9

\bibitem[{{Gwyn}(2008)}]{Gwyn-PASP08-MegaCam}
{Gwyn}, S.~D.~J. 2008, PASP, 120, 212

\bibitem[{{Hainline} {et~al.}(2011){Hainline}, {Blain}, {Smail}, {Alexander},
  {Armus}, {Chapman}, \& {Ivison}}]{Hain-ApJ11-mass-SMGs}
{Hainline}, L.~J., {Blain}, A.~W., {Smail}, I., {et~al.} 2011, ApJ, 740, 96

\bibitem[{H\"aring \& Rix(2004)}]{Haring-ApJL04-BHM}
H\"aring, N. \& Rix, H.~W. 2004, ApJ, 604, L89

\bibitem[{{Hoaglin} {et~al.}(1983){Hoaglin}, {Mosteller}, \&
  {Tukey}}]{Hoaglin-83}
{Hoaglin}, D.~C., {Mosteller}, F., \& {Tukey}, J.~W. 1983, {Understanding
  robust and exploratory data anlysis} (New York: Wiley)

\bibitem[{{Hopkins} {et~al.}(2009){Hopkins}, {Cox}, {Younger}, \&
  {Hernquist}}]{Hopk-ApJ09-discs}
{Hopkins}, P.~F., {Cox}, T.~J., {Younger}, J.~D., \& {Hernquist}, L. 2009, ApJ,
  691, 1168

\bibitem[{{Ilbert} {et~al.}(2006){Ilbert}, {Arnouts}, {McCracken},
  {Bolzonella}, {Bertin}, {Le F{\`e}vre}, {Mellier}, {Zamorani}, {Pell{\`o}},
  {Iovino}, {Tresse}, {Le Brun}, {Bottini}, {Garilli}, {Maccagni}, {Picat},
  {Scaramella}, {Scodeggio}, {Vettolani}, {Zanichelli}, {Adami}, {Bardelli},
  {Cappi}, {Charlot}, {Ciliegi}, {Contini}, {Cucciati}, {Foucaud}, {Franzetti},
  {Gavignaud}, {Guzzo}, {Marano}, {Marinoni}, {Mazure}, {Meneux}, {Merighi},
  {Paltani}, {Pollo}, {Pozzetti}, {Radovich}, {Zucca}, {Bondi}, {Bongiorno},
  {Busarello}, {de La Torre}, {Gregorini}, {Lamareille}, {Mathez}, {Merluzzi},
  {Ripepi}, {Rizzo}, \& {Vergani}}]{Ilbert-AA06-EGSphz}
{Ilbert}, O., {Arnouts}, S., {McCracken}, H.~J., {et~al.} 2006, A\&A, 457, 841

\bibitem[{{Ilbert} {et~al.}(2009){Ilbert}, {Capak}, {Salvato}, {Aussel},
  {McCracken}, {Sanders}, {Scoville}, {Kartaltepe}, {Arnouts}, {Le Floc'h},
  {Mobasher}, {Taniguchi}, {Lamareille}, {Leauthaud}, {Sasaki}, {Thompson},
  {Zamojski}, {Zamorani}, {Bardelli}, {Bolzonella}, {Bongiorno}, {Brusa},
  {Caputi}, {Carollo}, {Contini}, {Cook}, {Coppa}, {Cucciati}, {de la Torre},
  {de Ravel}, {Franzetti}, {Garilli}, {Hasinger}, {Iovino}, {Kampczyk},
  {Kneib}, {Knobel}, {Kovac}, {Le Borgne}, {Le Brun}, {F{\`e}vre}, {Lilly},
  {Looper}, {Maier}, {Mainieri}, {Mellier}, {Mignoli}, {Murayama}, {Pell{\`o}},
  {Peng}, {P{\'e}rez-Montero}, {Renzini}, {Ricciardelli}, {Schiminovich},
  {Scodeggio}, {Shioya}, {Silverman}, {Surace}, {Tanaka}, {Tasca}, {Tresse},
  {Vergani}, \& {Zucca}}]{Ilbert-ApJ09-COSMOS}
{Ilbert}, O., {Capak}, P., {Salvato}, M., {et~al.} 2009, ApJ, 690, 1236

\bibitem[{{Kajisawa} {et~al.}(2011){Kajisawa}, {Ichikawa}, {Tanaka}, {Yamada},
  {Akiyama}, {Suzuki}, {Tokoku}, {Katsuno Uchimoto}, {Konishi}, {Yoshikawa},
  {Nishimura}, {Omata}, {Ouchi}, {Iwata}, {Hamana}, \&
  {Onodera}}]{Kaji-PASJ11-MOIRCS}
{Kajisawa}, M., {Ichikawa}, T., {Tanaka}, I., {et~al.} 2011, PASJ, 63, 379

\bibitem[{{Kautsch} {et~al.}(2006){Kautsch}, {Grebel}, {Barazza}, \&
  {Gallagher}}]{Kautsch-AA06-Eon}
{Kautsch}, S.~J., {Grebel}, E.~K., {Barazza}, F.~D., \& {Gallagher}, III, J.~S.
  2006, A\&A, 445, 765

\bibitem[{{Kewley} {et~al.}(2001){Kewley}, {Dopita}, {Sutherland}, {Heisler},
  \& {Trevena}}]{Kewley-ApJ01-SBmodel}
{Kewley}, L.~J., {Dopita}, M.~A., {Sutherland}, R.~S., {Heisler}, C.~A., \&
  {Trevena}, J. 2001, ApJ, 556, 121

\bibitem[{{Koda} {et~al.}(2009){Koda}, {Milosavljevi{\'c}}, \&
  {Shapiro}}]{Koda-ApJ09-mergers}
{Koda}, J., {Milosavljevi{\'c}}, M., \& {Shapiro}, P.~R. 2009, ApJ, 696, 254

\bibitem[{{Kormendy} {et~al.}(2011){Kormendy}, {Bender}, \&
  {Cornell}}]{Kormendy-Nat11-SMBHbar}
{Kormendy}, J., {Bender}, R., \& {Cornell}, M.~E. 2011, Nature, 469, 374

\bibitem[{{Kormendy} \& {Kennicutt}(2004)}]{KeK-ARAA04-bulges}
{Kormendy}, J. \& {Kennicutt}, Jr., R.~C. 2004, ARA\&A, 42, 603

\bibitem[{Kormendy \& Richstone(1995)}]{Kormendy-ARAA95-BHs}
Kormendy, J. \& Richstone, D. 1995, ARA\&A, 33, 581

\bibitem[{{Kunth} {et~al.}(1988){Kunth}, {Maurogordato}, \&
  {Vigroux}}]{Kunth-AA88-BluGal}
{Kunth}, D., {Maurogordato}, S., \& {Vigroux}, L. 1988, A\&A, 10, 204

\bibitem[{{Laird} {et~al.}(2009){Laird}, {Nandra}, {Georgakakis}, {Aird},
  {Barmby}, {Conselice}, {Coil}, {Davis}, {Faber}, {Fazio}, {Guhathakurta},
  {Koo}, {Sarajedini}, \& {Willmer}}]{Laird-ApJS09-AEGISX}
{Laird}, E.~S., {Nandra}, K., {Georgakakis}, A., {et~al.} 2009, ApJ, 180, 102

\bibitem[{{Larson} {et~al.}(2011){Larson}, {Dunkley}, {Hinshaw}, {Komatsu},
  {Nolta}, {Bennett}, {Gold}, {Halpern}, {Hill}, {Jarosik}, {Kogut}, {Limon},
  {Meyer}, {Odegard}, {Page}, {Smith}, {Spergel}, {Tucker}, {Weiland},
  {Wollack}, \& {Wright}}]{Larson-ApJS11-WMAP7}
{Larson}, D., {Dunkley}, J., {Hinshaw}, G., {et~al.} 2011, ApJS, 192, 16

\bibitem[{{Lehmer} {et~al.}(2005){Lehmer}, {Brandt}, {Alexander}, {Bauer},
  {Schneider}, {Tozzi}, {Bergeron}, {Garmire}, {Giacconi}, {Gilli}, {Hasinger},
  {Hornschemeier}, {Koekemoer}, {Mainieri}, {Miyaji}, {Nonino}, {Rosati},
  {Silverman}, {Szokoly}, \& {Vignali}}]{Lehmer-ApJS05-ECDFS}
{Lehmer}, B.~D., {Brandt}, W.~N., {Alexander}, D.~M., {et~al.} 2005, ApJS, 161,
  21

\bibitem[{{Lilly} {et~al.}(2007){Lilly}, {Le F{\`e}vre}, {Renzini}, {Zamorani},
  {Scodeggio}, {Contini}, {Carollo}, {Hasinger}, {Kneib}, {Iovino}, {Le Brun},
  {Maier}, {Mainieri}, {Mignoli}, {Silverman}, {Tasca}, {Bolzonella},
  {Bongiorno}, {Bottini}, {Capak}, {Caputi}, {Cimatti}, {Cucciati}, {Daddi},
  {Feldmann}, {Franzetti}, {Garilli}, {Guzzo}, {Ilbert}, {Kampczyk}, {Kovac},
  {Lamareille}, {Leauthaud}, {Borgne}, {McCracken}, {Marinoni}, {Pello},
  {Ricciardelli}, {Scarlata}, {Vergani}, {Sanders}, {Schinnerer}, {Scoville},
  {Taniguchi}, {Arnouts}, {Aussel}, {Bardelli}, {Brusa}, {Cappi}, {Ciliegi},
  {Finoguenov}, {Foucaud}, {Franceschini}, {Halliday}, {Impey}, {Knobel},
  {Koekemoer}, {Kurk}, {Maccagni}, {Maddox}, {Marano}, {Marconi}, {Meneux},
  {Mobasher}, {Moreau}, {Peacock}, {Porciani}, {Pozzetti}, {Scaramella},
  {Schiminovich}, {Shopbell}, {Smail}, {Thompson}, {Tresse}, {Vettolani},
  {Zanichelli}, \& {Zucca}}]{Lilly-ApJS07-zCOSMOS}
{Lilly}, S.~J., {Le F{\`e}vre}, O., {Renzini}, A., {et~al.} 2007, ApJS, 172, 70

\bibitem[{{Loose} \& {Thuan}(1986{\natexlab{a}})}]{Loose-MitAG86-Morph}
{Loose}, H.-H. \& {Thuan}, F.~X. 1986{\natexlab{a}}, Mitteilungen der
  Astronomischen Gesellschaft Hamburg, 65, 231

\bibitem[{{Loose} \& {Thuan}(1986{\natexlab{b}})}]{Loose-ApJ86-SBs}
{Loose}, H.-H. \& {Thuan}, T.~X. 1986{\natexlab{b}}, ApJ, 309, 59

\bibitem[{{Lotz} {et~al.}(2008){Lotz}, {Davis}, {Faber}, {Guhathakurta},
  {Gwyn}, {Huang}, {Koo}, {Le Floc'h}, {Lin}, {Newman}, {Noeske}, {Papovich},
  {Willmer}, {Coil}, {Conselice}, {Cooper}, {Hopkins}, {Metevier}, {Primack},
  {Rieke}, \& {Weiner}}]{Lotz-APJ08-EGS}
{Lotz}, J.~M., {Davis}, M., {Faber}, S.~M., {et~al.} 2008, ApJ, 672, 177

\bibitem[{{Lotz} {et~al.}(2004){Lotz}, {Primack}, \& {Madau}}]{Lotz-AJ04-morph}
{Lotz}, J.~M., {Primack}, J., \& {Madau}, P. 2004, AJ, 128, 163

\bibitem[{{Magorrian} {et~al.}(1998){Magorrian}, {Tremaine}, {Richstone},
  {Bender}, {Bower}, {Dressler}, {Faber}, {Gebhardt}, {Green}, {Grillmair},
  {Kormendy}, \& {Lauer}}]{Magorrian-AJ98-BHs}
{Magorrian}, J., {Tremaine}, S., {Richstone}, D., {et~al.} 1998, AJ, 115, 2285

\bibitem[{{Marchesini} {et~al.}(2009){Marchesini}, {van Dokkum}, {F{\"o}rster
  Schreiber}, {Franx}, {Labb{\'e}}, \& {Wuyts}}]{March-ApJ09-SMF}
{Marchesini}, D., {van Dokkum}, P.~G., {F{\"o}rster Schreiber}, N.~M., {et~al.}
  2009, ApJ, 701, 1765

\bibitem[{{Mathur} {et~al.}(2012){Mathur}, {Fields}, {Peterson}, \&
  {Grupe}}]{Mathur-ApJ12-SMBHs}
{Mathur}, S., {Fields}, D., {Peterson}, B.~M., \& {Grupe}, D. 2012, ApJ, 754,
  146

\bibitem[{{McAlpine} {et~al.}(2011){McAlpine}, {Satyapal}, {Gliozzi}, {Cheung},
  {Sambruna}, \& {Eracleous}}]{McAlpine-ApJ11-NGC3367}
{McAlpine}, W., {Satyapal}, S., {Gliozzi}, M., {et~al.} 2011, ApJ, 728, 25

\bibitem[{{McCracken} {et~al.}(2012){McCracken}, {Milvang-Jensen}, {Dunlop},
  {Franx}, {Fynbo}, {Le F{\`e}vre}, {Holt}, {Caputi}, {Goranova}, {Buitrago},
  {Emerson}, {Freudling}, {Hudelot}, {L{\'o}pez-Sanjuan}, {Magnard}, {Mellier},
  {M{\o}ller}, {Nilsson}, {Sutherland}, {Tasca}, \&
  {Zabl}}]{McCrack-AA12-UVista}
{McCracken}, H.~J., {Milvang-Jensen}, B., {Dunlop}, J., {et~al.} 2012, A\&A,
  544, A156

\bibitem[{{McGrath} {et~al.}(2008){McGrath}, {Stockton}, {Canalizo}, {Iye}, \&
  {Maihara}}]{McGrath-ApJ08-Morph}
{McGrath}, E.~J., {Stockton}, A., {Canalizo}, G., {Iye}, M., \& {Maihara}, T.
  2008, ApJ, 682, 303

\bibitem[{{Mineo} {et~al.}(2012){Mineo}, {Gilfanov}, {Lehmer}, {Morrison}, \&
  {Sunyaev}}]{Mineo-arXiv12-LxSFR}
{Mineo}, S., {Gilfanov}, M., {Lehmer}, B.~D., {Morrison}, G.~E., \& {Sunyaev},
  R. 2012, ArXiv e-prints

\bibitem[{{Moster} {et~al.}(2011){Moster}, {Somerville}, {Newman}, \&
  {Rix}}]{Moster-ApJ11-CosVar}
{Moster}, B.~P., {Somerville}, R.~S., {Newman}, J.~A., \& {Rix}, H.-W. 2011,
  ApJ, 731, 113

\bibitem[{{Muzzin} {et~al.}(2009){Muzzin}, {Marchesini}, {van Dokkum},
  {Labb{\'e}}, {Kriek}, \& {Franx}}]{Muzz-ApJ09-NIR-Kgal}
{Muzzin}, A., {Marchesini}, D., {van Dokkum}, P.~G., {et~al.} 2009, ApJ, 701,
  1839

\bibitem[{{Noeske} {et~al.}(2007){Noeske}, {Weiner}, {Faber}, {Papovich},
  {Koo}, {Somerville}, {Bundy}, {Conselice}, {Newman}, {Schiminovich}, {Le
  Floc'h}, {Coil}, {Rieke}, {Lotz}, {Primack}, {Barmby}, {Cooper}, {Davis},
  {Ellis}, {Fazio}, {Guhathakurta}, {Huang}, {Kassin}, {Martin}, {Phillips},
  {Rich}, {Small}, {Willmer}, \& {Wilson}}]{Noeske-ApJ07-AEGIS}
{Noeske}, K.~G., {Weiner}, B.~J., {Faber}, S.~M., {et~al.} 2007, ApJ, 660, L43

\bibitem[{{Oesch} {et~al.}(2010){Oesch}, {Carollo}, {Feldmann}, {Hahn},
  {Lilly}, {Sargent}, {Scarlata}, {Aller}, {Aussel}, {Bolzonella}, {Bschorr},
  {Bundy}, {Capak}, {Ilbert}, {Kneib}, {Koekemoer}, {Kova{\v c}}, {Leauthaud},
  {Le Floc'h}, {Massey}, {McCracken}, {Pozzetti}, {Renzini}, {Rhodes},
  {Salvato}, {Sanders}, {Scoville}, {Sheth}, {Taniguchi}, \&
  {Thompson}}]{Oesch-ApJ10-COSMOS}
{Oesch}, P.~A., {Carollo}, C.~M., {Feldmann}, R., {et~al.} 2010, ApJ, 714, L47

\bibitem[{{Pannella} {et~al.}(2009){Pannella}, {Gabasch}, {Goranova}, {Drory},
  {Hopp}, {Noll}, {Saglia}, {Strazzullo}, \& {Bender}}]{Panella-ApJ09-COSMOS}
{Pannella}, M., {Gabasch}, A., {Goranova}, Y., {et~al.} 2009, ApJ, 701, 787

\bibitem[{Peng {et~al.}(2002)Peng, Ho, Impey, \& Rix}]{Peng-AJ02-GALFIT}
Peng, C.~Y., Ho, L.~C., Impey, C.~D., \& Rix, H.-W. 2002, AJ, 124, 266

\bibitem[{{Peterson} {et~al.}(2005){Peterson}, {Bentz}, {Desroches},
  {Filippenko}, {Ho}, {Kaspi}, {Laor}, {Maoz}, {Moran}, {Pogge}, \&
  {Quillen}}]{Peterson-ApJ05-NGC4395}
{Peterson}, B.~M., {Bentz}, M.~C., {Desroches}, L.-B., {et~al.} 2005, ApJ, 632,
  799

\bibitem[{{Reines} {et~al.}(2011){Reines}, {Sivakoff}, {Johnson}, \&
  {Brogan}}]{Reines-NAt11-H210}
{Reines}, A.~E., {Sivakoff}, G.~R., {Johnson}, K.~E., \& {Brogan}, C.~L. 2011,
  Nature, 470, 6

\bibitem[{{Robertson} {et~al.}(2006){Robertson}, {Bullock}, {Cox}, {Di Matteo},
  {Hernquist}, {Springel}, \& {Yoshida}}]{Robert-ApJ06-mergers}
{Robertson}, B., {Bullock}, J.~S., {Cox}, T.~J., {et~al.} 2006, ApJ, 645, 986

\bibitem[{Robertson {et~al.}(2004)Robertson, Yoshida, Springel, \&
  Hernquist}]{Robertson-ApJ04-disk}
Robertson, B., Yoshida, N., Springel, V., \& Hernquist, L. 2004, ApJ, 606, 32

\bibitem[{{Robertson} \& {Bullock}(2008)}]{Robert-ApJ08-disks}
{Robertson}, B.~E. \& {Bullock}, J.~S. 2008, ApJ, 685, L27

\bibitem[{{Sanders} {et~al.}(2007){Sanders}, {Salvato}, {Aussel}, {Ilbert},
  {Scoville}, {Surace}, {Frayer}, {Sheth}, {Helou}, {Brooke}, {Bhattacharya},
  {Yan}, {Kartaltepe}, {Barnes}, {Blain}, {Calzetti}, {Capak}, {Carilli},
  {Carollo}, {Comastri}, {Daddi}, {Ellis}, {Elvis}, {Fall}, {Franceschini},
  {Giavalisco}, {Hasinger}, {Impey}, {Koekemoer}, {Le F{\`e}vre}, {Lilly},
  {Liu}, {McCracken}, {Mobasher}, {Renzini}, {Rich}, {Schinnerer}, {Shopbell},
  {Taniguchi}, {Thompson}, {Urry}, \& {Williams}}]{Sanders-ApJS07-SCOSMOS}
{Sanders}, D.~B., {Salvato}, M., {Aussel}, H., {et~al.} 2007, ApJS, 172, 86

\bibitem[{{Satyapal} {et~al.}(2009){Satyapal}, {B{\"o}ker}, {Mcalpine},
  {Gliozzi}, {Abel}, \& {Heckman}}]{Satyapal-ApJ09-NGC4178}
{Satyapal}, S., {B{\"o}ker}, T., {Mcalpine}, W., {et~al.} 2009, ApJ, 704, 439

\bibitem[{{Satyapal} {et~al.}(2007){Satyapal}, {Vega}, {Heckman}, {O'Halloran},
  \& {Dudik}}]{Satyapal-ApJL07-NGC3621}
{Satyapal}, S., {Vega}, D., {Heckman}, T., {O'Halloran}, B., \& {Dudik}, R.
  2007, ApJ, 663, L9

\bibitem[{{Scarlata} {et~al.}(2007){Scarlata}, {Carollo}, {Lilly}, {Sargent},
  {Feldmann}, {Kampczyk}, {Porciani}, {Koekemoer}, {Scoville}, {Kneib},
  {Leauthaud}, {Massey}, {Rhodes}, {Tasca}, {Capak}, {Maier}, {McCracken},
  {Mobasher}, {Renzini}, {Taniguchi}, {Thompson}, {Sheth}, {Ajiki}, {Aussel},
  {Murayama}, {Sanders}, {Sasaki}, {Shioya}, \&
  {Takahashi}}]{Scarlata-ApJS07-ZEST}
{Scarlata}, C., {Carollo}, C.~M., {Lilly}, S., {et~al.} 2007, ApJS, 172, 406

\bibitem[{{Schawinski} {et~al.}(2010){Schawinski}, {Urry}, {Virani}, {Coppi},
  {Bamford}, {Treister}, {Lintott}, {Sarzi}, {Keel}, {Kaviraj}, {Cardamone},
  {Masters}, {Ross}, {Andreescu}, {Murray}, {Nichol}, {Raddick}, {Slosar},
  {Szalay}, {Thomas}, \& {Vandenberg}}]{Schaw-ApJ10-SMBH}
{Schawinski}, K., {Urry}, C.~M., {Virani}, S., {et~al.} 2010, ApJ, 711, 284

\bibitem[{{Scoville} {et~al.}(2007){Scoville}, {Abraham}, {Aussel}, {Barnes},
  {Benson}, {Blain}, {Calzetti}, {Comastri}, {Capak}, {Carilli}, {Carlstrom},
  {Carollo}, {Colbert}, {Daddi}, {Ellis}, {Elvis}, {Ewald}, {Fall},
  {Franceschini}, {Giavalisco}, {Green}, {Griffiths}, {Guzzo}, {Hasinger},
  {Impey}, {Kneib}, {Koda}, {Koekemoer}, {Lefevre}, {Lilly}, {Liu},
  {McCracken}, {Massey}, {Mellier}, {Miyazaki}, {Mobasher}, {Mould}, {Norman},
  {Refregier}, {Renzini}, {Rhodes}, {Rich}, {Sanders}, {Schiminovich},
  {Schinnerer}, {Scodeggio}, {Sheth}, {Shopbell}, {Taniguchi}, {Tyson}, {Urry},
  {Van Waerbeke}, {Vettolani}, {White}, \& {Yan}}]{Scov-ApJS07-COSMOS}
{Scoville}, N., {Abraham}, R.~G., {Aussel}, H., {et~al.} 2007, ApJS, 172, 38

\bibitem[{{Secrest} {et~al.}(2012){Secrest}, {Satyapal}, {Gliozzi}, {Cheung},
  {Seth}, \& {B{\"o}ker}}]{Secrest-ApJ12-NGC4178}
{Secrest}, N.~J., {Satyapal}, S., {Gliozzi}, M., {et~al.} 2012, ApJ, 753, 38

\bibitem[{{S\'ersic}(1968)}]{Sersic-68}
{S\'ersic}, J.~L. 1968, {Atlas de galaxias australes} (Cordoba, Argentina:
  Observatorio Astronomico)

\bibitem[{{Shen} \& {Sellwood}(2003)}]{Shen-pc03-warps}
{Shen}, J. \& {Sellwood}, J.~A. 2003, in American Astronomical Society Meeting
  Abstracts, 1353

\bibitem[{{Silk} \& {Mamon}(2012)}]{Silk-RAA12-galaxies}
{Silk}, J. \& {Mamon}, G.~A. 2012, Research in Astronomy and Astrophysics, 12,
  917

\bibitem[{{Simmons} {et~al.}(2012){Simmons}, {Lintott}, {Schawinski}, {Moran},
  {Han}, {Kaviraj}, {Masters}, {Urry}, {Willett}, {Bamford}, \&
  {Nichol}}]{Simmons-MNRAS12-Gzoo}
{Simmons}, B.~D., {Lintott}, C., {Schawinski}, K., {et~al.} 2012, MNRAS, 429,
  2199

\bibitem[{{Springel} \& {Hernquist}(2005)}]{Spring-ApJ05-spirals}
{Springel}, V. \& {Hernquist}, L. 2005, ApJ, 622, L9

\bibitem[{{Suh} {et~al.}(2010){Suh}, {Jeong}, {Oh}, {Yi}, {Ferreras}, \&
  {Schawinski}}]{Suh-ApJS10-SDSS}
{Suh}, H., {Jeong}, H., {Oh}, K., {et~al.} 2010, ApJS, 187, 374

\bibitem[{{Sutherland} \& {Saunders}(1992)}]{SuthSau-MNRAS92-likel}
{Sutherland}, W. \& {Saunders}, W. 1992, MNRAS, 259, 413

\bibitem[{{Szokoly} {et~al.}(2004){Szokoly}, {Bergeron}, {Hasinger}, {Lehmann},
  {Kewley}, {Mainieri}, {Nonino}, {Rosati}, {Giacconi}, {Gilli}, {Gilmozzi},
  {Norman}, {Romaniello}, {Schreier}, {Tozzi}, {Wang}, {Zheng}, \&
  {Zirm}}]{Szok-ApJS04-XcritAGN}
{Szokoly}, G.~P., {Bergeron}, J., {Hasinger}, G., {et~al.} 2004, ApJS, 155, 271

\bibitem[{{Taniguchi} {et~al.}(2007){Taniguchi}, {Scoville}, {Murayama},
  {Sanders}, {Mobasher}, {Aussel}, {Capak}, {Ajiki}, {Miyazaki}, {Komiyama},
  {Shioya}, {Nagao}, {Sasaki}, {Koda}, {Carilli}, {Giavalisco}, {Guzzo},
  {Hasinger}, {Impey}, {LeFevre}, {Lilly}, {Renzini}, {Rich}, {Schinnerer},
  {Shopbell}, {Kaifu}, {Karoji}, {Arimoto}, {Okamura}, \&
  {Ohta}}]{Tanig-ApJS07-COSMOS}
{Taniguchi}, Y., {Scoville}, N., {Murayama}, T., {et~al.} 2007, ApJS, 172, 9

\bibitem[{{Thomas} {et~al.}(2010){Thomas}, {Maraston}, {Schawinski}, {Sarzi},
  \& {Silk}}]{Thomas-MNRAS10-SelReg}
{Thomas}, D., {Maraston}, C., {Schawinski}, K., {Sarzi}, M., \& {Silk}, J.
  2010, MNRAS, 404, 1775

\bibitem[{{Thornton} {et~al.}(2008){Thornton}, {Barth}, {Ho}, {Rutledge}, \&
  {Greene}}]{Thornton-ApJ08-POX52}
{Thornton}, C.~E., {Barth}, A.~J., {Ho}, L.~C., {Rutledge}, R.~E., \& {Greene},
  J.~E. 2008, ApJ, 686, 892

\bibitem[{{Toth} \& {Ostriker}(1992)}]{Toth-ApJ92-disks}
{Toth}, G. \& {Ostriker}, J.~P. 1992, ApJ, 389, 5

\bibitem[{{Tremaine} {et~al.}(2002){Tremaine}, {Gebhardt}, {Bender}, {Bower},
  {Dressler}, {Faber}, {Filippenko}, {Green}, {Grillmair}, {Ho}, {Kormendy},
  {Lauer}, {Magorrian}, {Pinkney}, \& {Richstone}}]{Tremaine-ApJ02-BHsigma}
{Tremaine}, S., {Gebhardt}, K., {Bender}, R., {et~al.} 2002, ApJ, 574, 740

\bibitem[{{van der Wel} {et~al.}(2011){van der Wel}, {Rix}, {Wuyts}, {McGrath},
  {Koekemoer}, {Bell}, {Holden}, {Robaina}, \& {McIntosh}}]{vdW-ApJ11-diskgal}
{van der Wel}, A., {Rix}, H.-W., {Wuyts}, S., {et~al.} 2011, ApJ, 730, 38

\bibitem[{{Wang} {et~al.}(2010){Wang}, {Cowie}, {Barger}, {Keenan}, \&
  {Ting}}]{Wang-ApJS10-GOODSNKs}
{Wang}, W.-H., {Cowie}, L.~L., {Barger}, A.~J., {Keenan}, R.~C., \& {Ting},
  H.-C. 2010, ApJS, 187, 251

\bibitem[{{Williams} {et~al.}(2009){Williams}, {Quadri}, {Franx}, {van Dokkum},
  \& {Labb{\'e}}}]{Williams-ApJ09-quiegal}
{Williams}, R.~J., {Quadri}, R.~F., {Franx}, M., {van Dokkum}, P., \&
  {Labb{\'e}}, I. 2009, ApJ, 691, 1879

\bibitem[{{Wirth} {et~al.}(2004){Wirth}, {Willmer}, {Amico}, {Chaffee},
  {Goodrich}, {Kwok}, {Lyke}, {Mader}, {Tran}, {Barger}, {Cowie}, {Capak},
  {Coil}, {Cooper}, {Conrad}, {Davis}, {Faber}, {Hu}, {Koo}, {Le Mignant},
  {Newman}, \& {Songaila}}]{Wirth-AJ04-TKRS}
{Wirth}, G.~D., {Willmer}, C.~N.~A., {Amico}, P., {et~al.} 2004, AJ, 127, 3121

\bibitem[{{Wolf} {et~al.}(2008){Wolf}, {Hildebrandt}, {Taylor}, \&
  {Meisenheimer}}]{Wolf-AA08-COMBO}
{Wolf}, C., {Hildebrandt}, H., {Taylor}, E.~N., \& {Meisenheimer}, K. 2008,
  A\&A, 492, 933

\bibitem[{{Wolf} {et~al.}(2004){Wolf}, {Meisenheimer}, {Kleinheinrich},
  {Borch}, {Dye}, {Gray}, {Wisotzki}, {Bell}, {Rix}, {Cimatti}, {Hasinger}, \&
  {Szokoly}}]{Wolf-AA04-COMBO}
{Wolf}, C., {Meisenheimer}, K., {Kleinheinrich}, M., {et~al.} 2004, A\&A, 421,
  913

\bibitem[{{Wuyts} {et~al.}(2011){Wuyts}, {F{\"o}rster Schreiber}, {van der
  Wel}, {Magnelli}, {Guo}, {Genzel}, {Lutz}, {Aussel}, {Barro}, {Berta},
  {Cava}, {Graci{\'a}-Carpio}, {Hathi}, {Huang}, {Kocevski}, {Koekemoer},
  {Lee}, {Le Floc'h}, {McGrath}, {Nordon}, {Popesso}, {Pozzi}, {Riguccini},
  {Rodighiero}, {Saintonge}, \& {Tacconi}}]{Wuyts-ApJ11-CANDELS}
{Wuyts}, S., {F{\"o}rster Schreiber}, N.~M., {van der Wel}, A., {et~al.} 2011,
  ApJ, 742, 96

\bibitem[{{Wuyts} {et~al.}(2007){Wuyts}, {Labb{\'e}}, {Franx}, {Rudnick}, {van
  Dokkum}, {Fazio}, {F{\"o}rster Schreiber}, {Huang}, {Moorwood}, {Rix},
  {R{\"o}ttgering}, \& {van der Werf}}]{Wuyts-ApJ07-IRAC}
{Wuyts}, S., {Labb{\'e}}, I., {Franx}, M., {et~al.} 2007, ApJ, 655, 51

\bibitem[{{Xue} {et~al.}(2011){Xue}, {Luo}, {Brandt}, {Bauer}, {Lehmer},
  {Broos}, {Schneider}, {Alexander}, {Brusa}, {Comastri}, {Fabian}, {Gilli},
  {Hasinger}, {Hornschemeier}, {Koekemoer}, {Liu}, {Mainieri}, {Paolillo},
  {Rafferty}, {Rosati}, {Shemmer}, {Silverman}, {Smail}, {Tozzi}, \&
  {Vignali}}]{Xue-ApJS11-CDFS}
{Xue}, Y.~Q., {Luo}, B., {Brandt}, W.~N., {et~al.} 2011, ApJS, 195, 10

\end{thebibliography}

\clearpage

\begin{deluxetable}{lrrccc}
\tabletypesize{\small}
\tablecolumns{6}
\tablewidth{0pc}
\tablecaption{ACS galaxy catalogue survey fields.\label{tab:surveys}}
\tablehead{
 \colhead{Survey}      &
 \colhead{RA}          &
 \colhead{Dec}         &
 \colhead{area}        &
 \colhead{filters}     &
 \colhead{pixel scale} \\
 \colhead{}            &
 \colhead{J2000}       &
 \colhead{J2000}       &
 \colhead{deg$^2$}     &
 \colhead{}            &
 \colhead{arcsec\,pix$^{-1}$}}
\startdata
COSMOS     & 10:00:28 & +02:12:21 &  1.8   & F814W           & 0.05 \\
AEGIS      & 14:17:00 & +52:30:00 &  0.197 & F606W \& F814W  & 0.03 \\
GEMS       & 03:32:25 & -27:48:50 &  0.21  & F606W \& F850LP & 0.03 \\
GOODS-S    & 03:32:30 & -27:48:20 &  0.07  & F606W \& F850LP & 0.03 \\
GOODS-N    & 12:36:55 & +62:14:15 &  0.07  & F606W \& F775W  & 0.03
\enddata
\end{deluxetable}

\begin{deluxetable}{lrrrcrr}
\tabletypesize{\small}
\tablecolumns{6}
\tablewidth{0pc}
\tablecaption{catalogue selection statistics split by field ($0.4\leq z \leq 1.0$).\label{tab:n-summ}}
\tablehead{
 \colhead{Field}               &
 \colhead{total}               &
 \colhead{spec-$z$}            &
 \colhead{$n \leq 1.5$}        &
 \colhead{$1.5 < n \leq 3.0$}  &
 \colhead{$n > 3.0$}           }
\startdata
COSMOS       & 31\,714 &  3\,116 &  14\,139 & 7\,259 & 10\,316 \\
AEGIS        &  2\,848 &  1\,451 &   1\,588 &    576 &     684 \\
GEMS         &  3\,595 &  1\,382 &   2\,267 &    793 &     535 \\
GOODS-S      &     852 &     524 &      482 &    199 &     171 \\
GOODS-N      &     843 &     648 &      749 &     74 &      20 \\[1ex]
\emph{total} & 39\,852 &  7\,121 &  19\,225 & 8\,901 & 11\,726 \\
\enddata
\end{deluxetable}

\begin{deluxetable}{crrr}
\tabletypesize{\small}
\tablecolumns{6}
\tablewidth{0pc}
\tablecaption{Morphological classification comparison of the COSMOS sources
              using ZEST \citep{Scarlata-ApJS07-ZEST} and \textsc{Galfit} (this work).
              \label{tab:zurich-comp}}
\tablehead{
 \colhead{ZEST type}           &
 \colhead{$n \leq 1.5$}        &
 \colhead{$1.5 < n \leq 3.0$}  &
 \colhead{$n > 3.0$}           }
\startdata
 1.0       &      13 &     130 &  2\,596 \\
 2.0       &      24 &     158 &  1\,966 \\
 2.1       &     570 &  2\,892 &  2\,760 \\
 2.2       &  4\,162 &  2\,530 &  1\,082 \\
 2.3       &  7\,115 &     723 &     698 \\
 3.0       &  1\,796 &     497 &     392 \\[0.5ex]
\emph{total}  & 13\,680 &  6\,930 &  9\,494
\enddata
\end{deluxetable}
\clearpage

\begin{deluxetable}{llc rrc cc}
\tabletypesize{\scriptsize}
\tablecolumns{7}
\tablewidth{0pc}
\tablecaption{X-ray selected AGN.\label{tab:AGN-X}}
\tablehead{
 \colhead{ID}               &
 \colhead{Morph ID}         &
 \colhead{Survey}           & 
 \colhead{$z_\mrm{spec}$}   & 
 \colhead{$z_\mrm{phot}$}   &
 \colhead{$L_\mrm{X}$}      & 
 \colhead{HR}               & 
 \colhead{AGN Type}         }
\startdata
1  & 13033960    & AEGIS-X     & 0.763 & \nodata & 5.352$\times 10^{42}$   &  -0.36  &  I \\
2  & 20006036    & XMM-COSMOS  & 0.661 & 0.6978  & 5.448$\times 10^{43}$   &   0.47  & II \\
3  & 20016006    & C-COSMOS    & 0.932 & \nodata & 1.181$\times 10^{43}$   &   0.23  &  I \\
4  & 20016854    & C-COSMOS    & 0.853 & \nodata & 1.152$\times 10^{43}$   &   1.0   &  I \\
5  & 20038166    & C-COSMOS    & 0.654 & \nodata & 2.157$\times 10^{42}$   &  -1.0   & II \\
6  & 20043608    & C-COSMOS    & 0.497 & 0.474   & 1.248$\times 10^{42}$   &   1.0   &  I \\
7  & 20067818    & C-COSMOS    & 0.689 & \nodata & 3.044$\times 10^{42}$   &   1.0   &  I \\
8  & 20091912    & C-COSMOS    & 0.984 & \nodata & 7.940$\times 10^{42}$   &  -1.0   & II \\
9  & 20096276    & C-COSMOS    & 0.847 & \nodata & 9.176$\times 10^{42}$   &  -0.29  & II \\
10 & 20101857    & C-COSMOS    & 0.651 & 0.6328  & 5.340$\times 10^{42}$   &  -0.17  &  I \\
11 & 20116906    & C-COSMOS    & 0.717 & \nodata & 2.955$\times 10^{42}$   &   1.0   &  I \\
12 & 20124624    & C-COSMOS    & 0.934 & \nodata & 6.306$\times 10^{42}$   &   1.0   &  I \\
13 & 20145070    & XMM-COSMOS  & 0.804 & \nodata & 5.517$\times 10^{43}$   &  -0.52  &  I \\
14 & 20183465    & C-COSMOS    & 0.728 & \nodata & 7.158$\times 10^{42}$   &  -1.0   & II \\
15 & 20188370    & XMM-COSMOS  & 0.406 & \nodata & 2.053$\times 10^{44}$   &  -0.27  &  I \\
16 & 20195289    & XMM-COSMOS  & 0.968 & \nodata & 2.762$\times 10^{43}$   &  -0.42  &  I(II \textit{spec}) \\
17 & 50007691    & CDF-N       & 0.77  & 0.7517  & 3.416$\times 10^{41}$   &  -0.078 & II \\
18 & 50016371    & CDF-N       & 0.56  & 0.6645  & 7.635$\times 10^{41}$   &  -0.052 & II \\
19 & 90002361    & E-CDF-S     & 0.673 & \nodata & 5.615$\times 10^{41}$   &   0.162 & II \\
20 & 90003063    & E-CDF-S     & 0.755 & \nodata & 6.739$\times 10^{42}$   &  -0.26  &  I \\
21 & 90004178    & E-CDF-S     & 0.960 & \nodata & 6.621$\times 10^{42}$   &   0.050 & II \\
22 & 90004955    & E-CDF-S     & 0.633 & \nodata & 1.605$\times 10^{43}$   &  -0.049 & II \\
23 & 90011317    & CDF-S       & 0.622 & \nodata & 6.281$\times 10^{42}$   &   0.550 & II \\
24 & 90011476    & E-CDF-S     & 0.572 & \nodata & 2.262$\times 10^{43}$   &  -0.059 & II \\
25 & 90014273    & CDF-S       & 0.750 & \nodata & 5.598$\times 10^{42}$   &   0.386 & II \\
26 & 90021223    & CDF-S       & 0.507 & \nodata & 1.943$\times 10^{42}$   &   0.299 & II \\
27 & 90021708    & CDF-S       & 0.628 & \nodata & 2.865$\times 10^{41}$   &   0.089 & II \\
28 & 90028848    & E-CDF-S     & 0.686 & 0.72    & 8.066$\times 10^{42}$   &   0.039 & II \\
29 & 90033648    & CDF-S       & 0.789 & 0.784   & 2.821$\times 10^{42}$   &  -0.104 & II \\
30 & 90038381    & E-CDF-S     & 0.554 & \nodata & 2.090$\times 10^{42}$   &   0.423 & II \\
\enddata
\vspace{-0.6cm}
\tablecomments{Column (1) indicates the catalogue object ID.
               Column (2) lists the original X-ray catalogue of each source. 
               Columns (3)--(4) refer to the spectroscopic and photometric redshift.
               Column (5) is the rest-frame luminosity expressed in erg\,s$^{-1}$. 
               Column (6) is the hardness ratio. 
               Column (7) indicates the AGN type.}
\end{deluxetable} 
\clearpage

\begin{deluxetable}{lcccccc}
\tabletypesize{\scriptsize}
\tablecolumns{7}
\tablewidth{0pc}
\tablecaption{Morphological parameters from surface brightness profile fitting using
              a single S\'ersic profile or a combination of a S\'ersic profile plus
              an exponential disc.\label{tab:morphfit}}
\tablehead{
 \colhead{ID}         &
 \colhead{ID MORPH}   &
 \colhead{$\mu_e$}    &
 \colhead{$R_e$}      &
 \colhead{$n$}        &
 \colhead{$\mu_d^0$}  &
 \colhead{$h_d$}      \\ 
 \colhead{}                  &
 \colhead{}                  &
 \colhead{(mag/arcsec$^2$)}  &
 \colhead{(arcsec)}          & 
 \colhead{}                  &
 \colhead{(mag/arcsec$^2$)}  &
 \colhead{(arcsec)}          }
\startdata
1  & 13033960 & $21.972\pm 0.023$ & $0.250\pm 0.004$ & $0.87\pm 0.02$ & $21.530\pm 0.048$ & $0.89\pm 0.03$ \\
2  & 20006036 & $22.831\pm 0.038$ & $0.538\pm 0.020$ & $0.44\pm 0.04$ &     \nodata       &  \nodata       \\
3  & 20016006 & $22.166\pm 0.005$ & $0.416\pm 0.003$ & $0.80\pm 0.01$ &     \nodata       &  \nodata       \\
4  & 20016854 & $20.469\pm 0.029$ & $0.106\pm 0.002$ & $0.62\pm 0.02$ & $21.004\pm 0.028$ & $0.31\pm 0.01$ \\
5  & 20038166 & $24.051\pm 0.021$ & $0.371\pm 0.005$ & $1.16\pm 0.03$ &     \nodata       &  \nodata       \\
6  & 20043608 & $23.278\pm 0.024$ & $1.168\pm 0.013$ & $1.25\pm 0.03$ &     \nodata       &  \nodata       \\
7  & 20067818 & $21.464\pm 0.007$ & $0.206\pm 0.001$ & $0.78\pm 0.01$ &     \nodata       &  \nodata       \\
8  & 20091912 & $23.834\pm 0.005$ & $0.923\pm 0.006$ & $0.93\pm 0.01$ &     \nodata       &  \nodata       \\
9  & 20096276 & $21.776\pm 0.013$ & $0.190\pm 0.002$ & $0.82\pm 0.01$ &     \nodata       &  \nodata       \\
10 & 20101857 & $21.466\pm 0.055$ & $0.112\pm 0.004$ & $0.71\pm 0.05$ & $20.647\pm 0.010$ & $0.68\pm 0.01$ \\
11 & 20116906 & $21.277\pm 0.009$ & $0.280\pm 0.002$ & $0.73\pm 0.01$ &     \nodata       &  \nodata       \\
12 & 20124624 & $23.386\pm 0.017$ & $0.664\pm 0.008$ & $0.92\pm 0.01$ &     \nodata       &  \nodata       \\
13 & 20145070 & $23.733\pm 0.038$ & $0.139\pm 0.003$ & $0.55\pm 0.03$ & $22.080\pm 0.013$ & $0.92\pm 0.01$ \\
14 & 20183465 & $21.737\pm 0.011$ & $0.447\pm 0.004$ & $0.85\pm 0.01$ &     \nodata       &  \nodata       \\
15 & 20188370 & $24.525\pm 0.009$ & $0.612\pm 0.007$ & $0.95\pm 0.02$ &     \nodata       &  \nodata       \\
16 & 20195289 & $23.153\pm 0.023$ & $0.474\pm 0.011$ & $0.55\pm 0.02$ &     \nodata       &  \nodata       \\
17 & 50007691 & $21.653\pm 0.022$ & $0.197\pm 0.002$ & $0.57\pm 0.01$ & $20.945\pm 0.023$ & $0.34\pm 0.01$ \\
18 & 50016371 & $20.086\pm 0.013$ & $0.099\pm 0.001$ & $0.61\pm 0.01$ & $20.180\pm 0.005$ & $0.32\pm 0.01$ \\
19 & 90002361 & $21.995\pm 0.159$ & $0.178\pm 0.022$ & $0.33\pm 0.07$ & $21.077\pm 0.024$ & $0.78\pm 0.01$ \\
20 & 90003063 & $21.541\pm 0.011$ & $0.426\pm 0.005$ & $0.55\pm 0.02$ &     \nodata       &  \nodata       \\
21 & 90004178 & $23.584\pm 0.086$ & $0.534\pm 0.030$ & $0.84\pm 0.11$ &     \nodata       &  \nodata       \\
22 & 90004955 & $22.945\pm 0.047$ & $0.319\pm 0.009$ & $1.06\pm 0.05$ &     \nodata       &  \nodata       \\
23 & 90011317 & $19.546\pm 0.003$ & $0.246\pm 0.001$ & $0.77\pm 0.01$ &     \nodata       &  \nodata       \\
24 & 90011476 & $19.900\pm 0.004$ & $0.245\pm 0.001$ & $1.19\pm 0.01$ &     \nodata       &  \nodata       \\
25 & 90014273 & $20.428\pm 0.014$ & $0.467\pm 0.006$ & $0.94\pm 0.01$ &     \nodata       &  \nodata       \\
26 & 90021223 & $21.382\pm 0.601$ & $0.149\pm 0.044$ & $1.07\pm 0.18$ &     \nodata       &  \nodata       \\  
27 & 90021708 & $21.235\pm 0.009$ & $0.202\pm 0.002$ & $0.84\pm 0.01$ &     \nodata       &  \nodata       \\
28 & 90028848 & $23.103\pm 0.110$ & $0.519\pm 0.038$ & $0.96\pm 0.11$ &     \nodata       &  \nodata       \\    
29 & 90033648 & $21.371\pm 0.402$ & $0.065\pm 0.018$ & $0.53\pm 0.18$ & $20.546\pm 0.015$ & $0.43\pm 0.01$ \\
30 & 90038381 & $21.211\pm 0.008$ & $0.374\pm 0.002$ & $1.37\pm 0.01$ &     \nodata       &  \nodata       \\
 \enddata
\vspace{-0.6cm}
\tablecomments{Column (1) indicates the catalogue object ID. Columns (2)--(4) refer to the S\'ersic profile 
               parameters: sufrace britghness, $\mu_e$, scale length, $R_e$, and the S\'ersic index $n$.
               Columns (5)--(6) are the corresponding disc profile coefficients.}
\end{deluxetable}
\clearpage

\clearpage

\clearpage

\clearpage

\clearpage

\clearpage

\clearpage

\begin{figure}
\includegraphics[width=4cm]{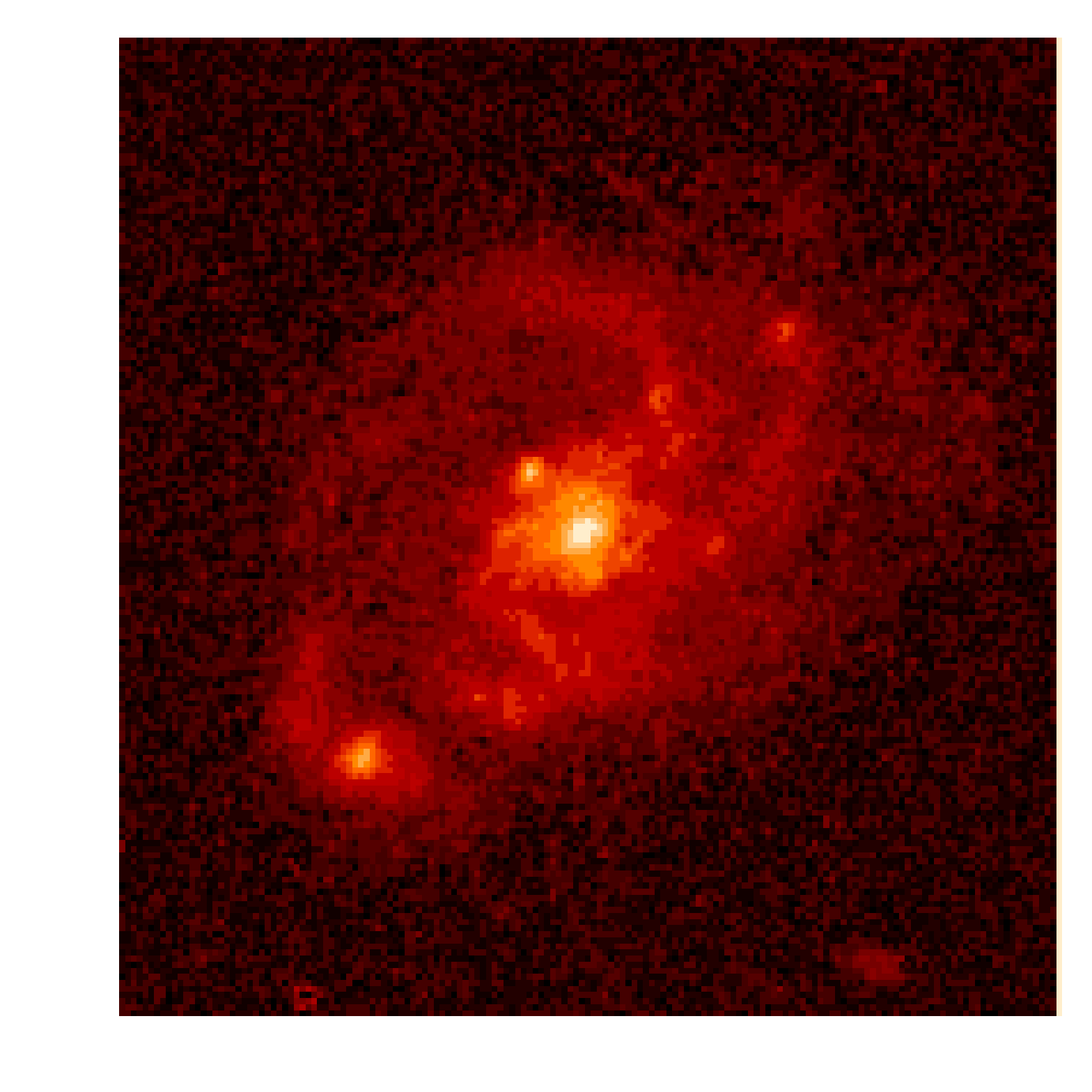}
\includegraphics[width=4cm]{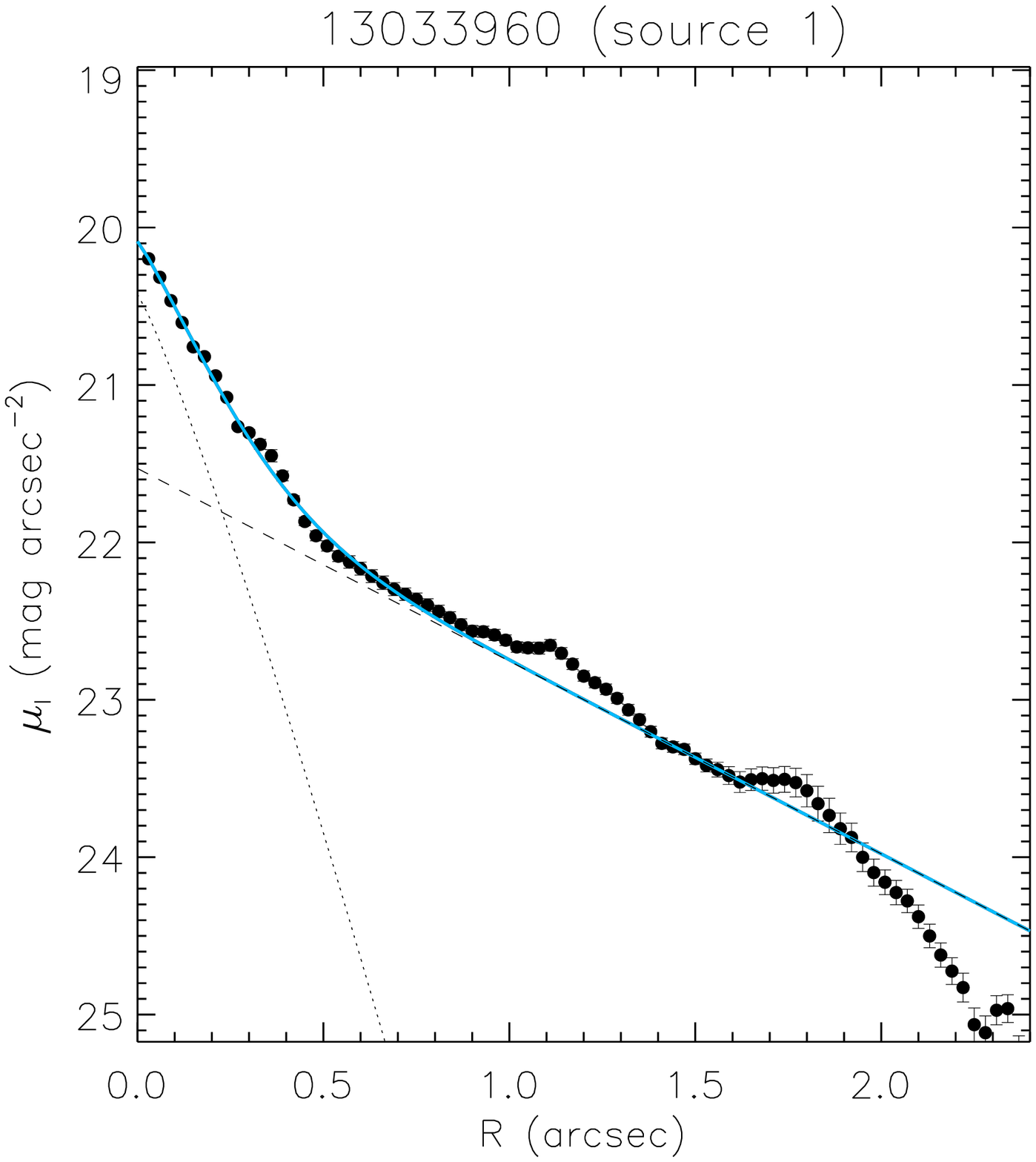}
\includegraphics[width=4cm]{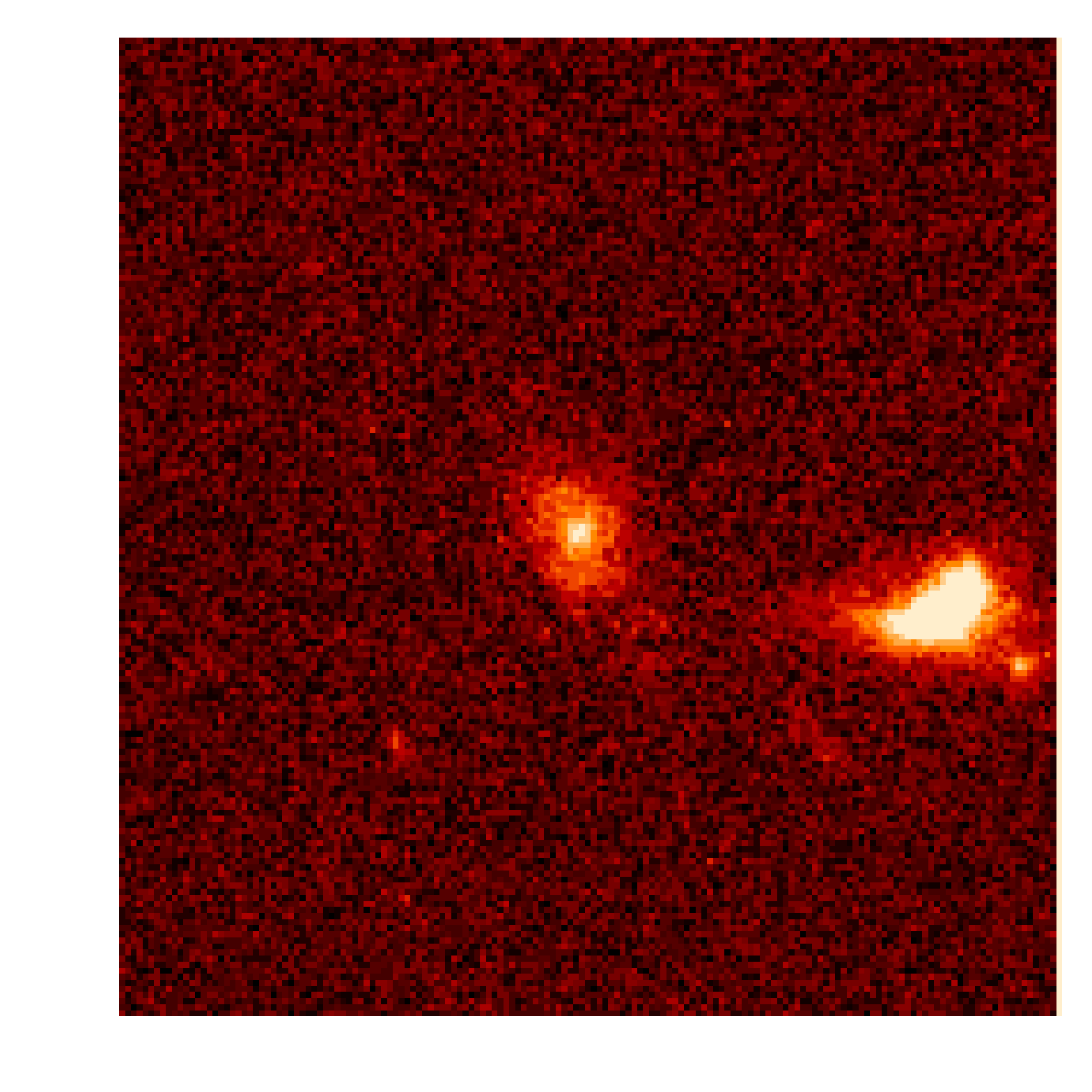}
\includegraphics[width=4cm]{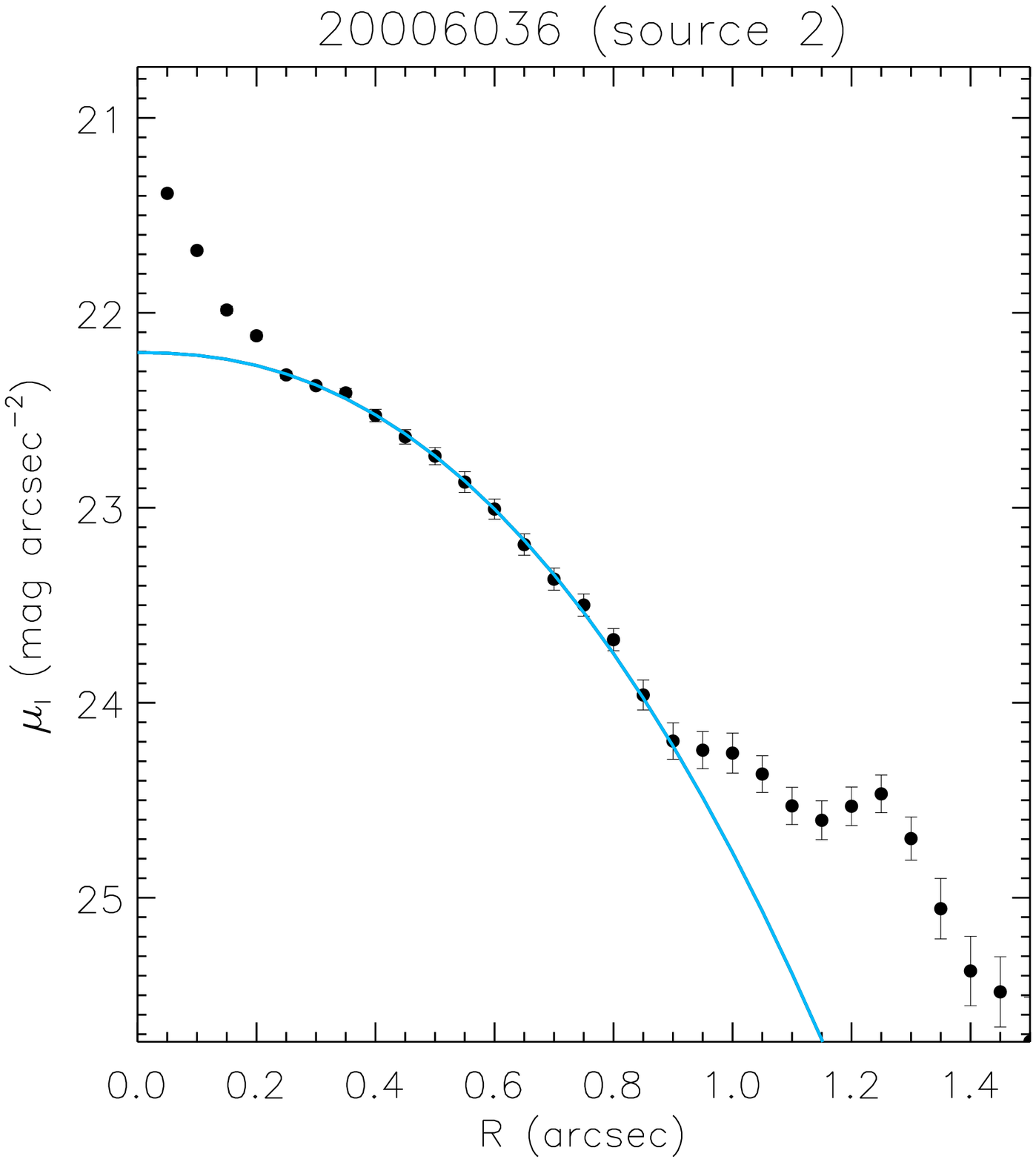}
\includegraphics[width=4cm]{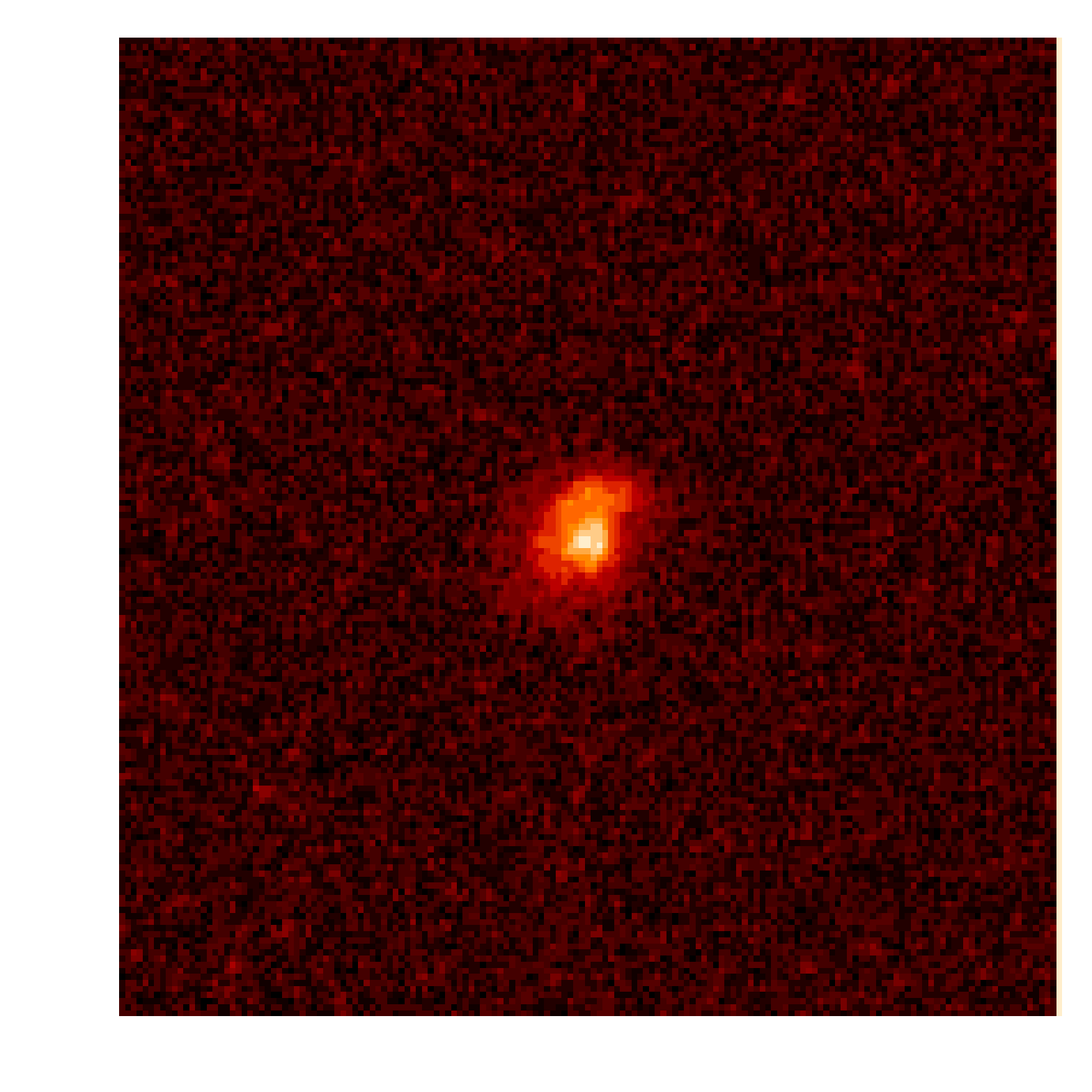}
\includegraphics[width=4cm]{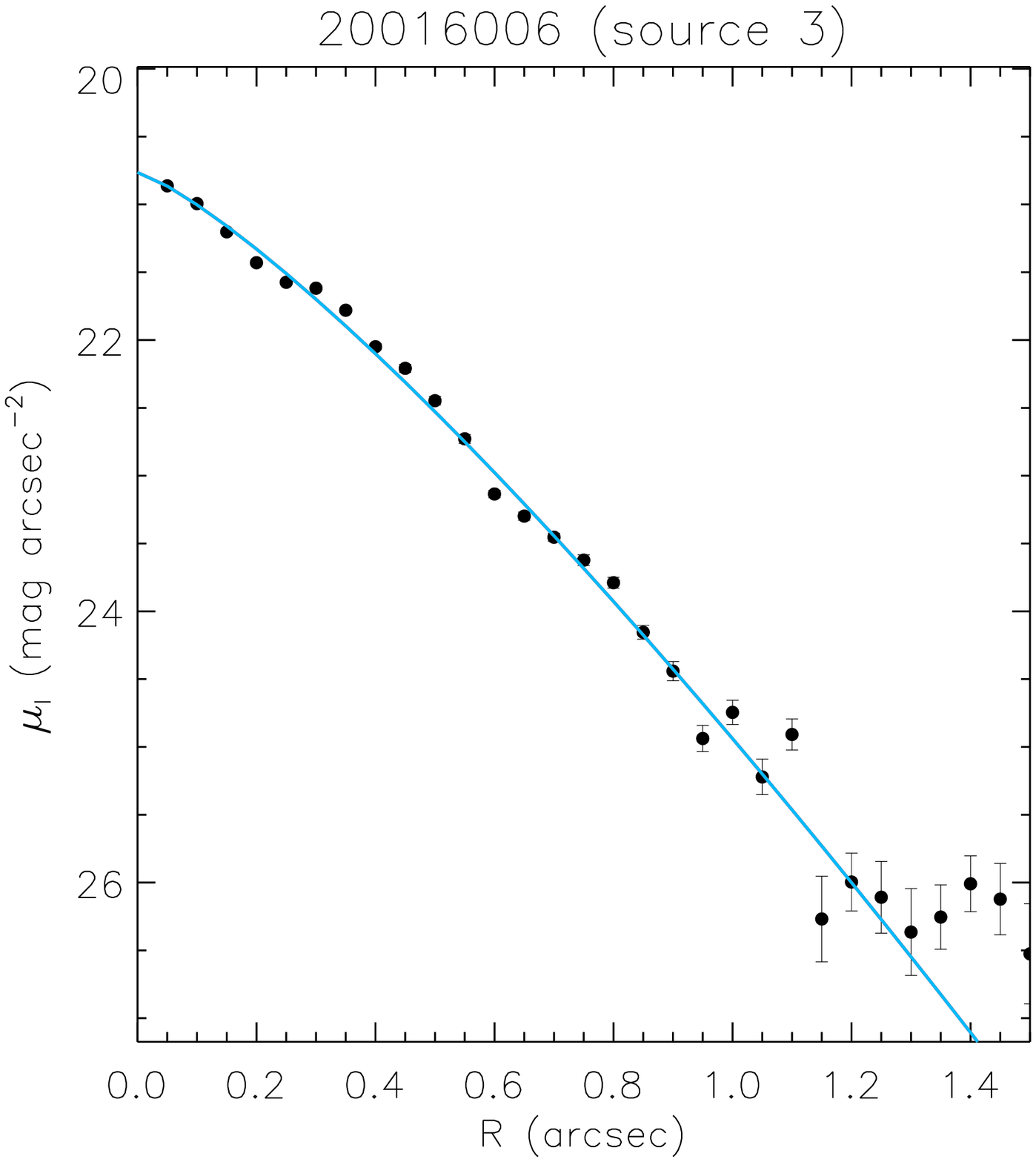}
\includegraphics[width=4cm]{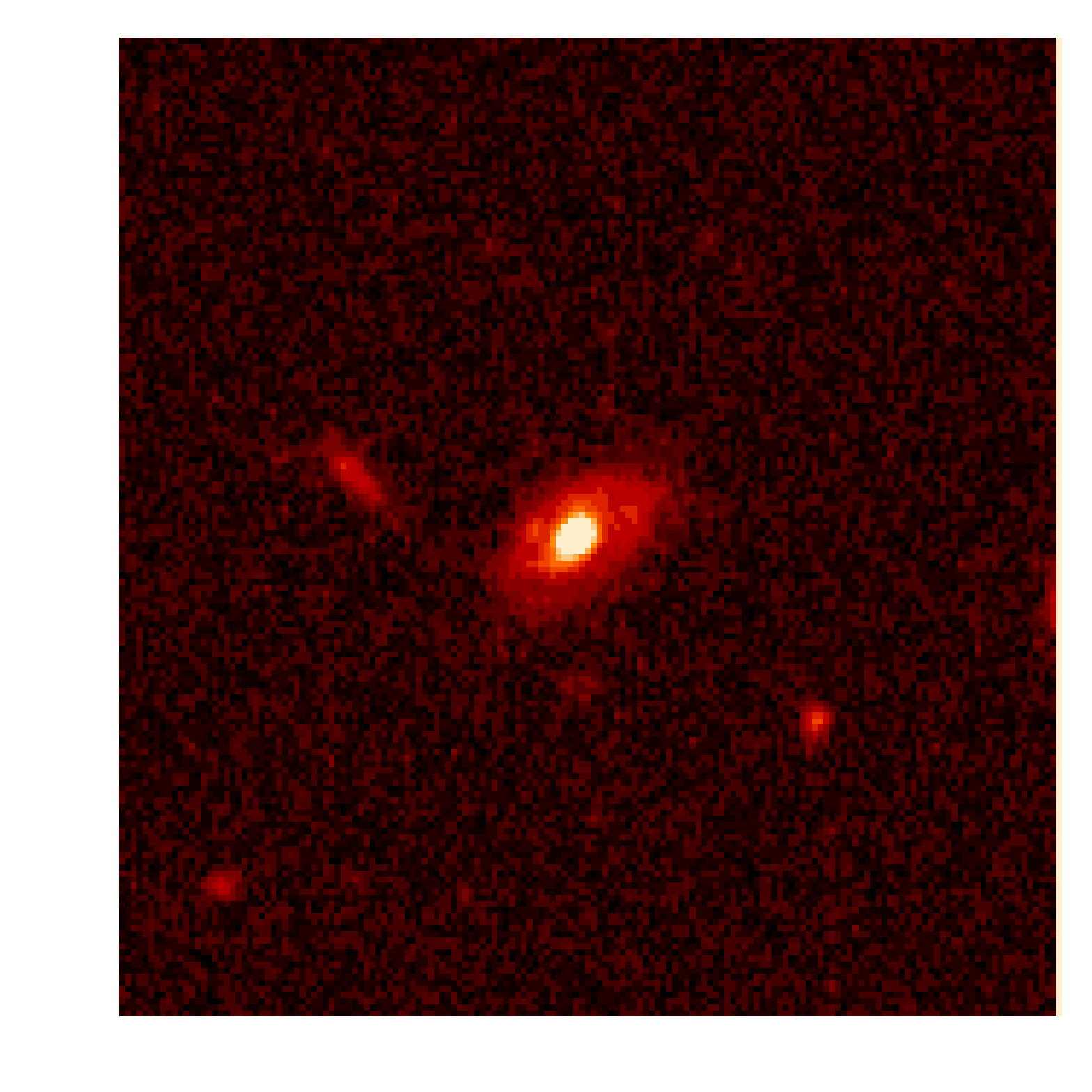}
\includegraphics[width=4cm]{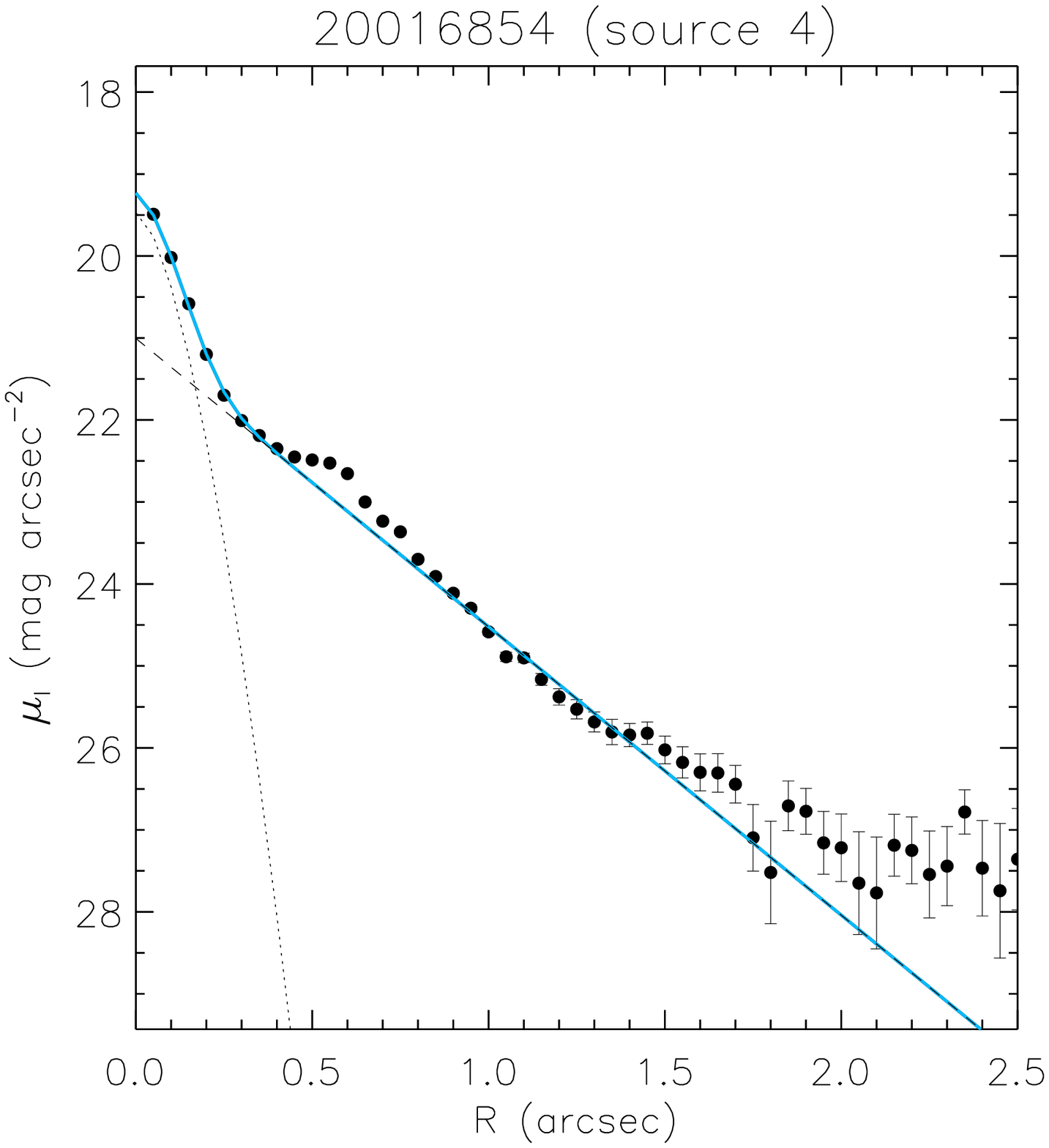}
\includegraphics[width=4cm]{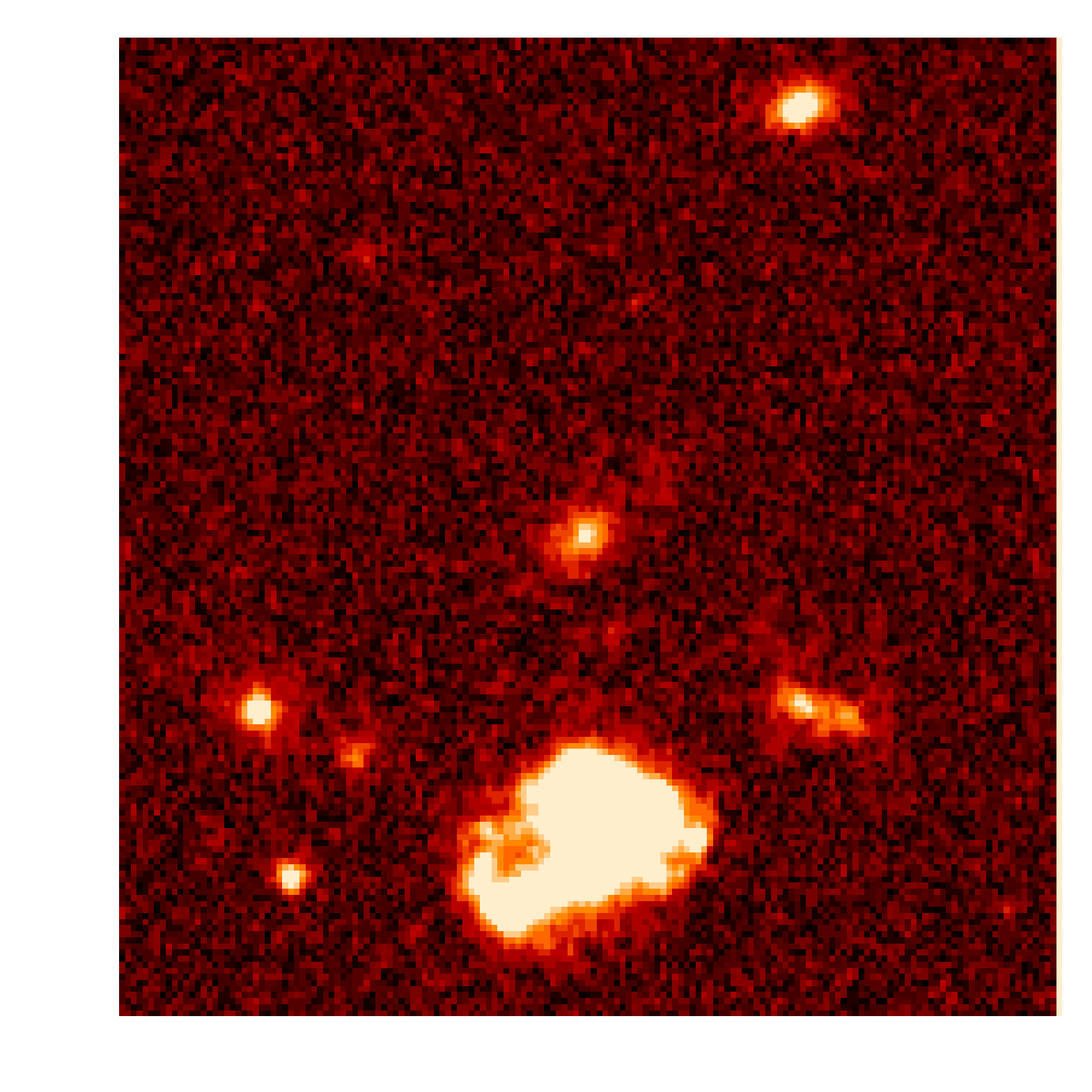}
\includegraphics[width=4cm]{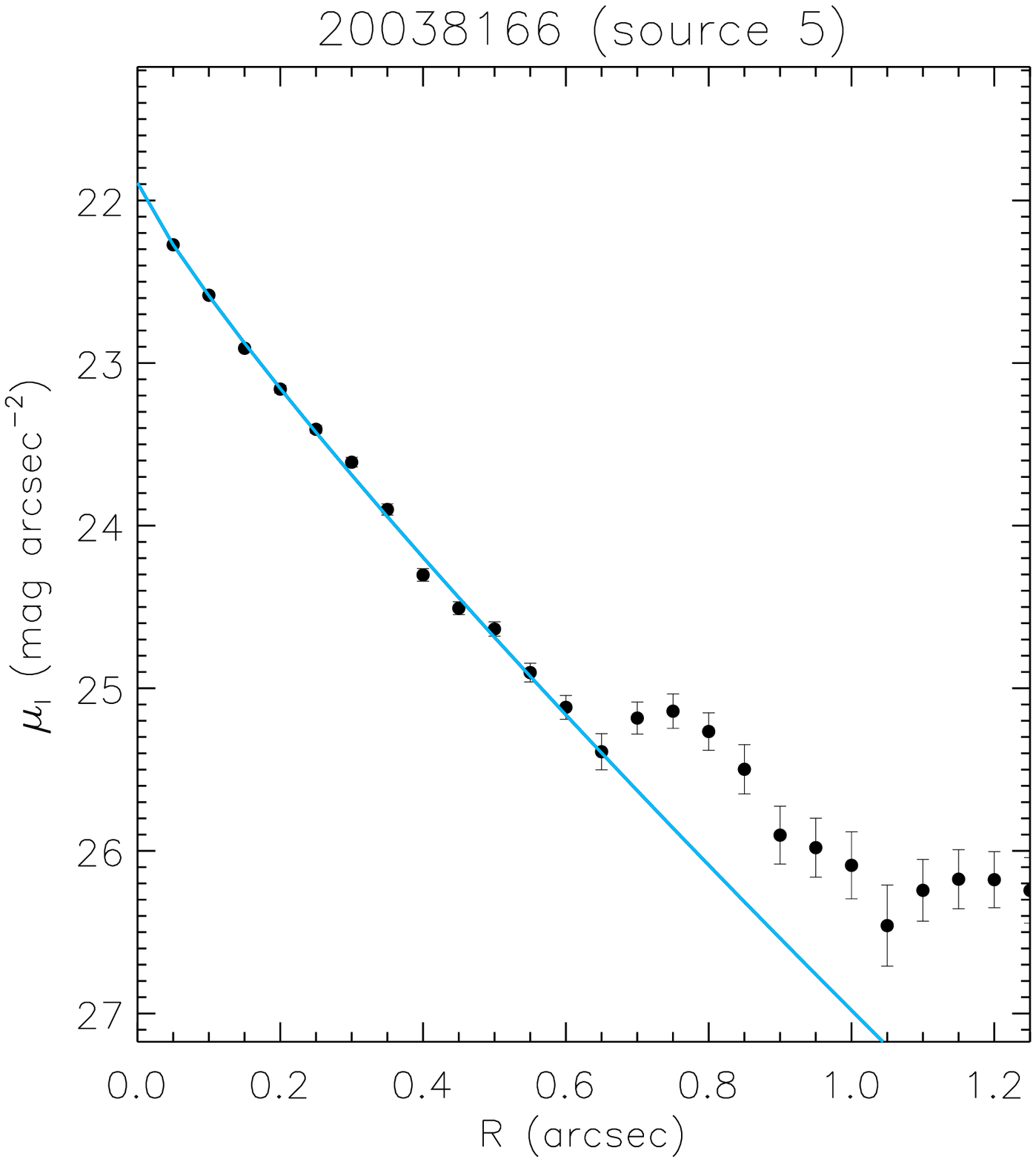}
\includegraphics[width=4cm]{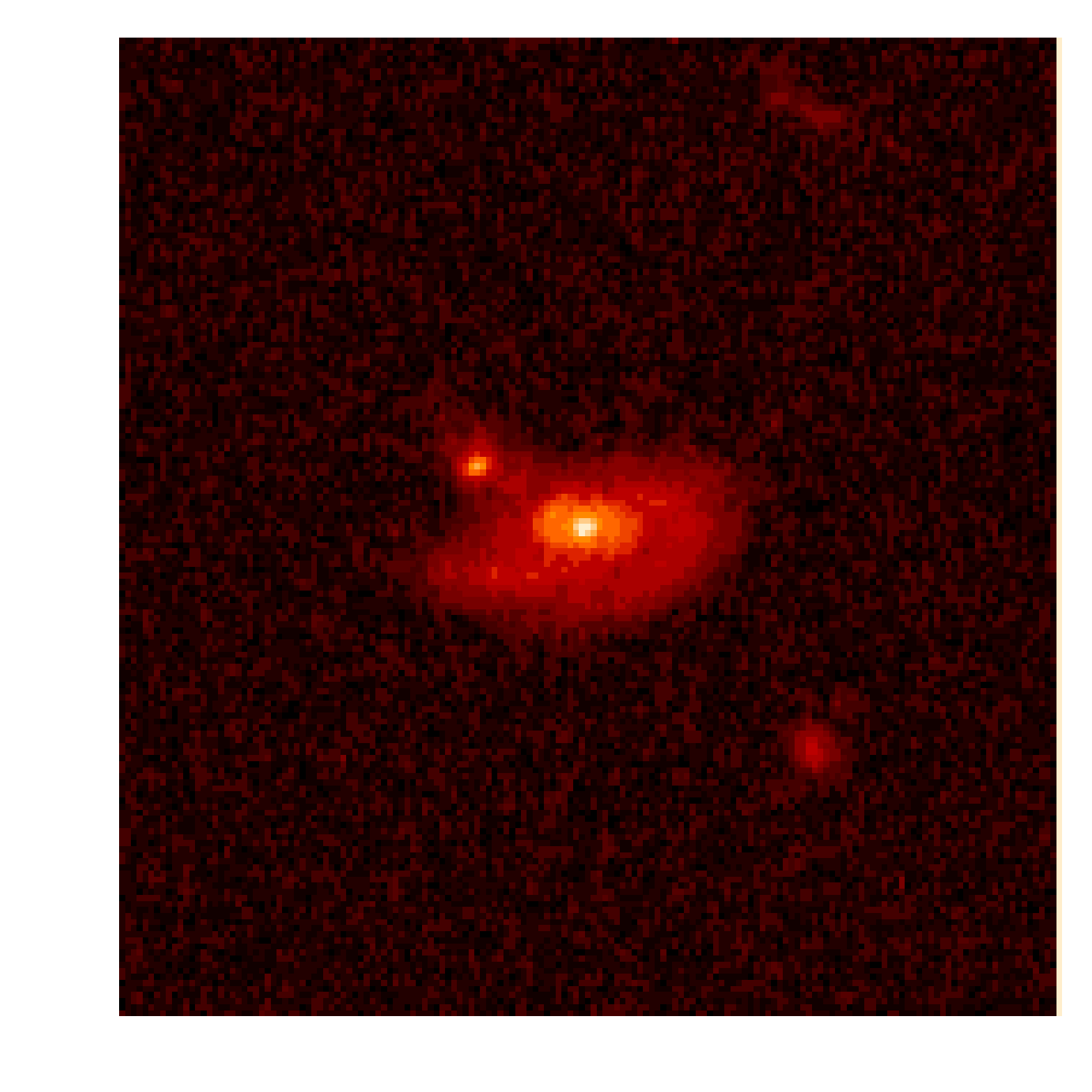}
\includegraphics[width=4cm]{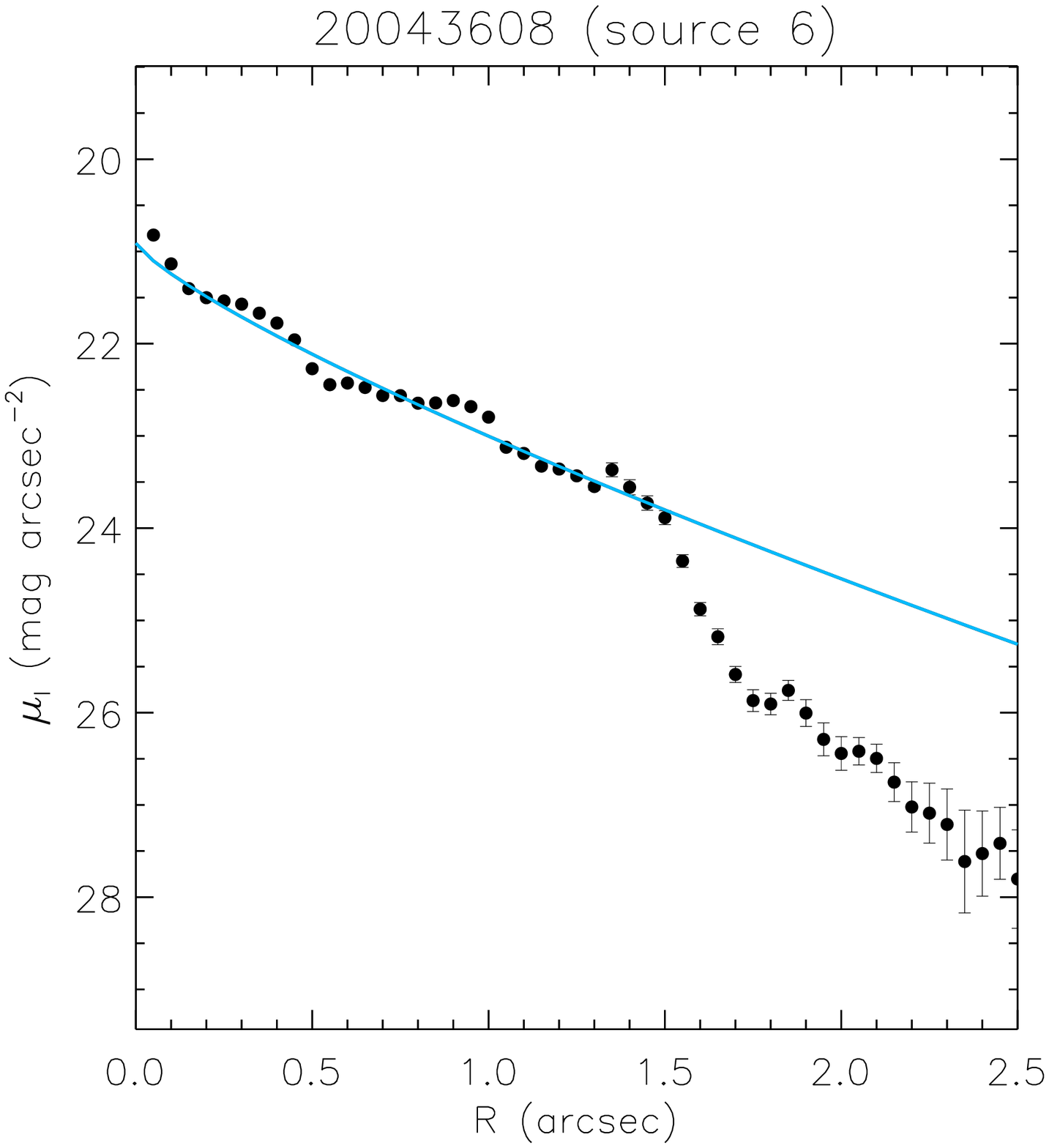}
\includegraphics[width=4cm]{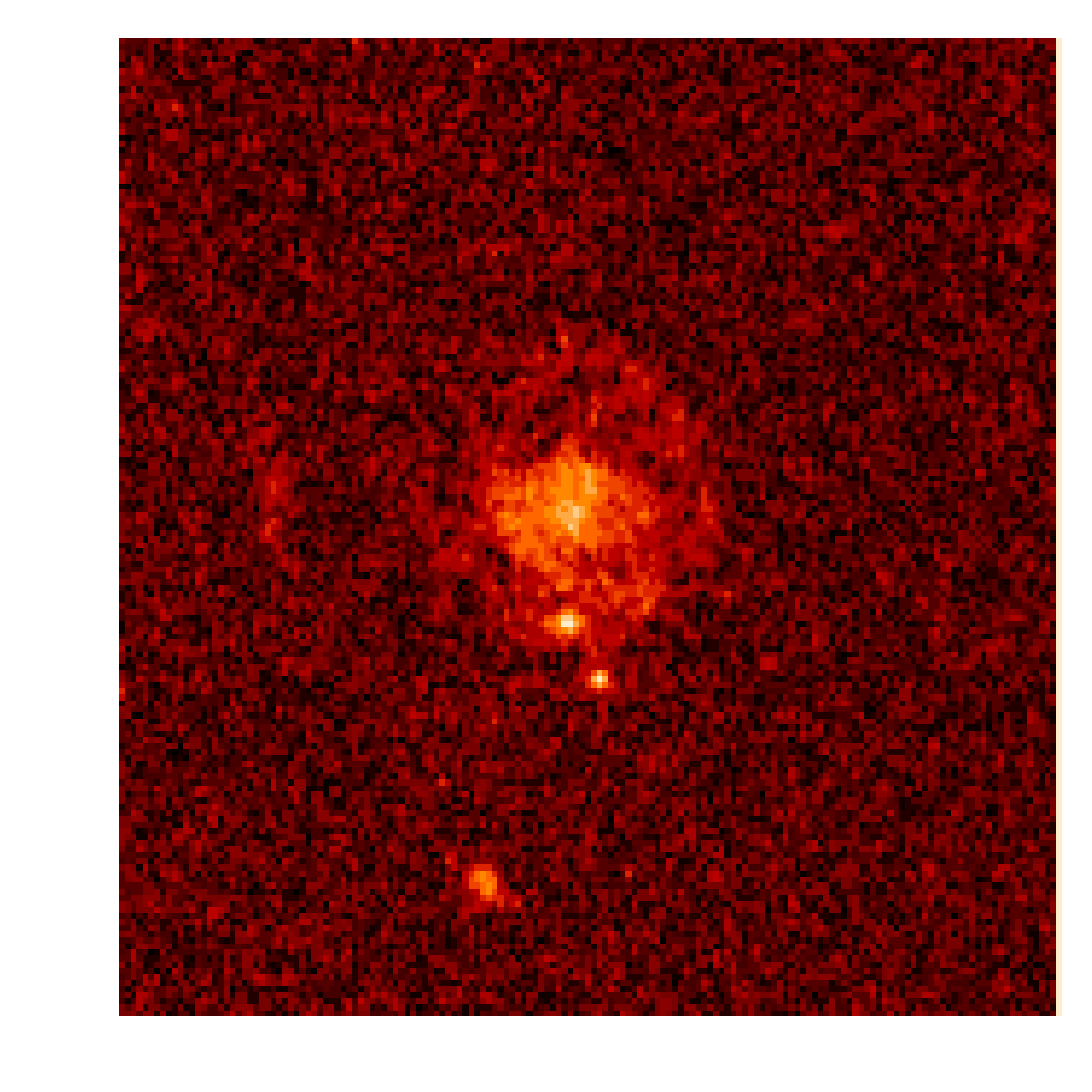}
\includegraphics[width=4cm]{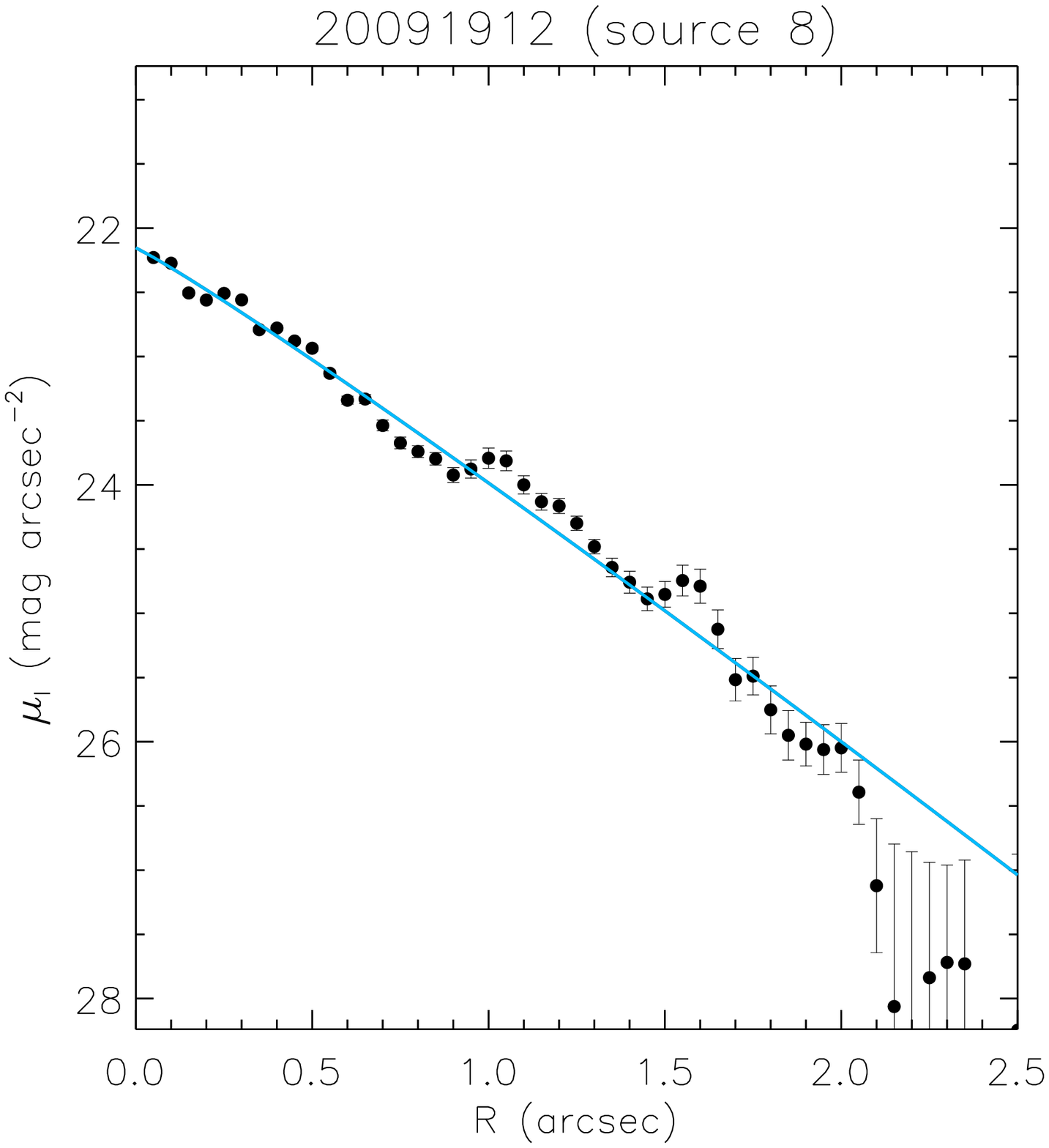}
\includegraphics[width=4cm]{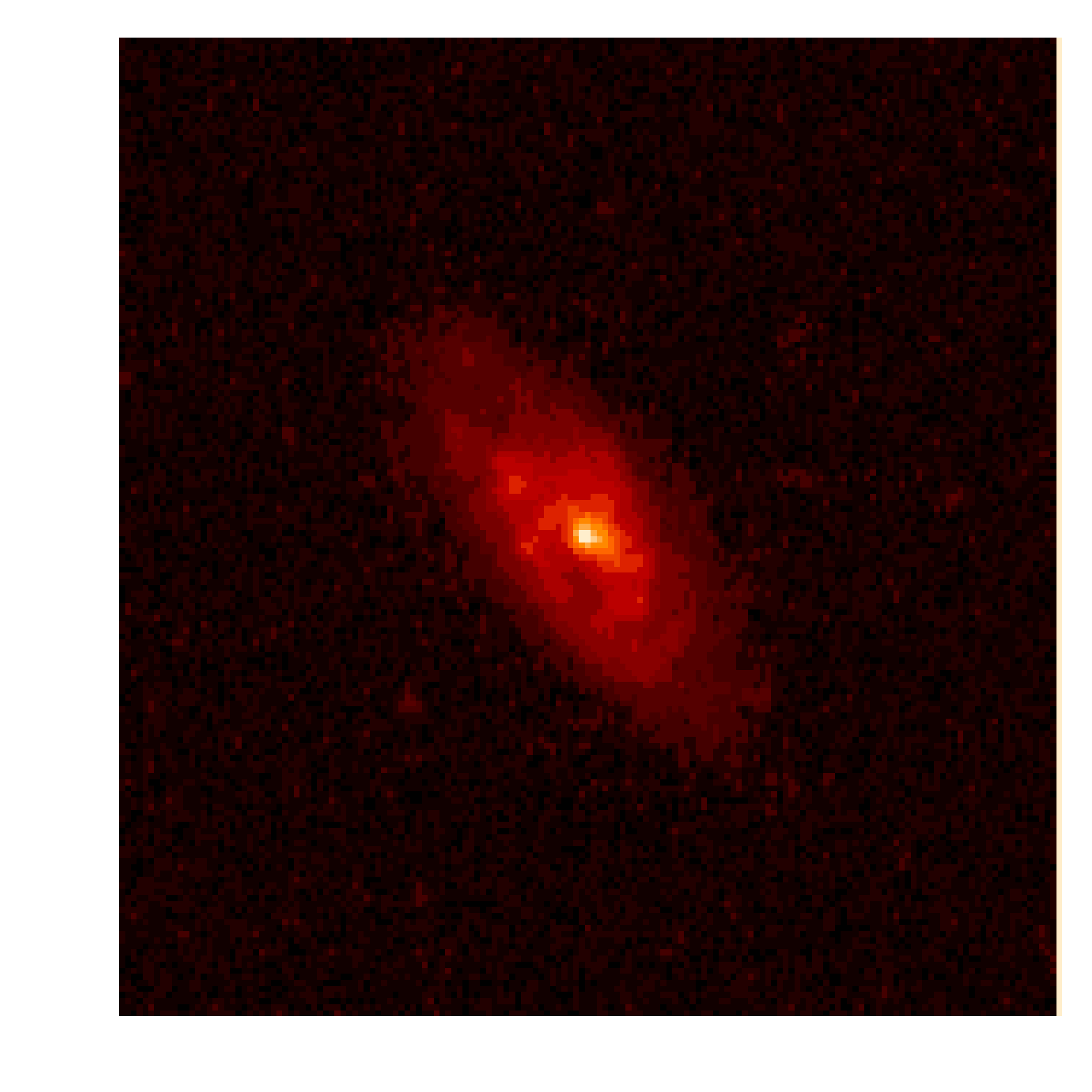}
\includegraphics[width=4cm]{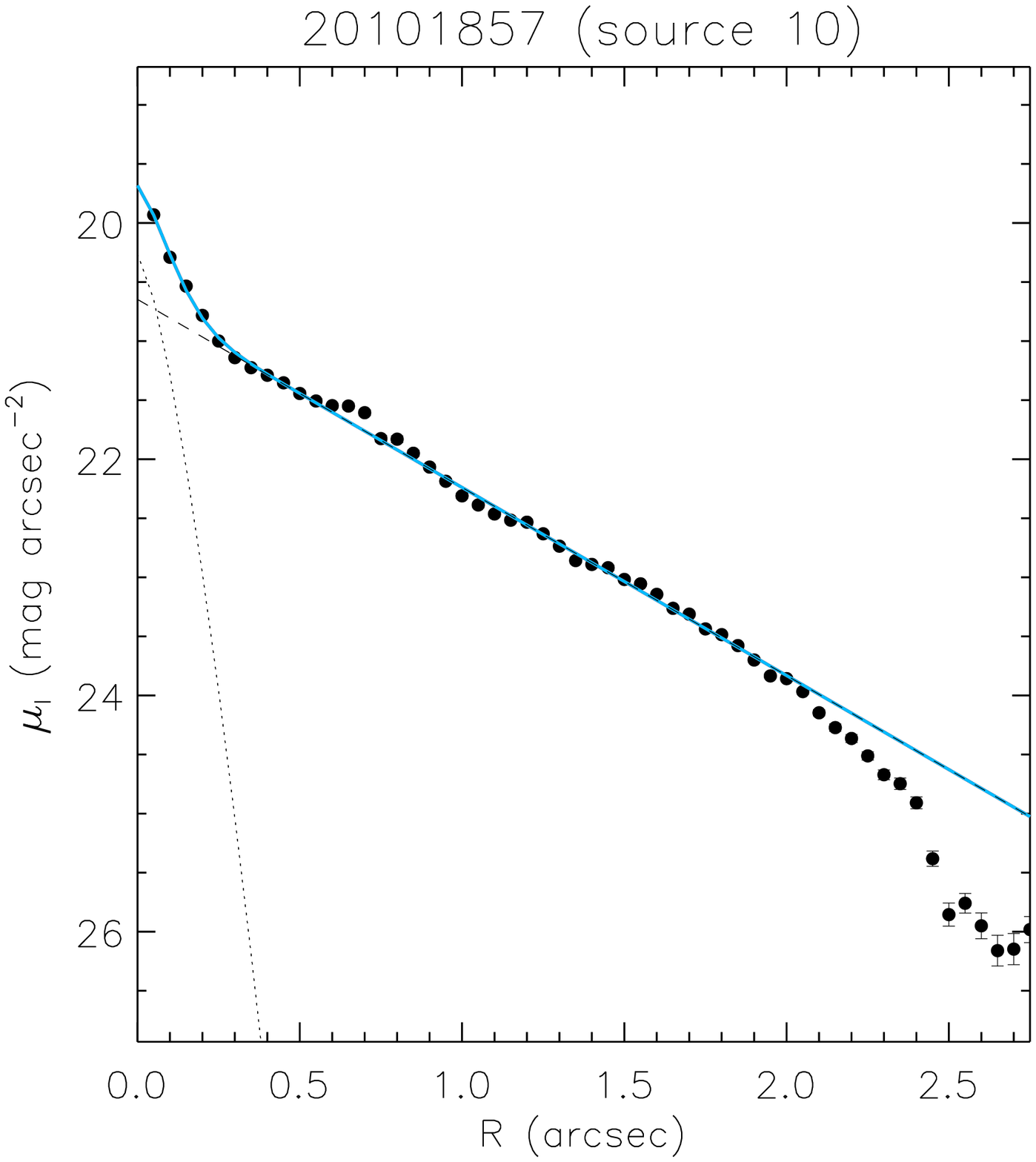}
\caption{HST/ACS images and surface brightness profile of the AGN 
         bulgeless host galaxy candidates with $n < 1.5$ and disc/irregular morphology.}
\label{fig:AGN-prof}
\end{figure}
\clearpage

\addtocounter{figure}{-1}
\begin{figure}
\includegraphics[width=4cm]{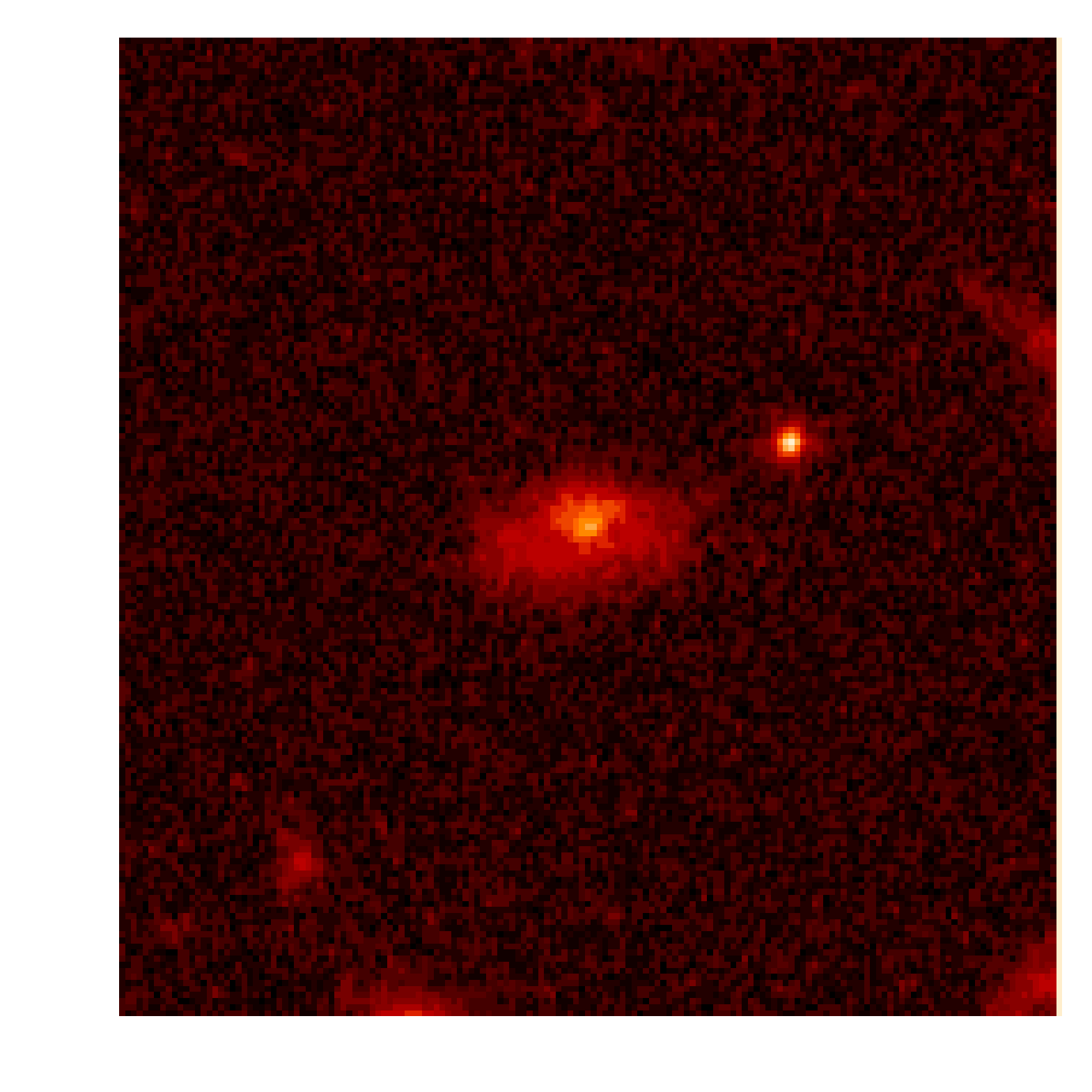}
\includegraphics[width=4cm]{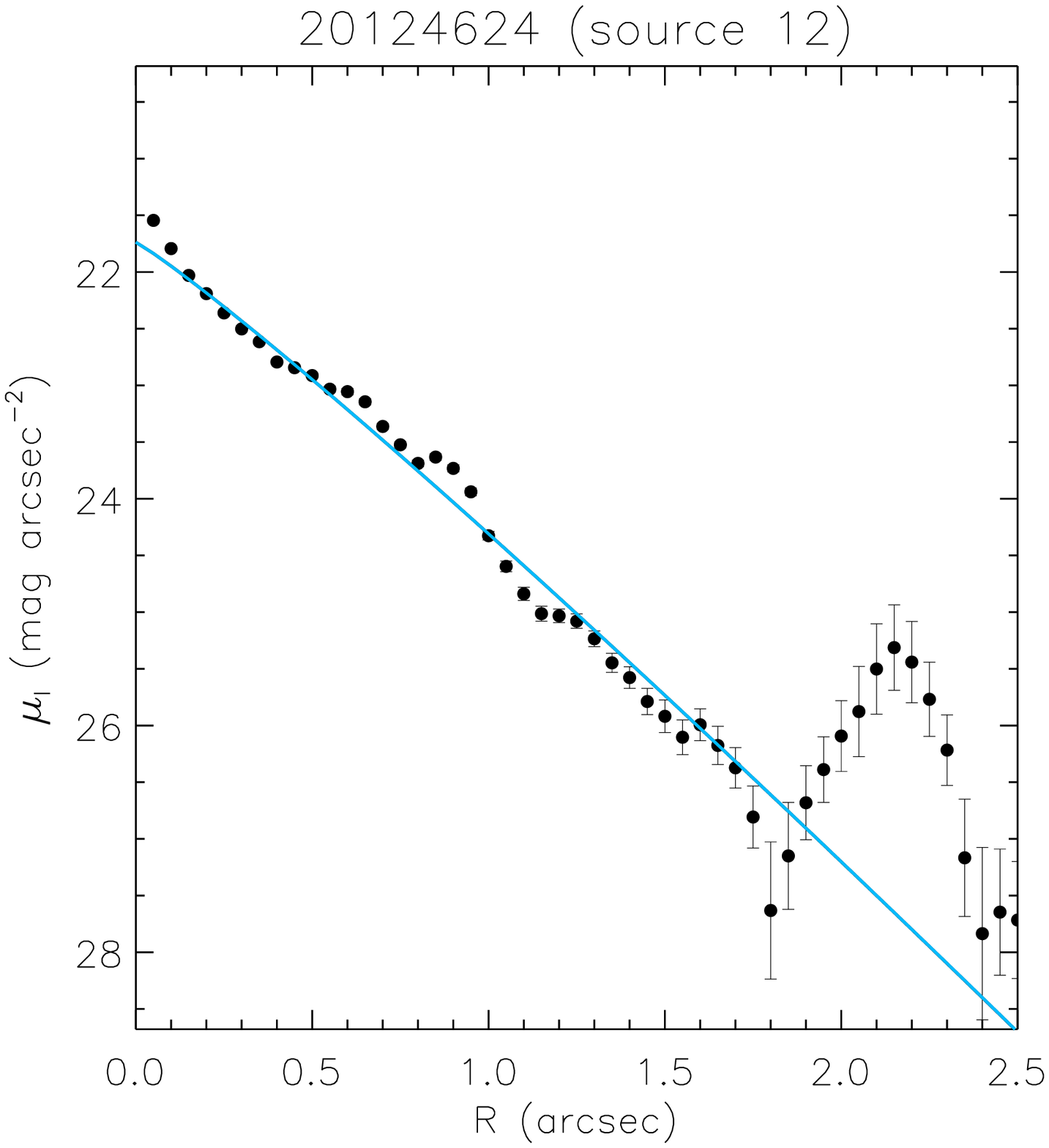}
\includegraphics[width=4cm]{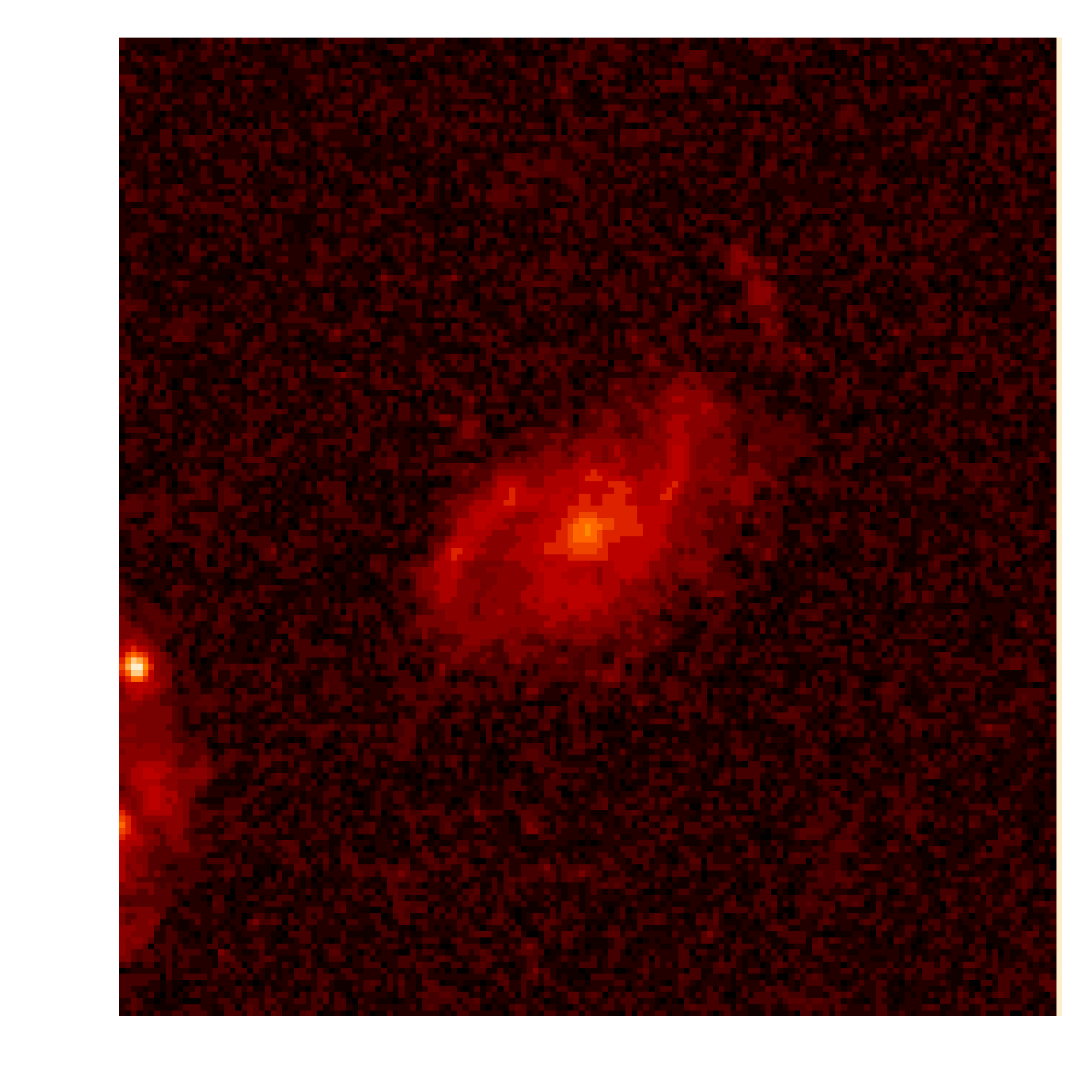}
\includegraphics[width=4cm]{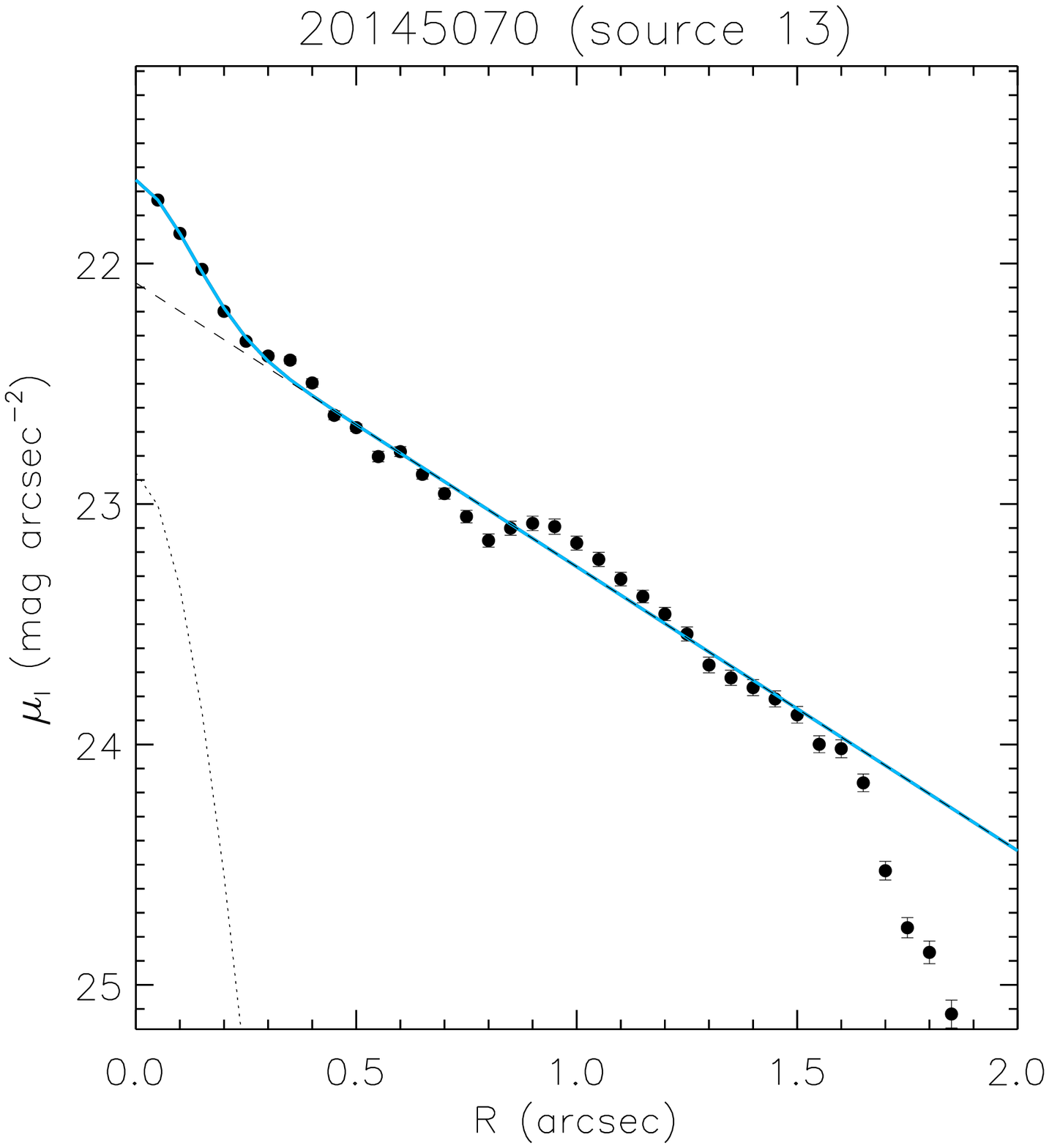}
\includegraphics[width=4cm]{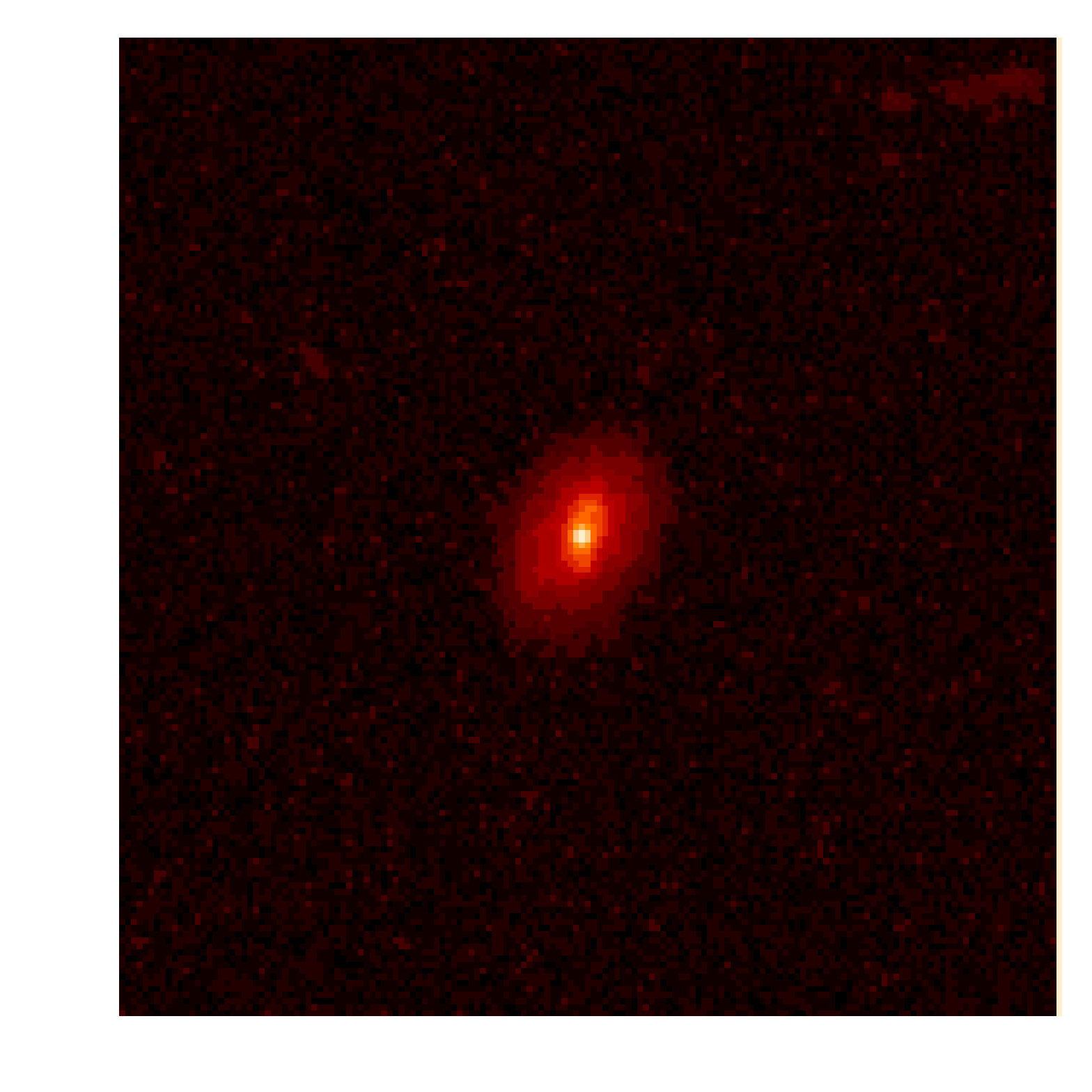}
\includegraphics[width=4cm]{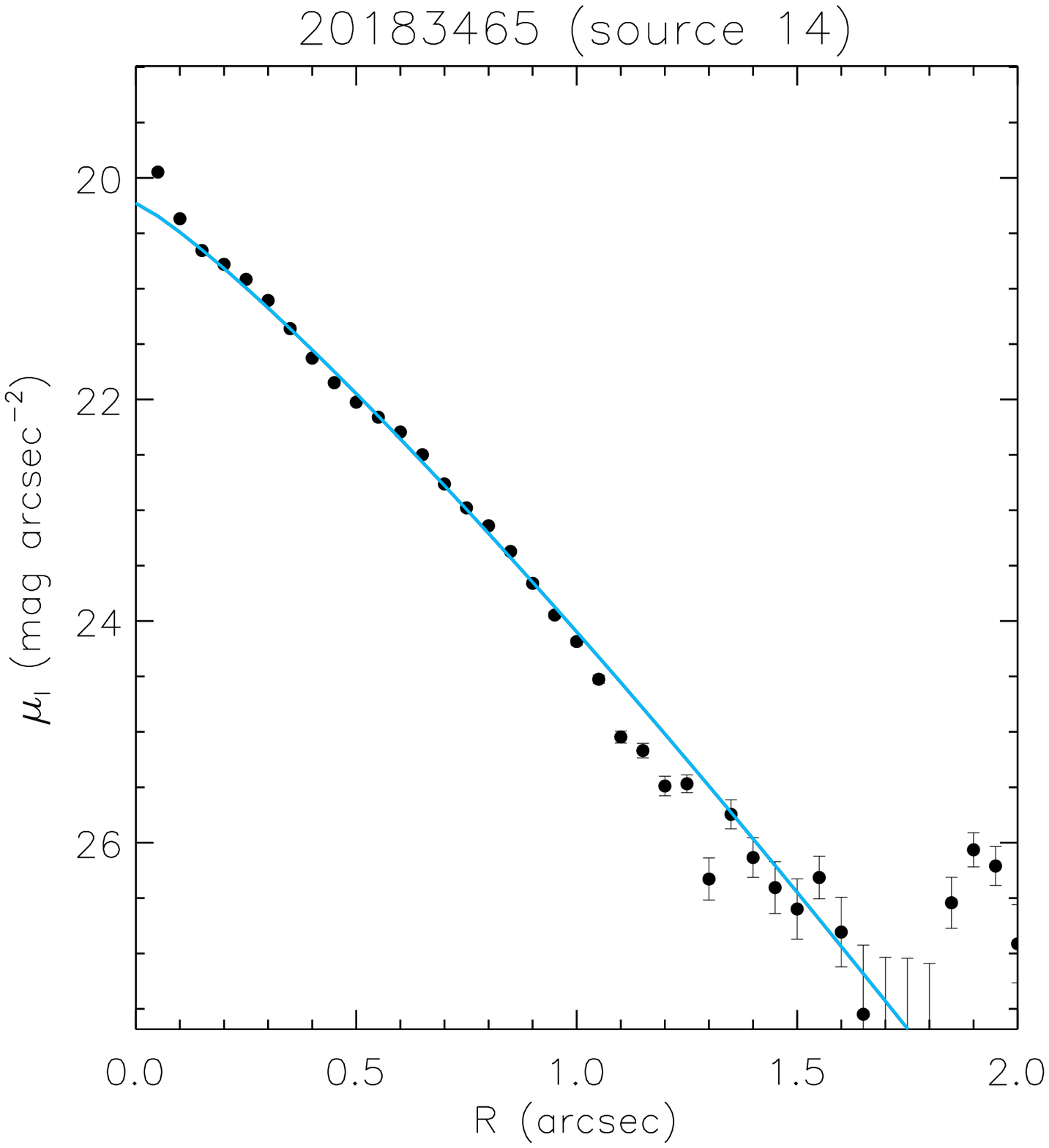}
\includegraphics[width=4cm]{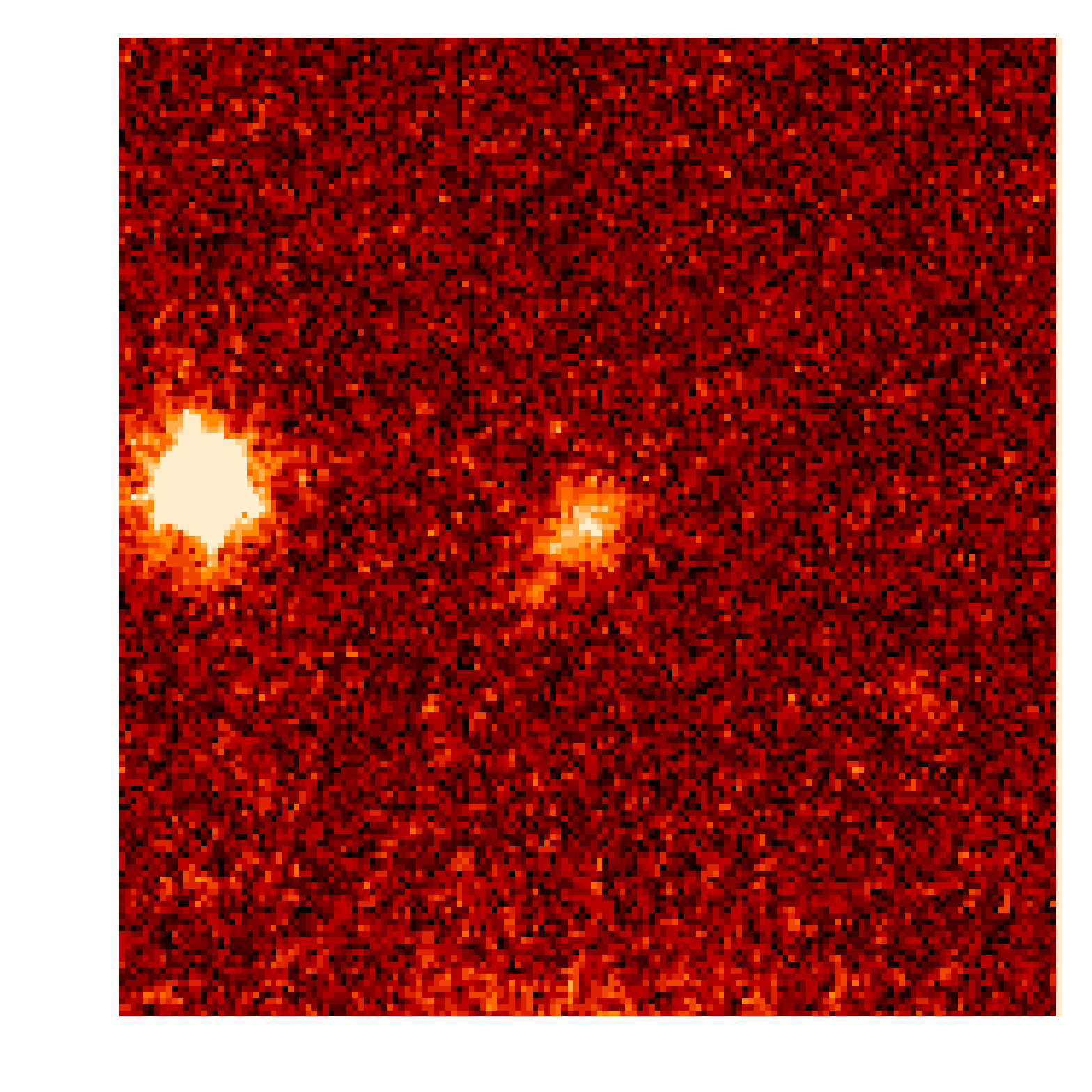}
\includegraphics[width=4cm]{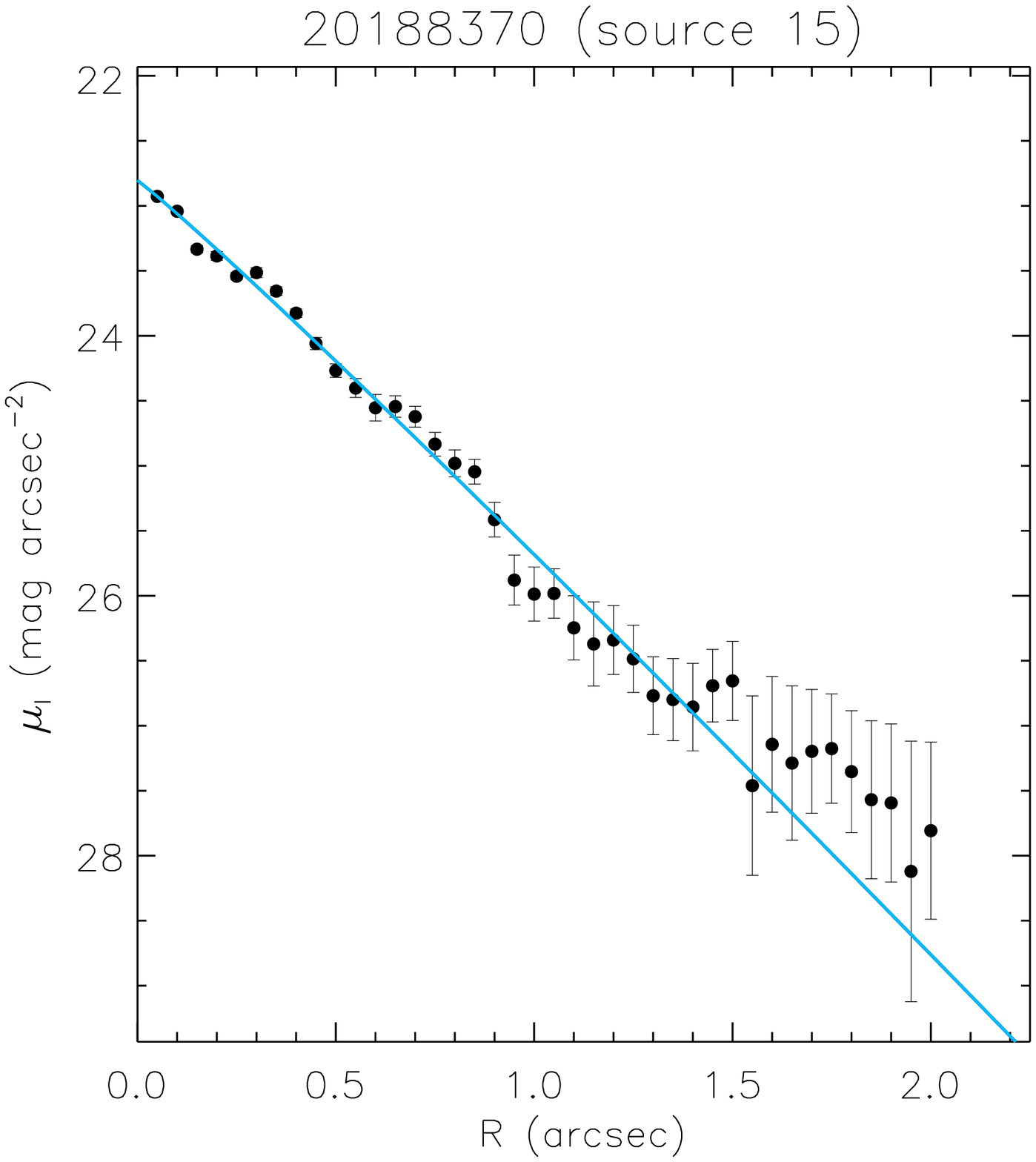}
\includegraphics[width=4cm]{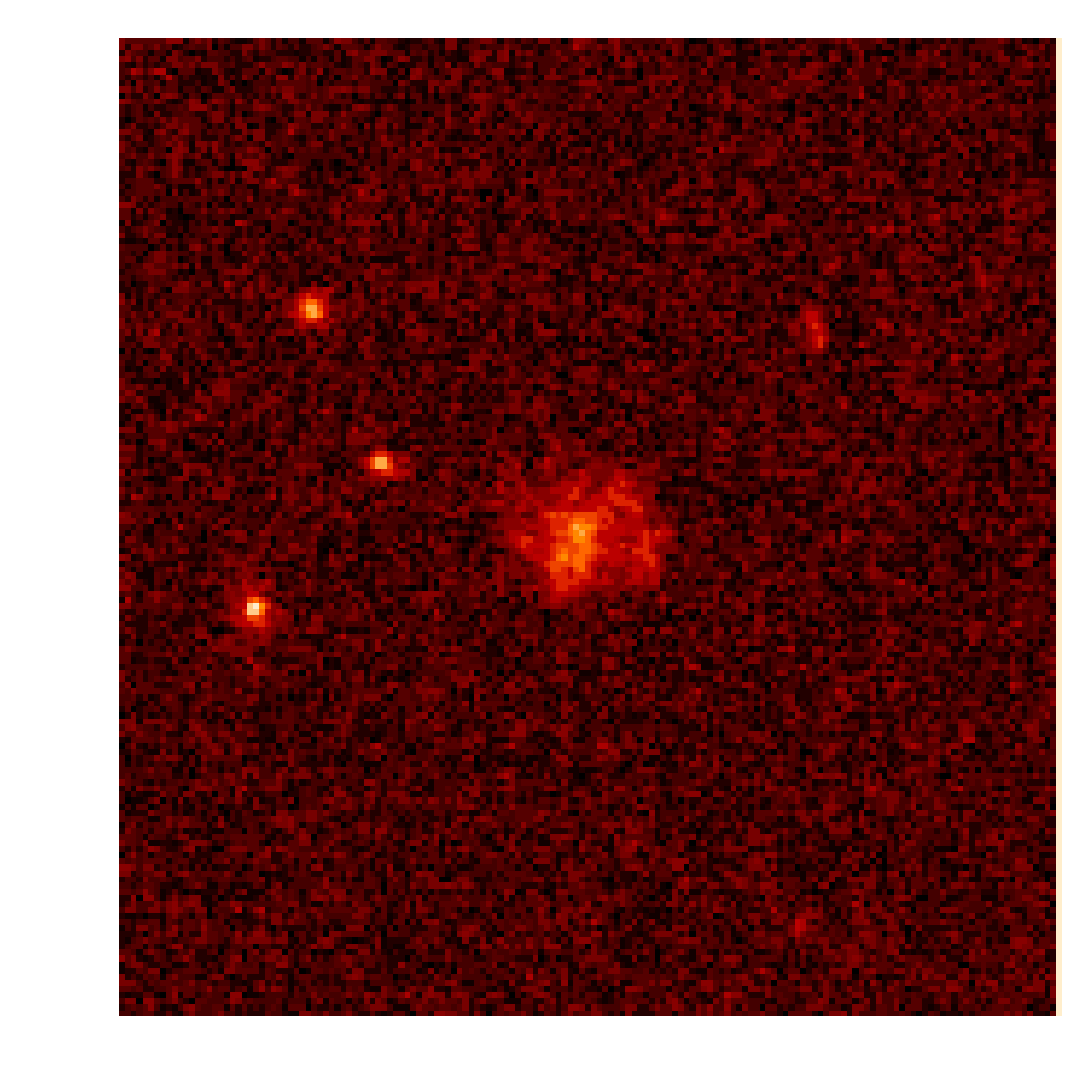}
\includegraphics[width=4cm]{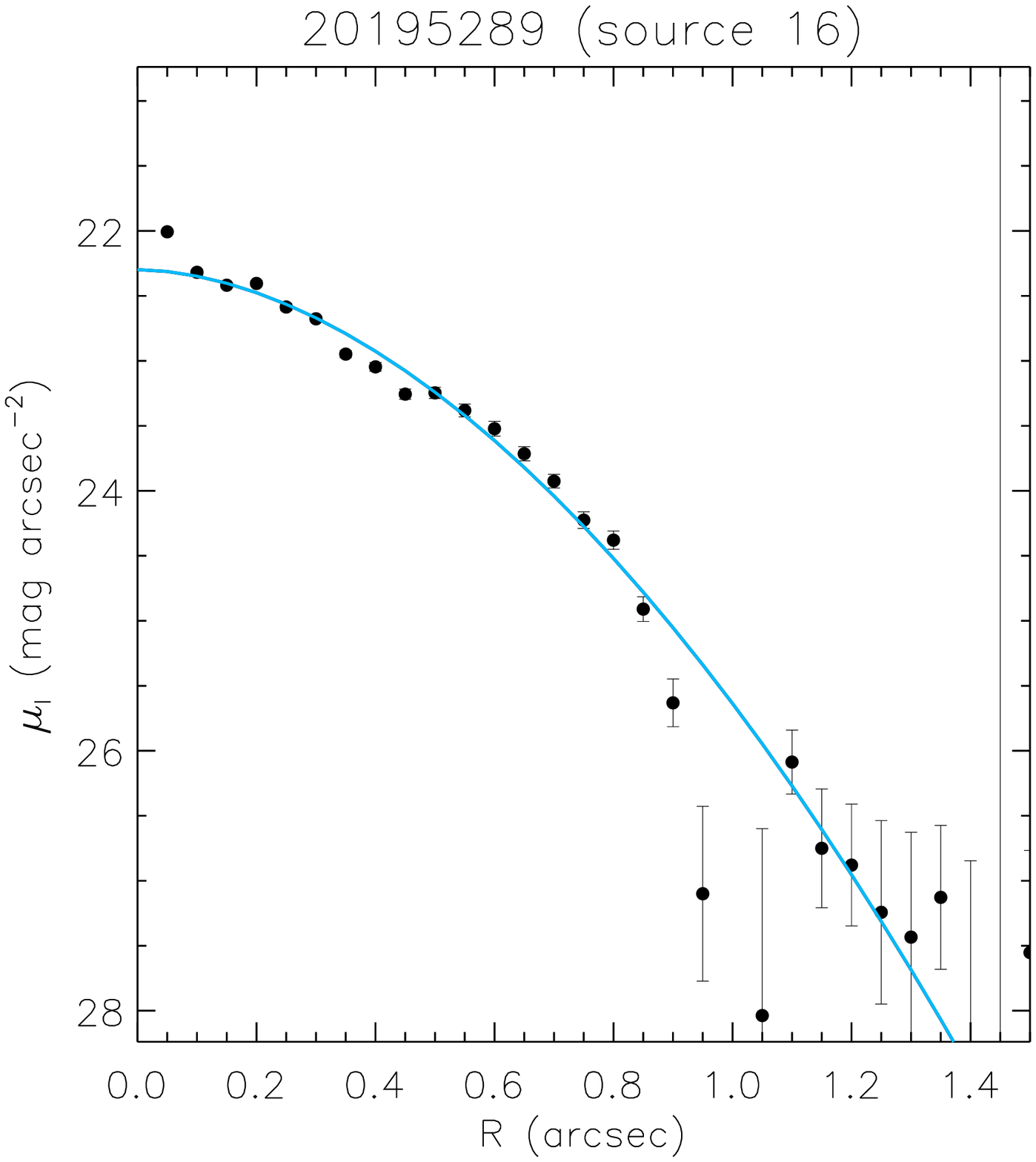}
\includegraphics[width=4cm]{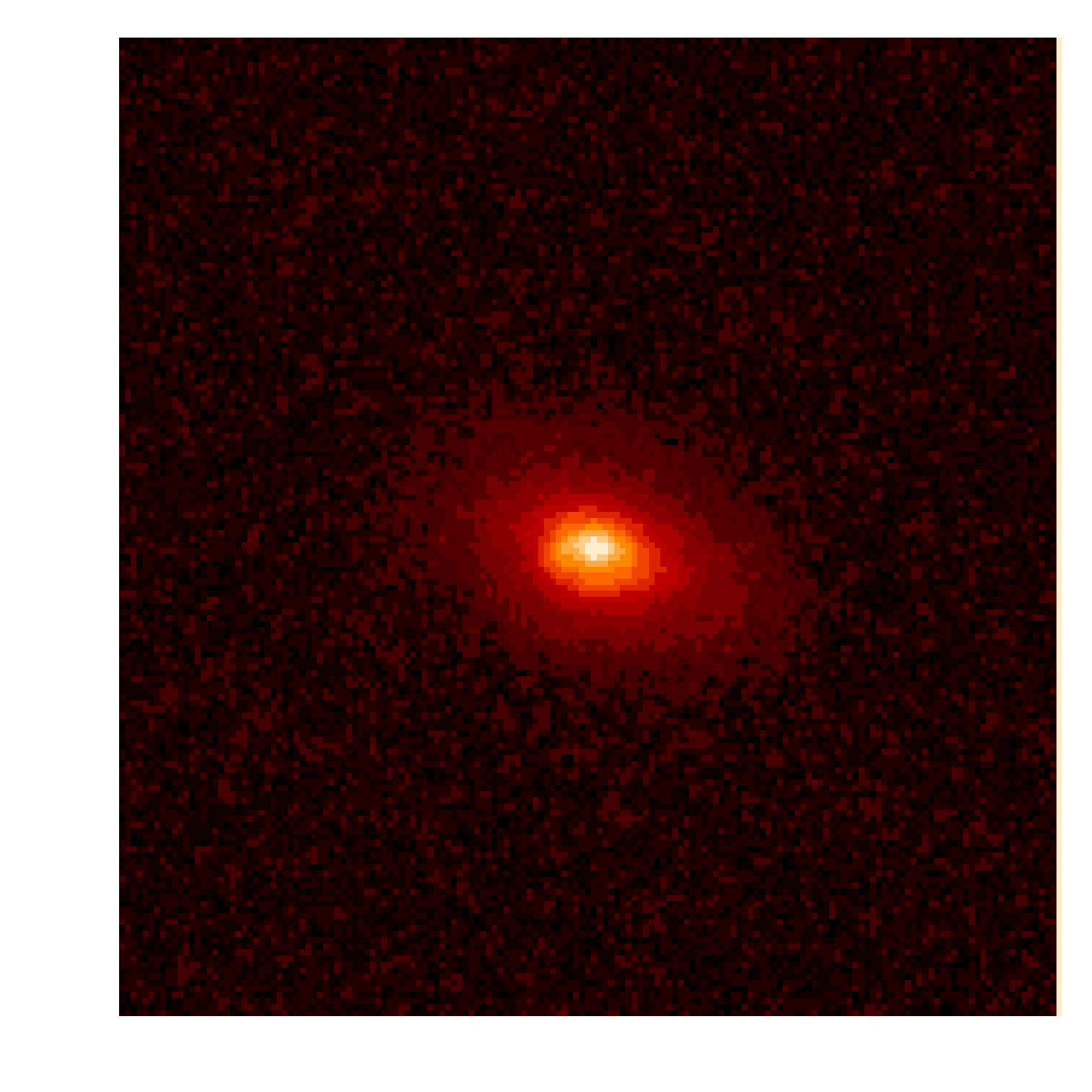}
\includegraphics[width=4cm]{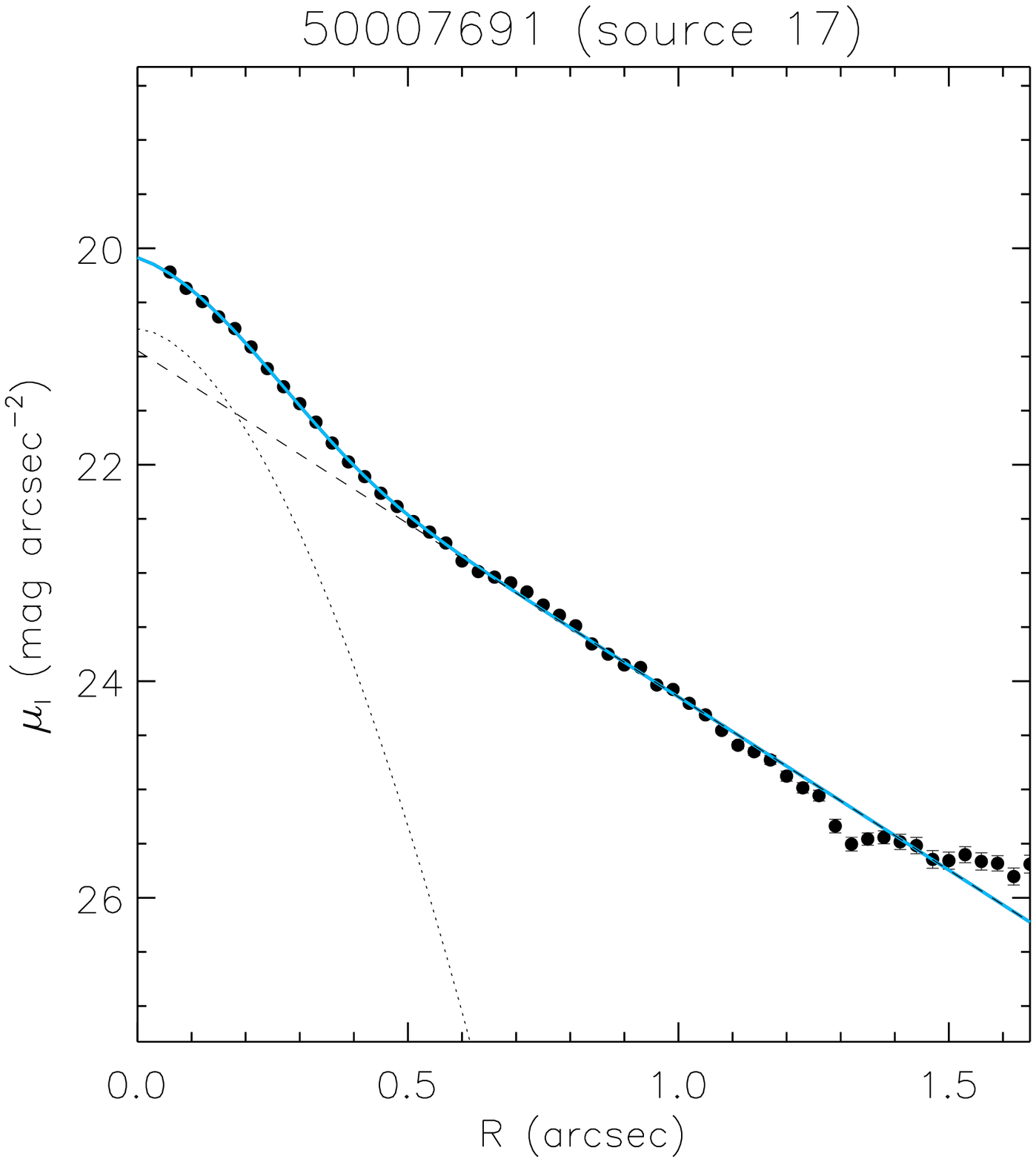}
\includegraphics[width=4cm]{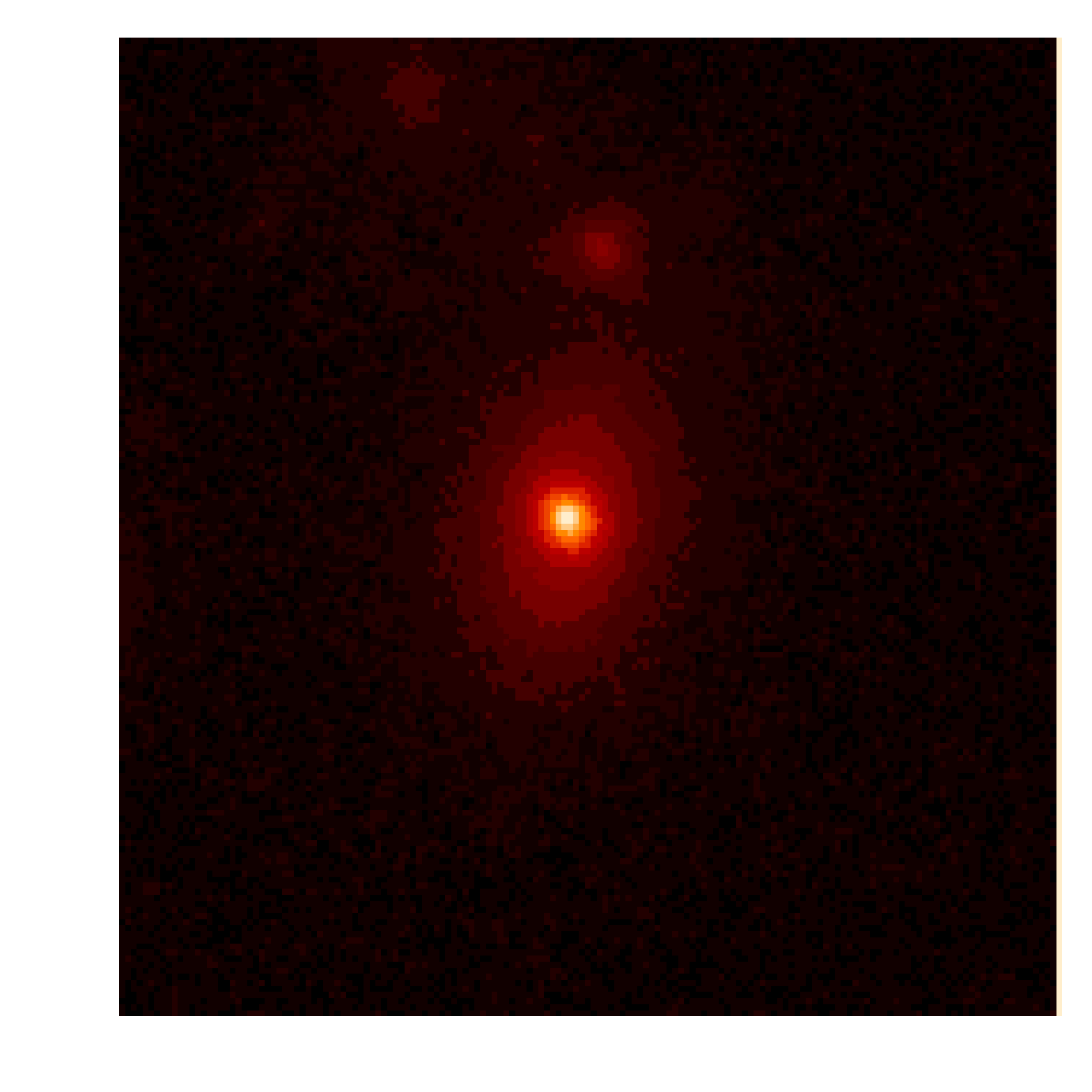}
\includegraphics[width=4cm]{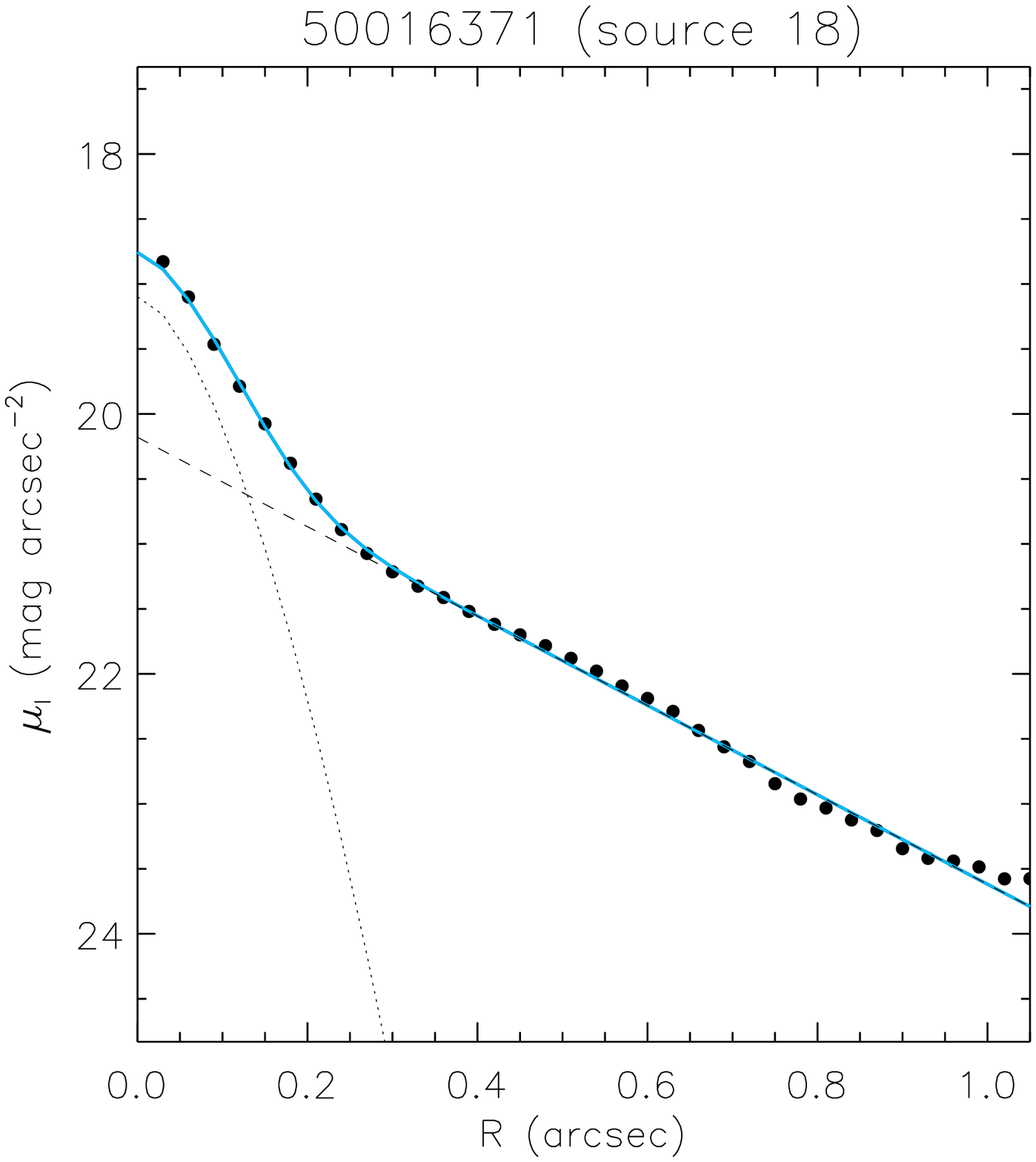}
\includegraphics[width=4cm]{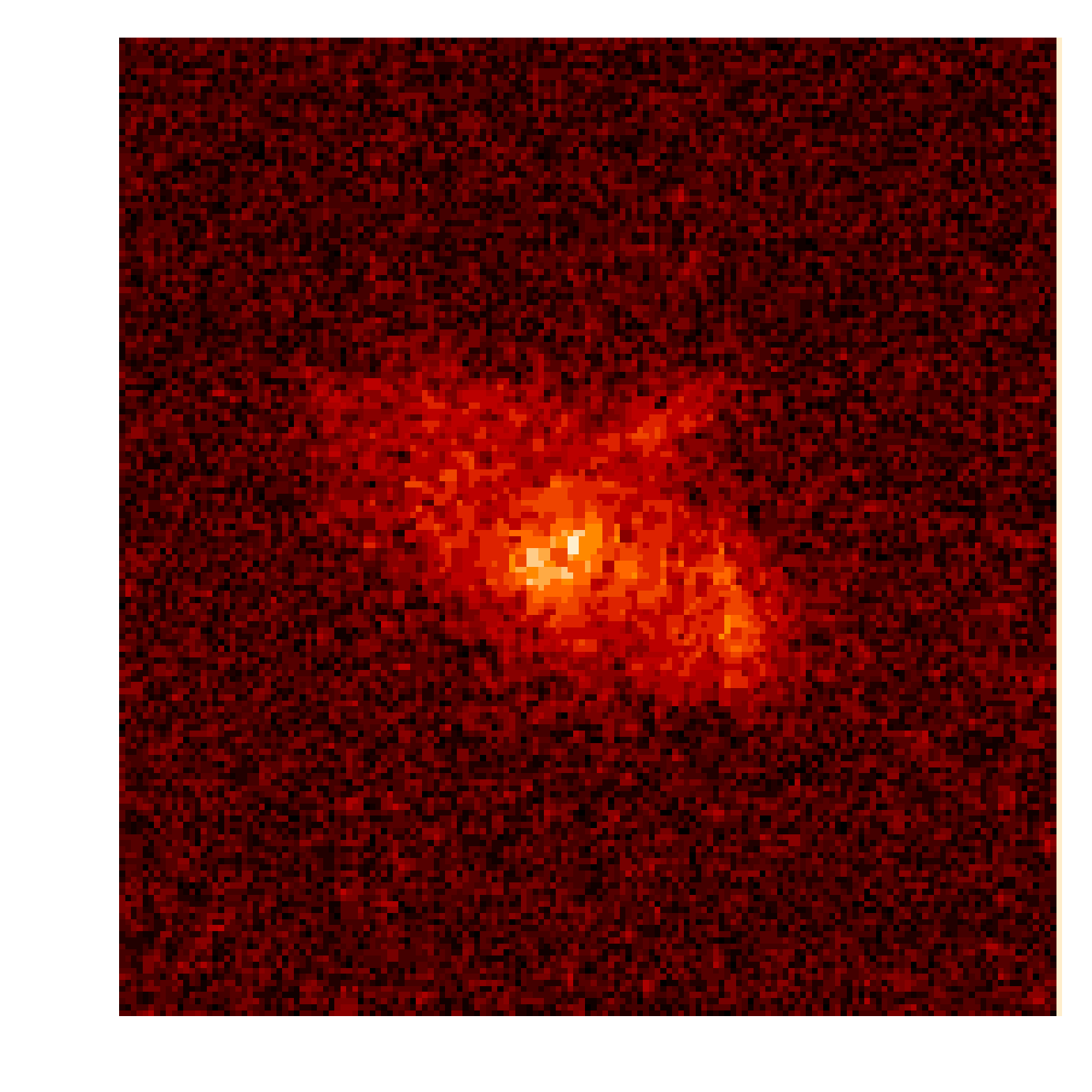}
\includegraphics[width=4cm]{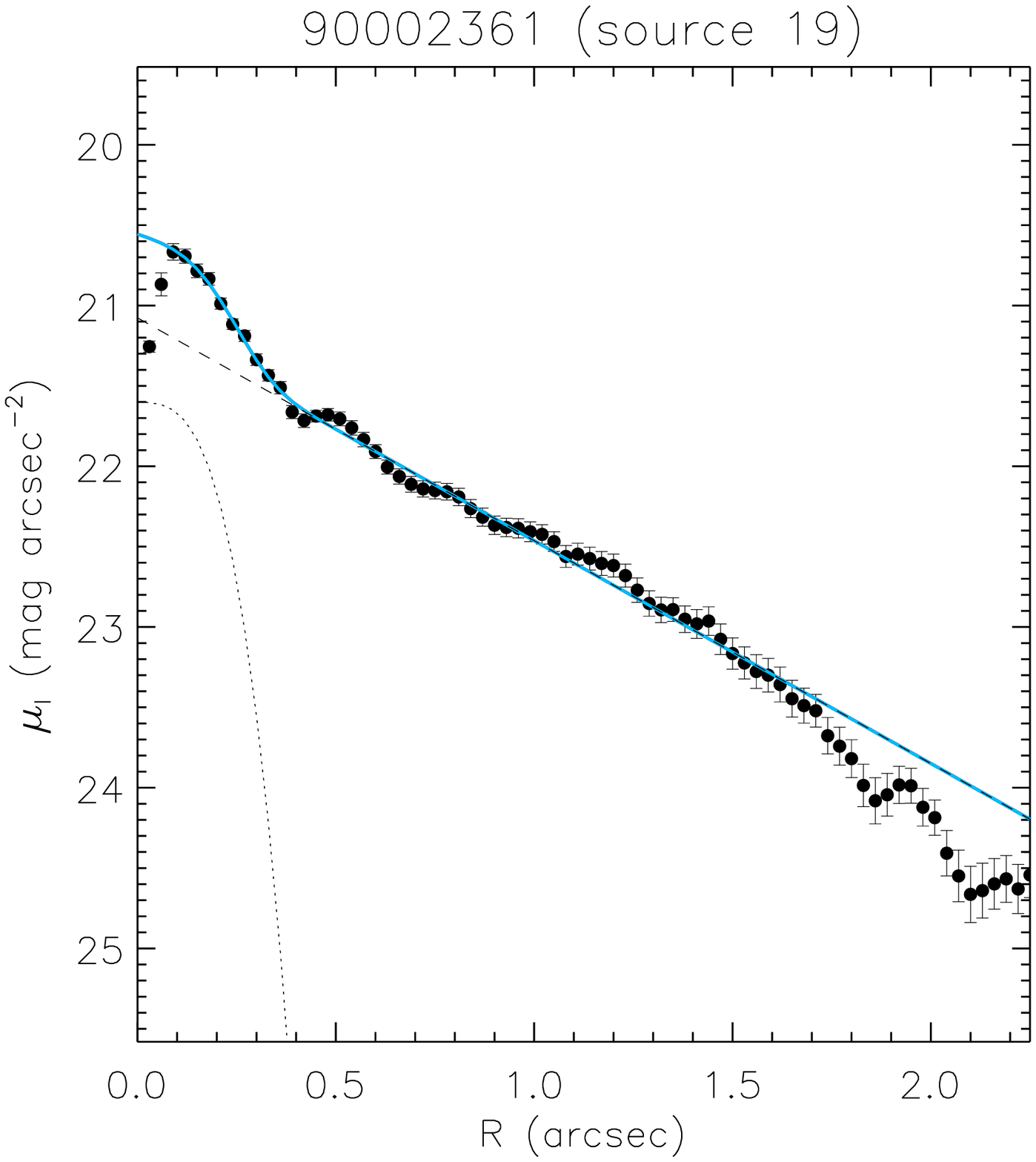}
\caption{(\emph{continued}) HST/ACS images and surface brightness profile of the AGN 
         bulgeless host galaxy candidates with $n < 1.5$ and disc/irregular morphology.}
\end{figure}
\clearpage

\addtocounter{figure}{-1}
\begin{figure}
\includegraphics[width=4cm]{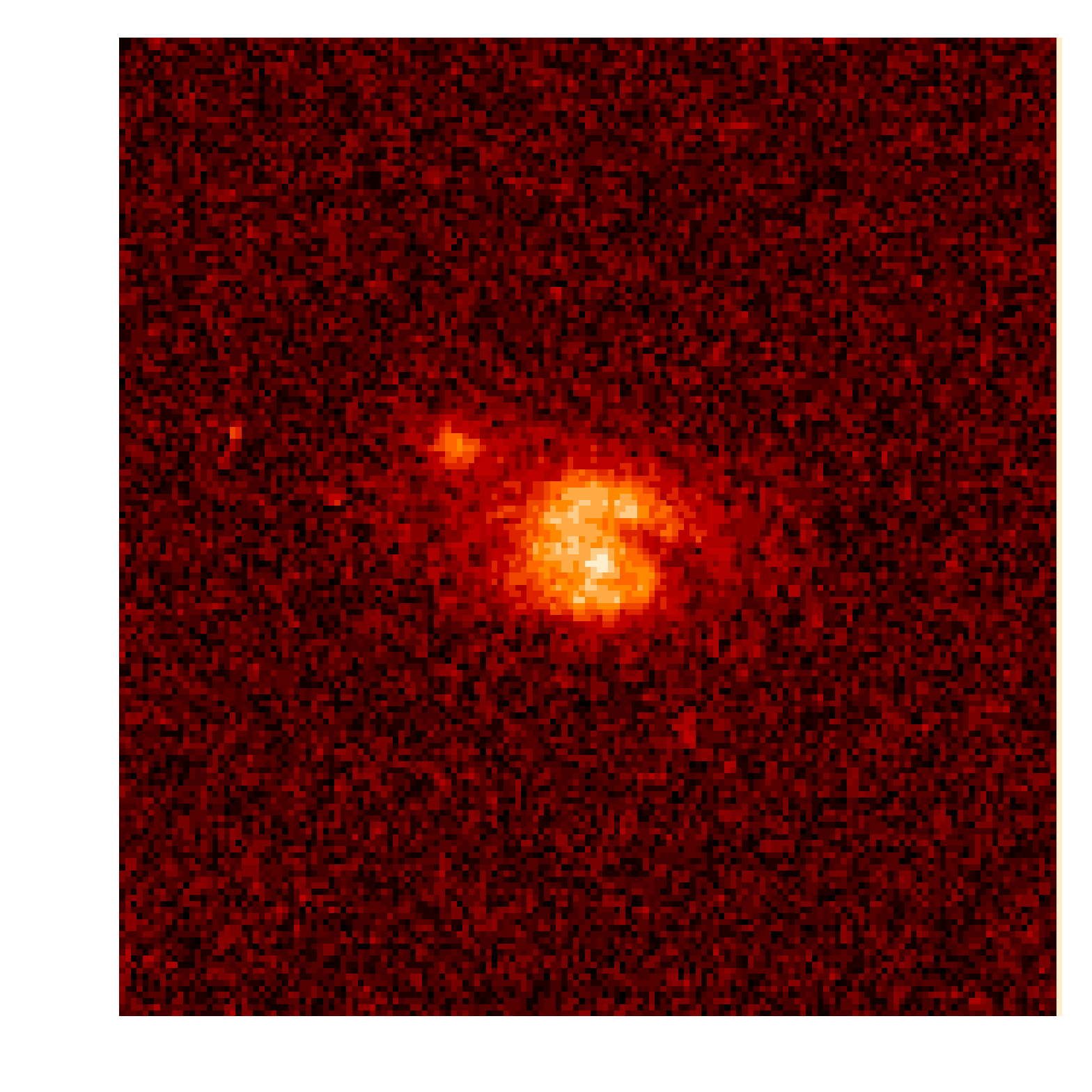}
\includegraphics[width=4cm]{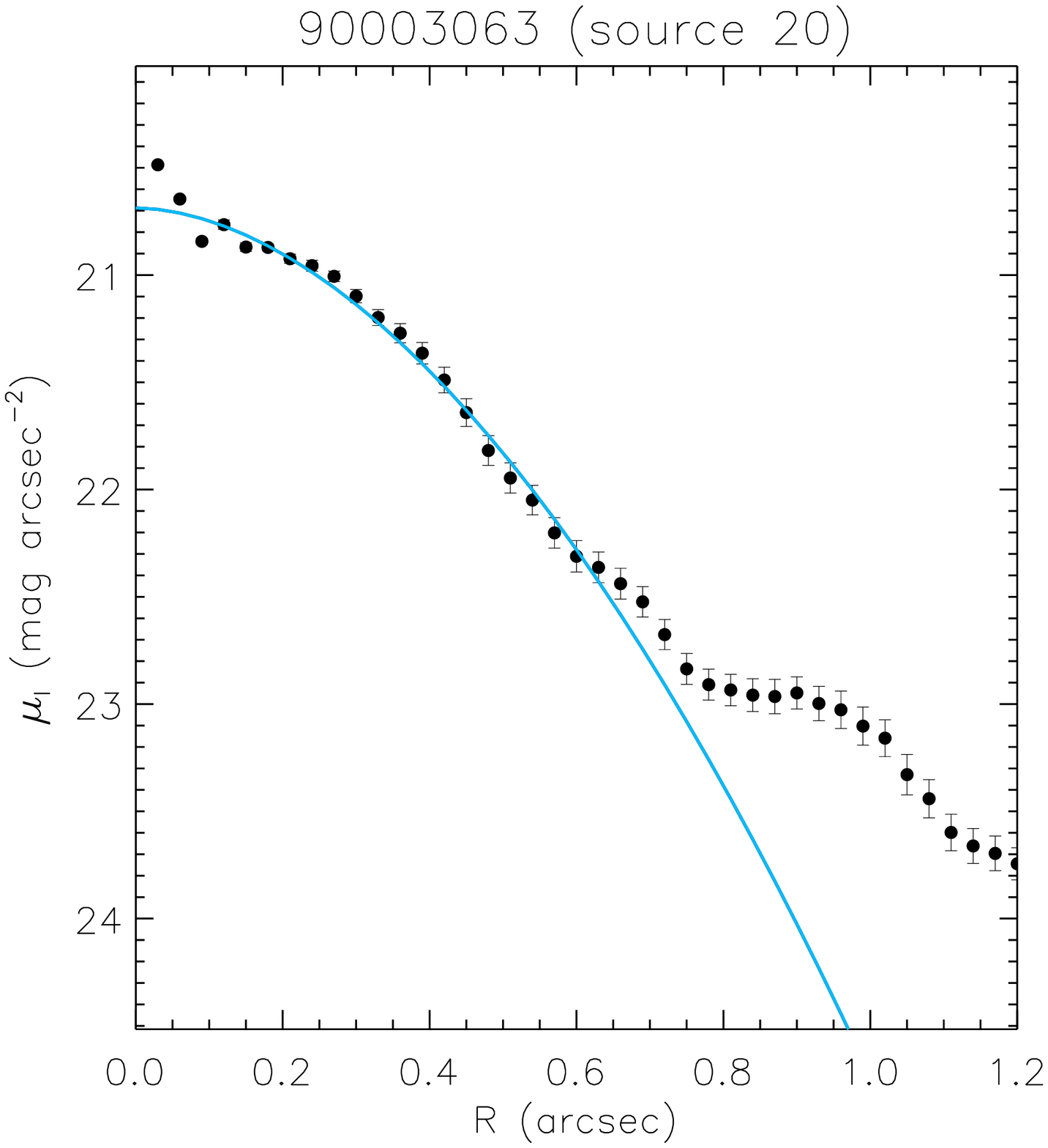}
\includegraphics[width=4cm]{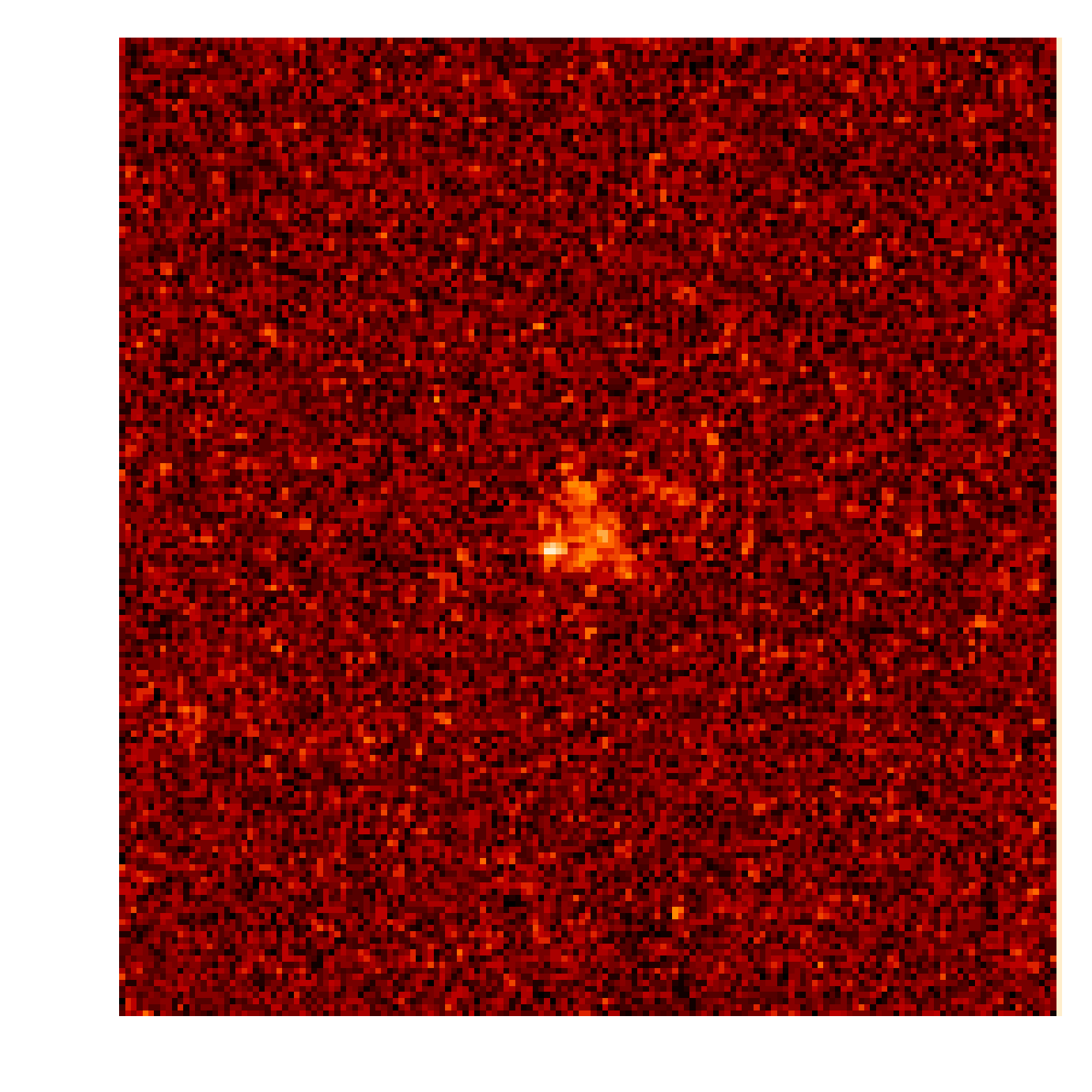}
\includegraphics[width=4cm]{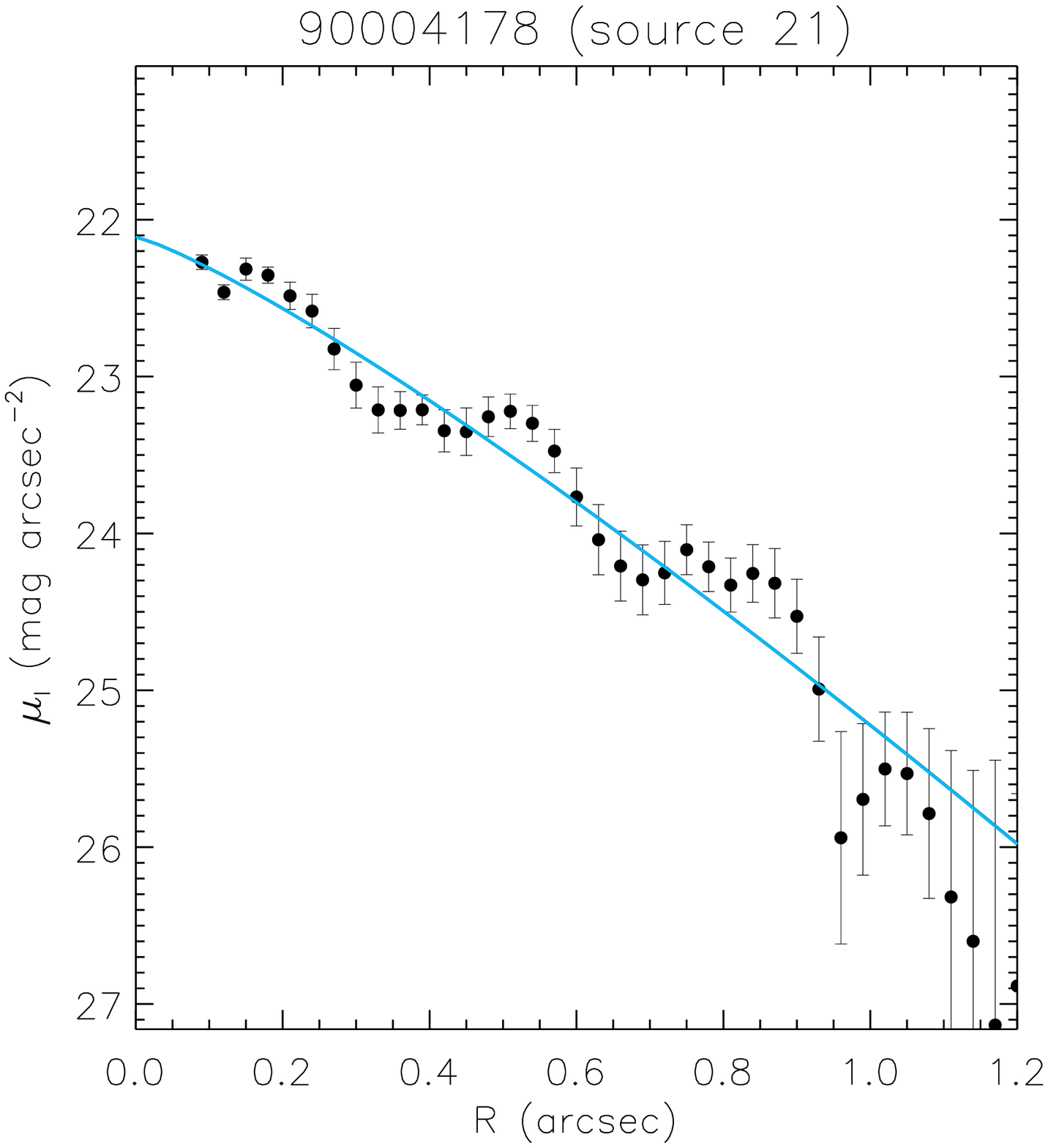}
\includegraphics[width=4cm]{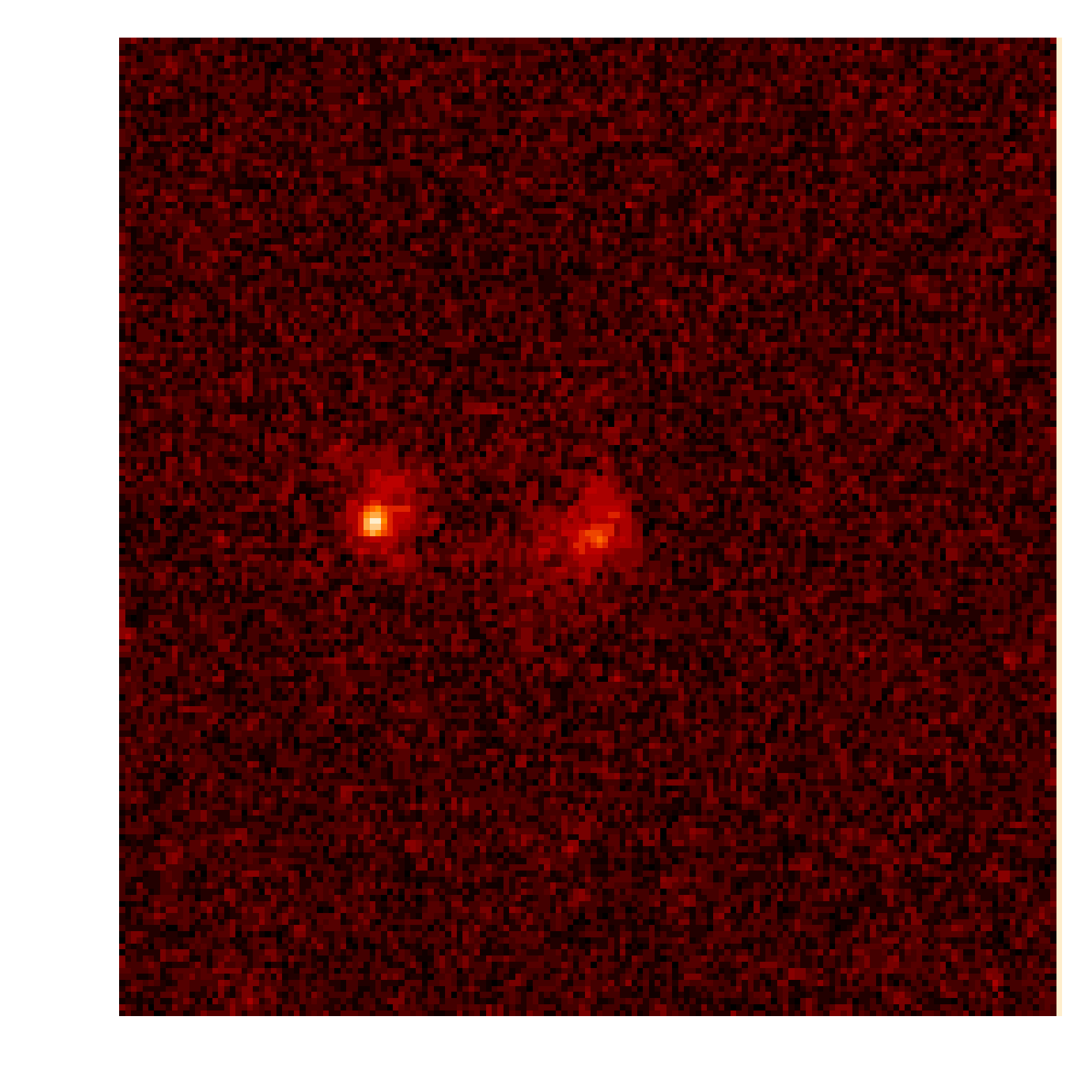}
\includegraphics[width=4cm]{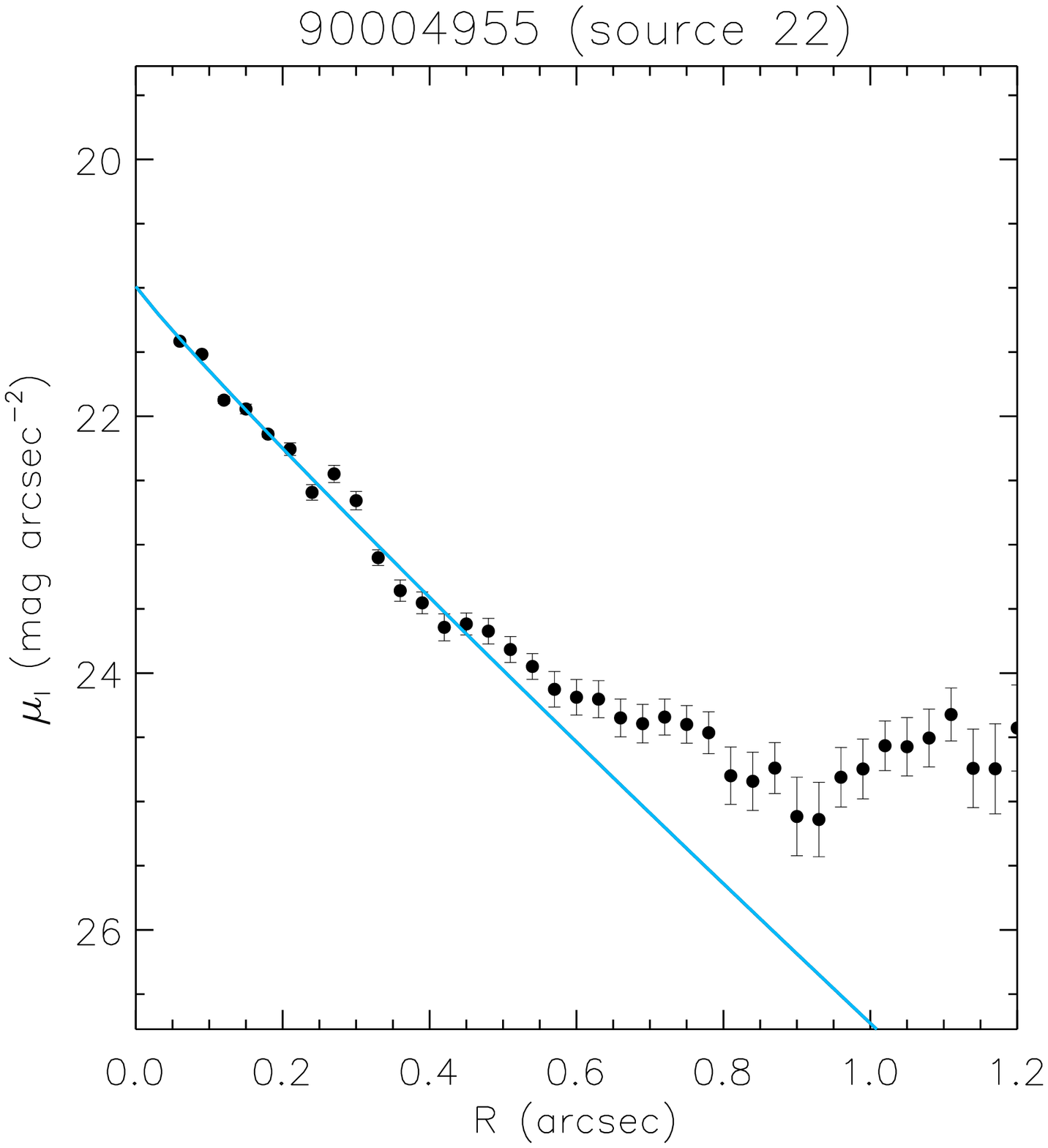}
\includegraphics[width=4cm]{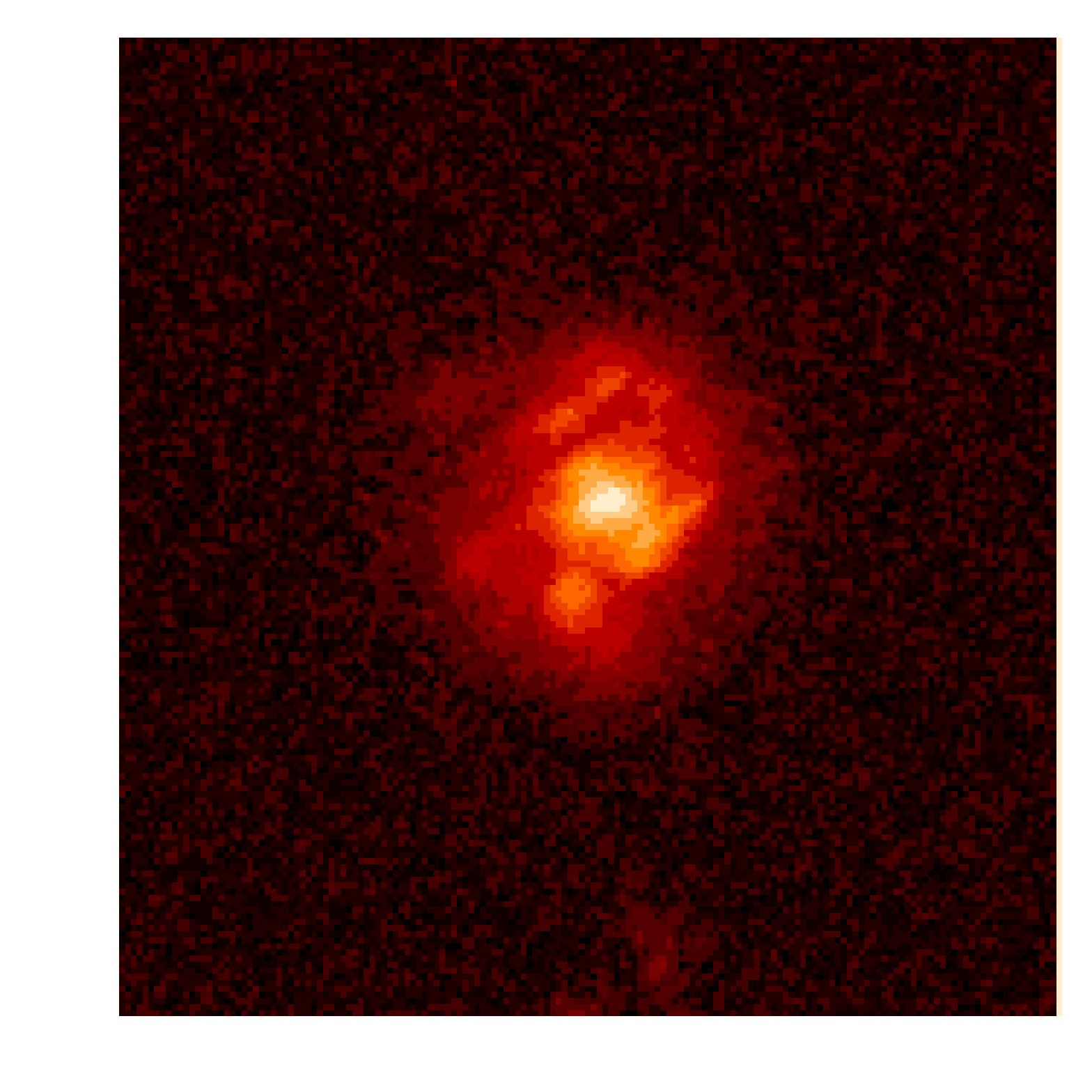}
\includegraphics[width=4cm]{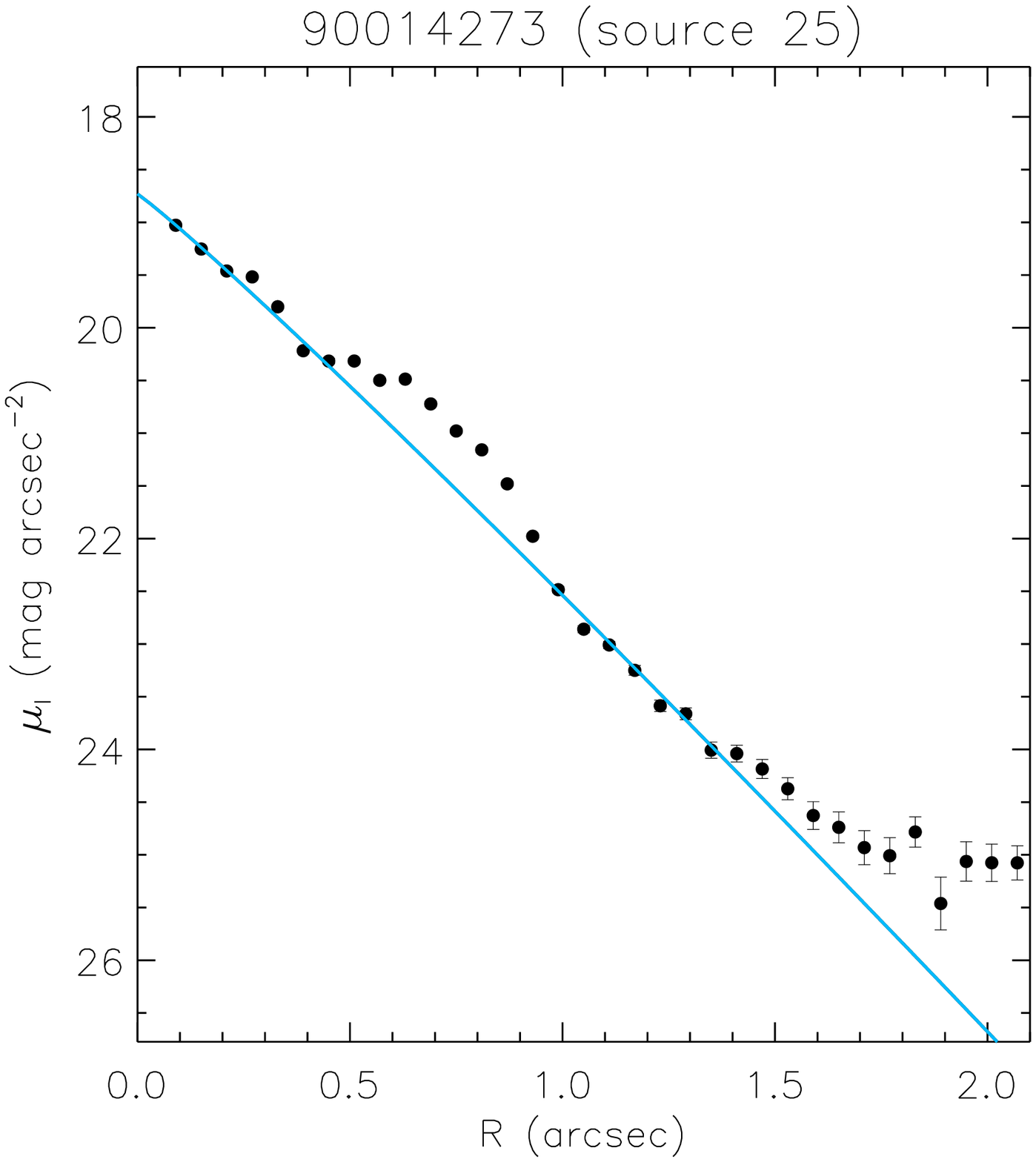}
\includegraphics[width=4cm]{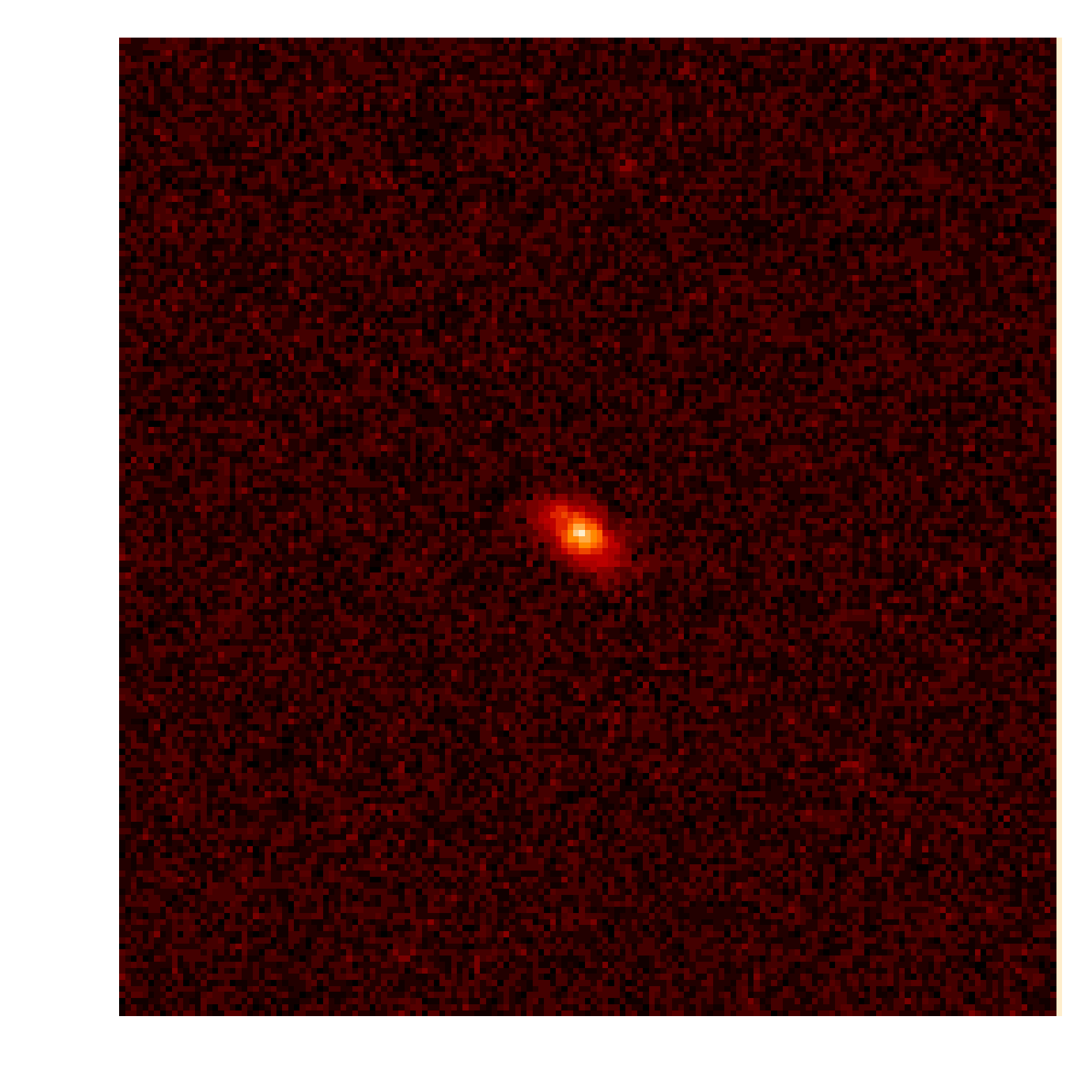}
\includegraphics[width=4cm]{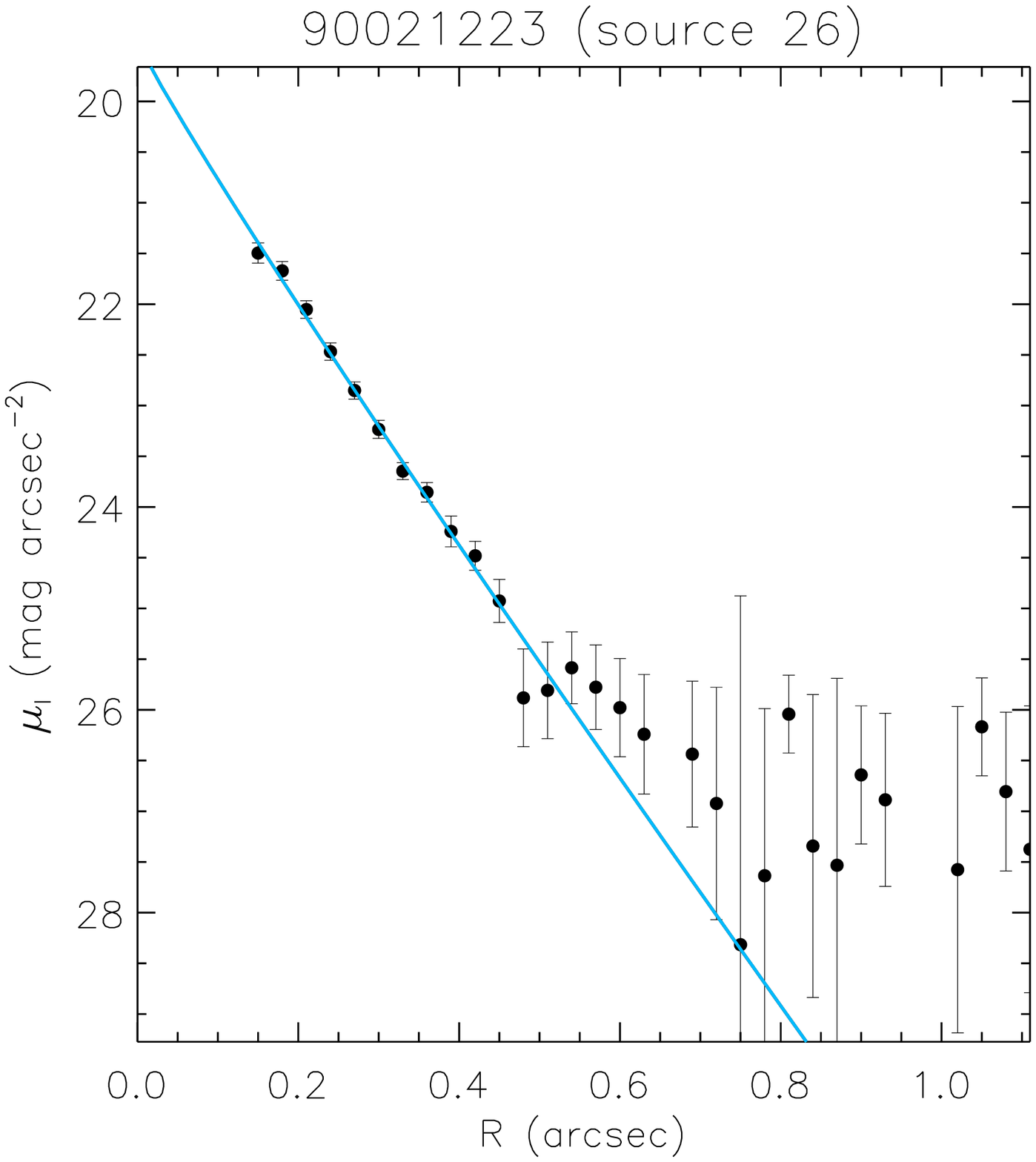}
\includegraphics[width=4cm]{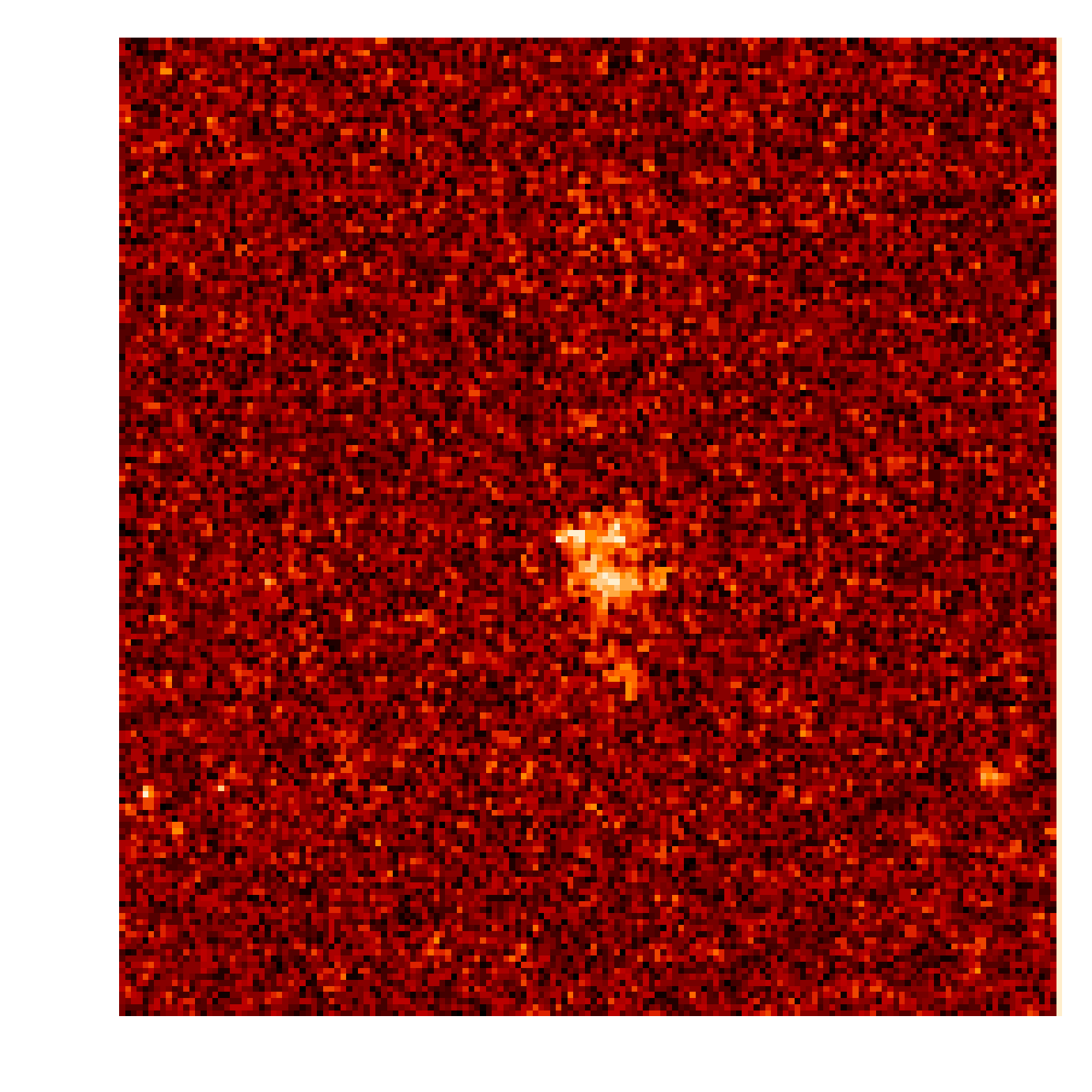}
\includegraphics[width=4cm]{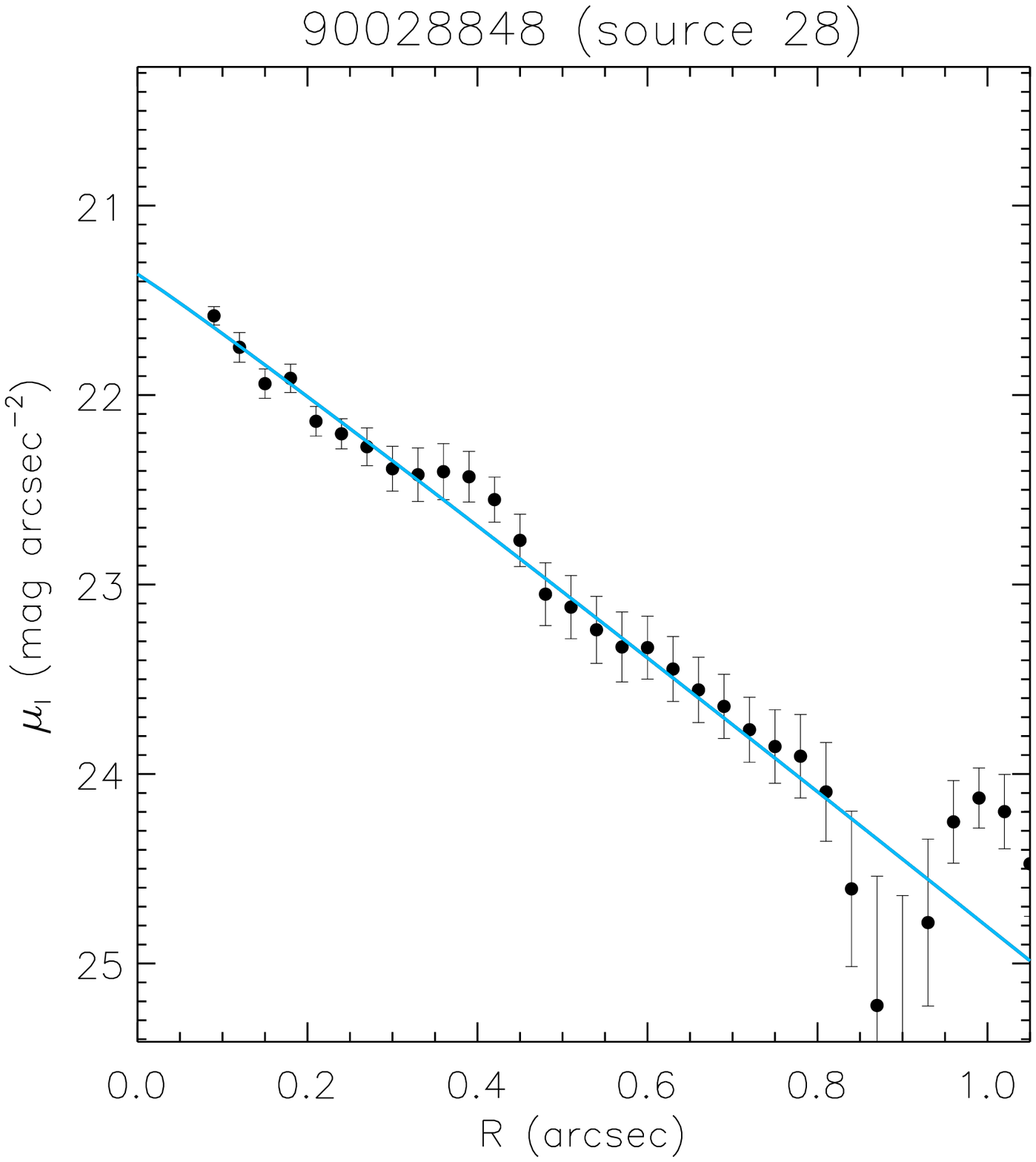}
\includegraphics[width=4cm]{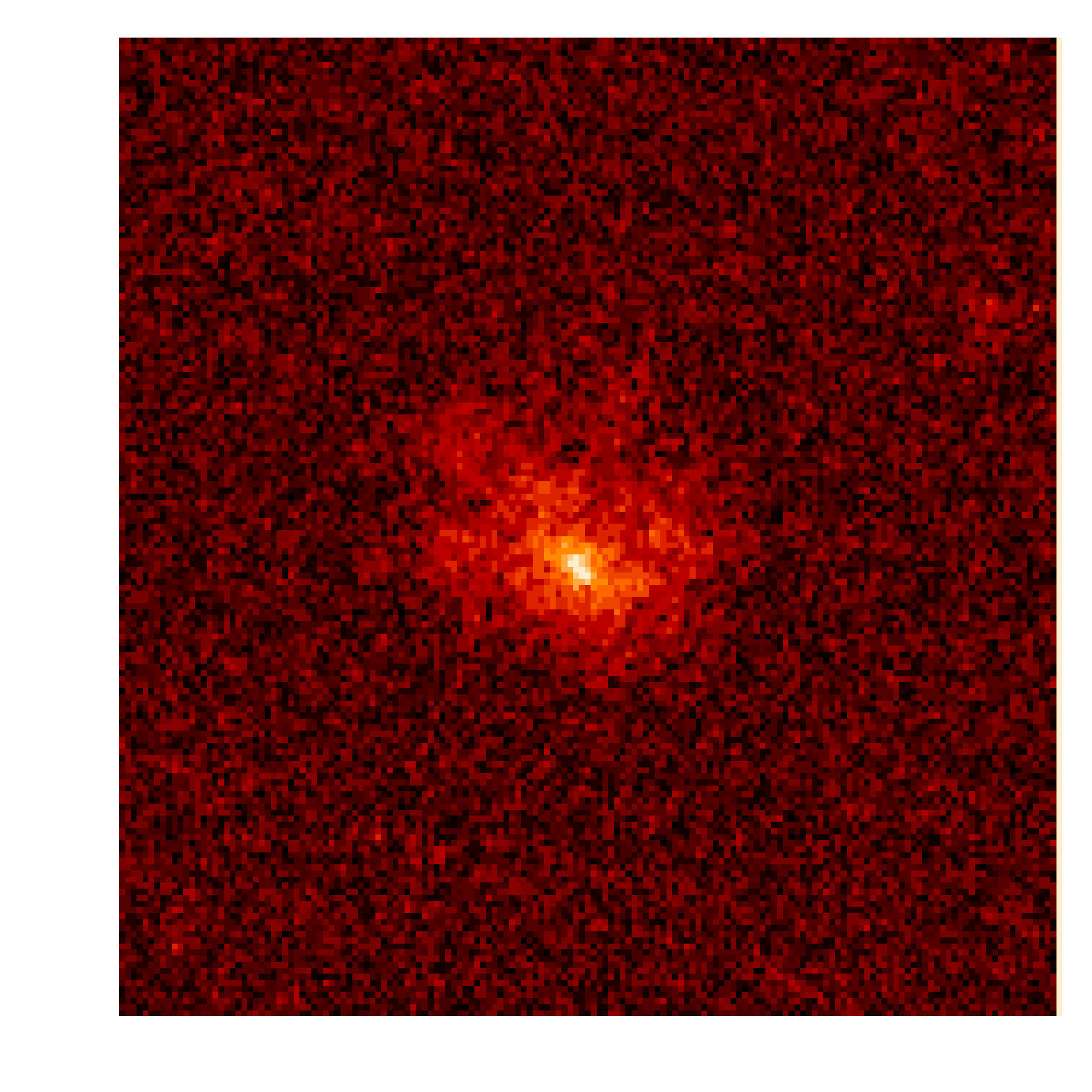}
\includegraphics[width=4cm]{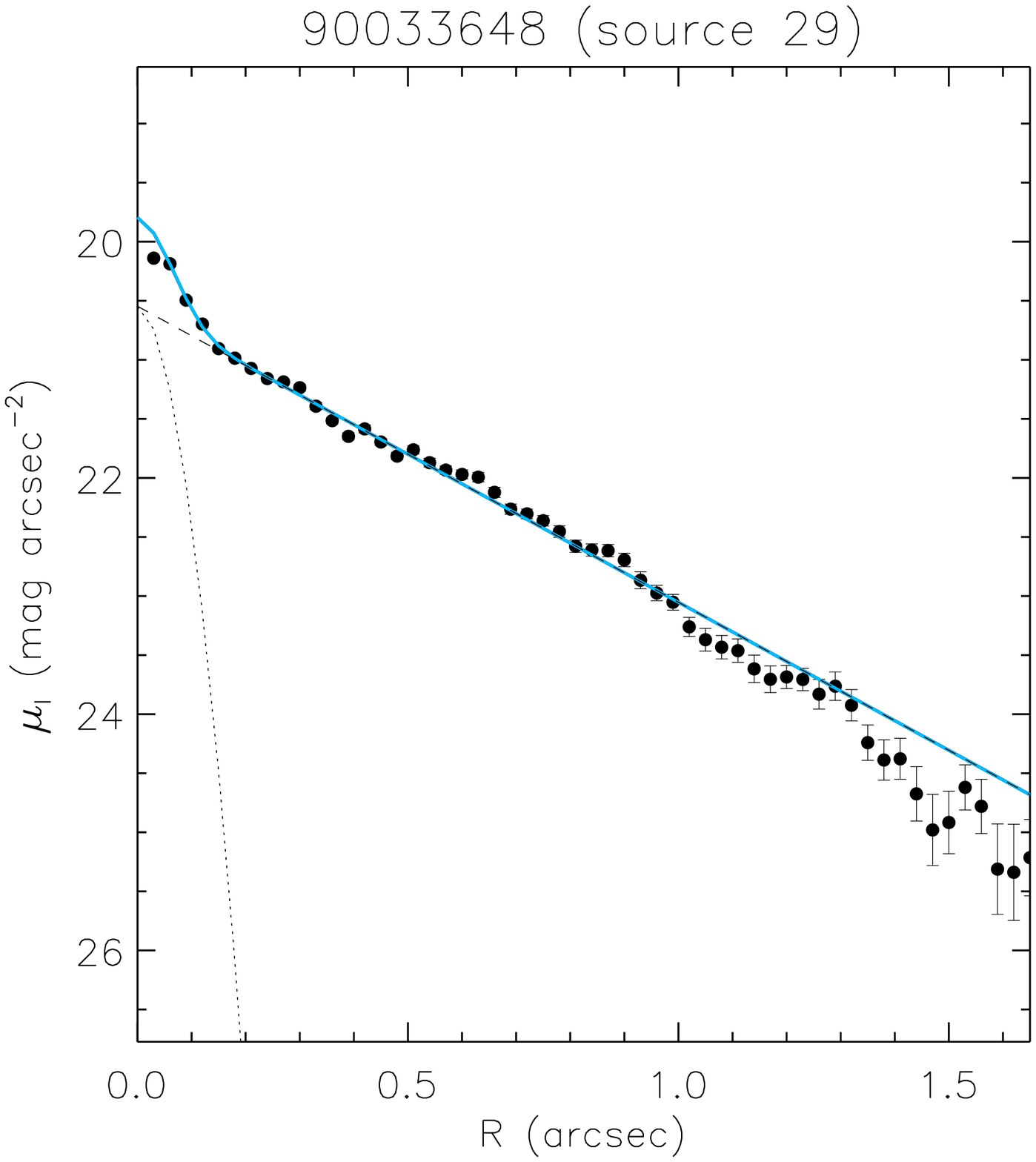}
\includegraphics[width=4cm]{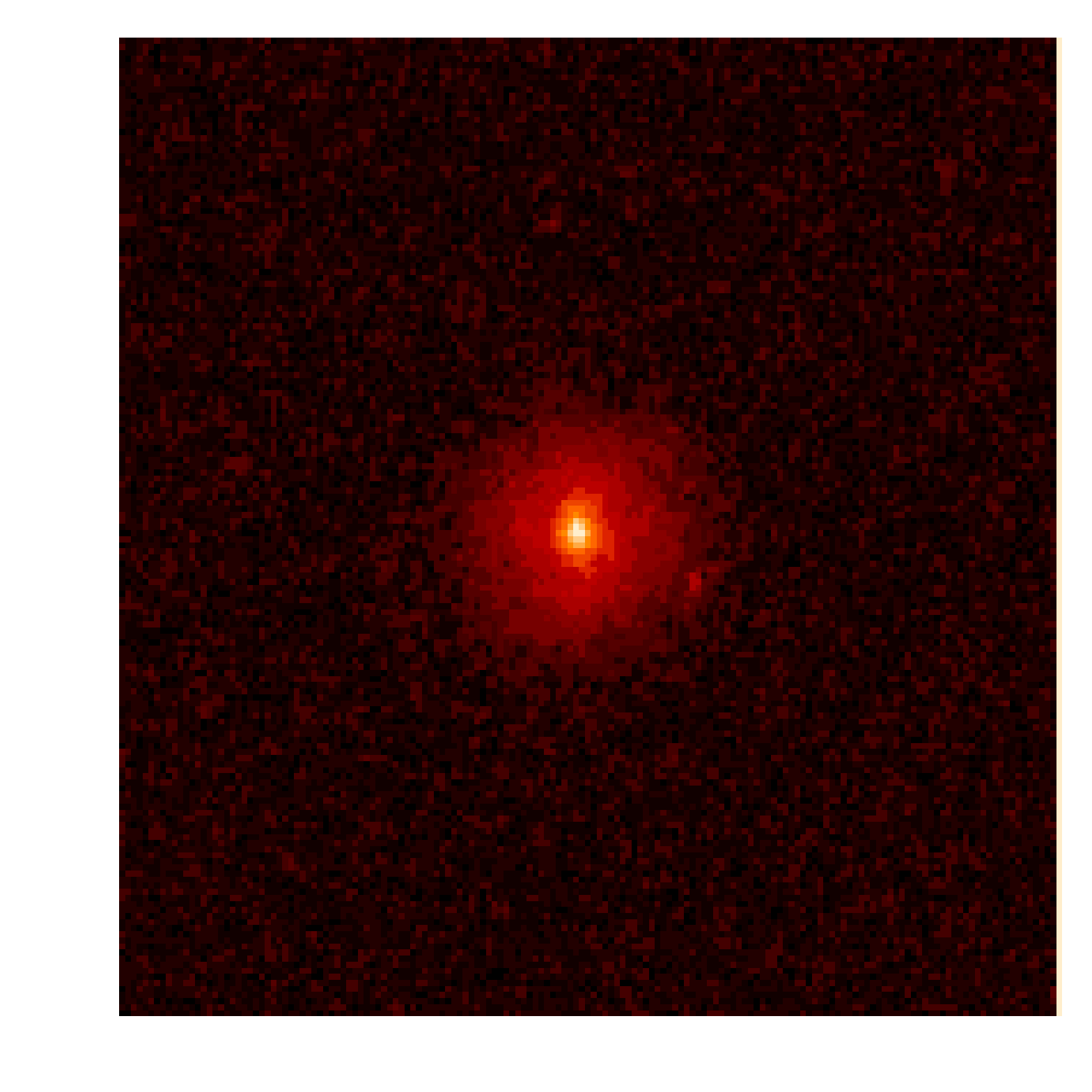}
\includegraphics[width=4cm]{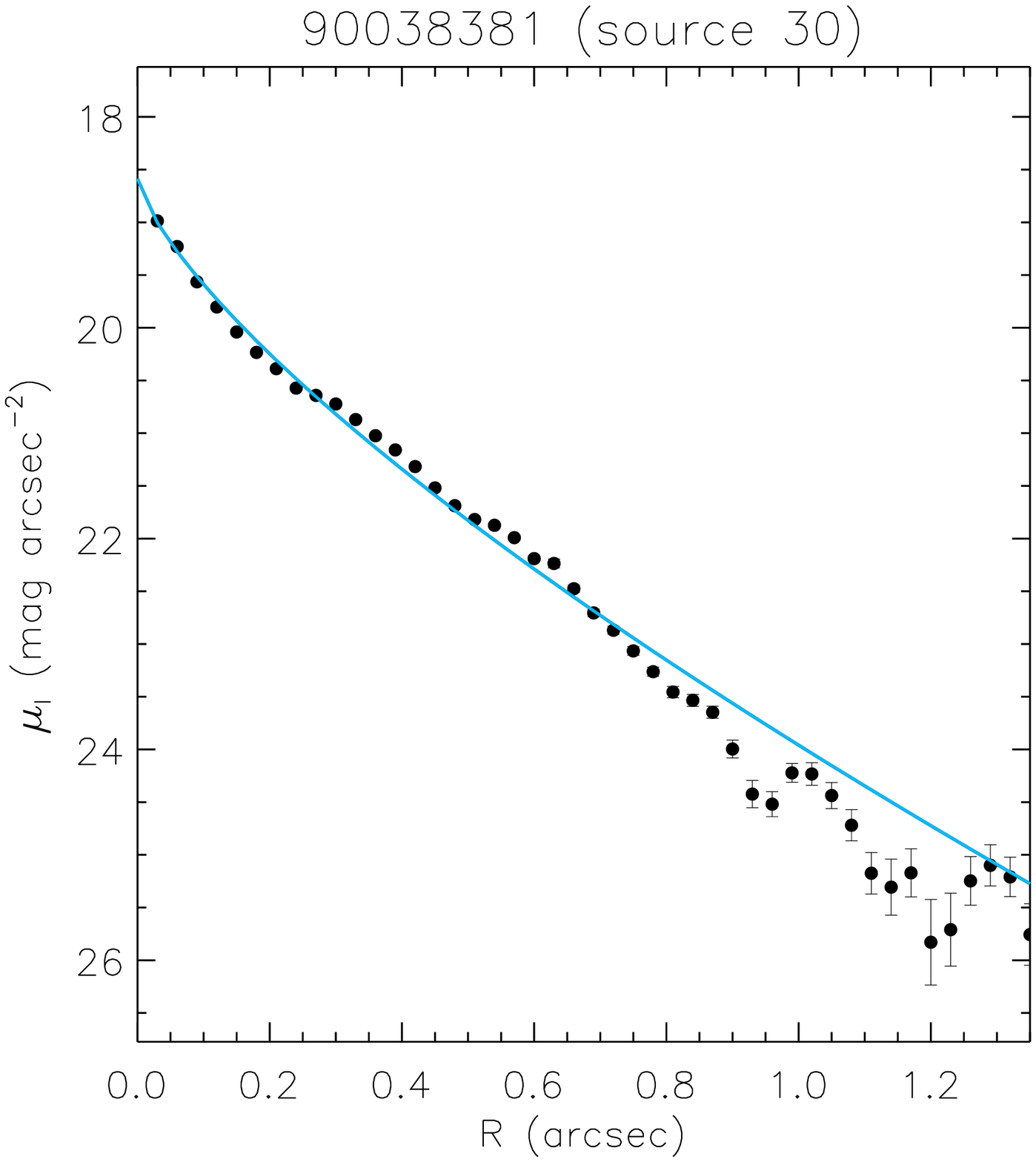}
\caption{(\emph{continued}) HST/ACS images and surface brightness profile of the AGN 
         bulgeless host galaxy candidates with $n < 1.5$ and disc/irregular morphology.}
\end{figure}
\clearpage

\begin{figure}
\includegraphics[width=4cm]{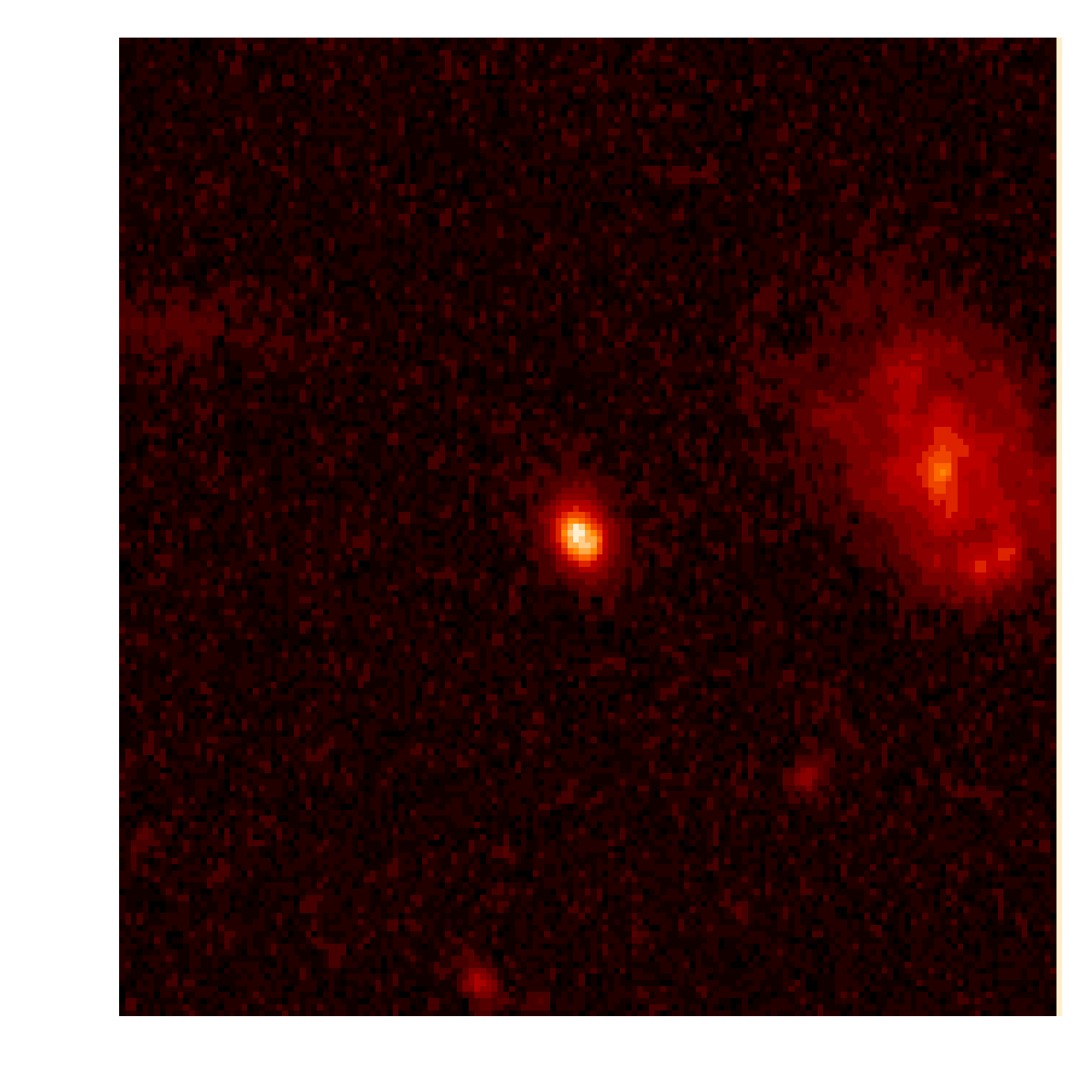}
\includegraphics[width=4cm]{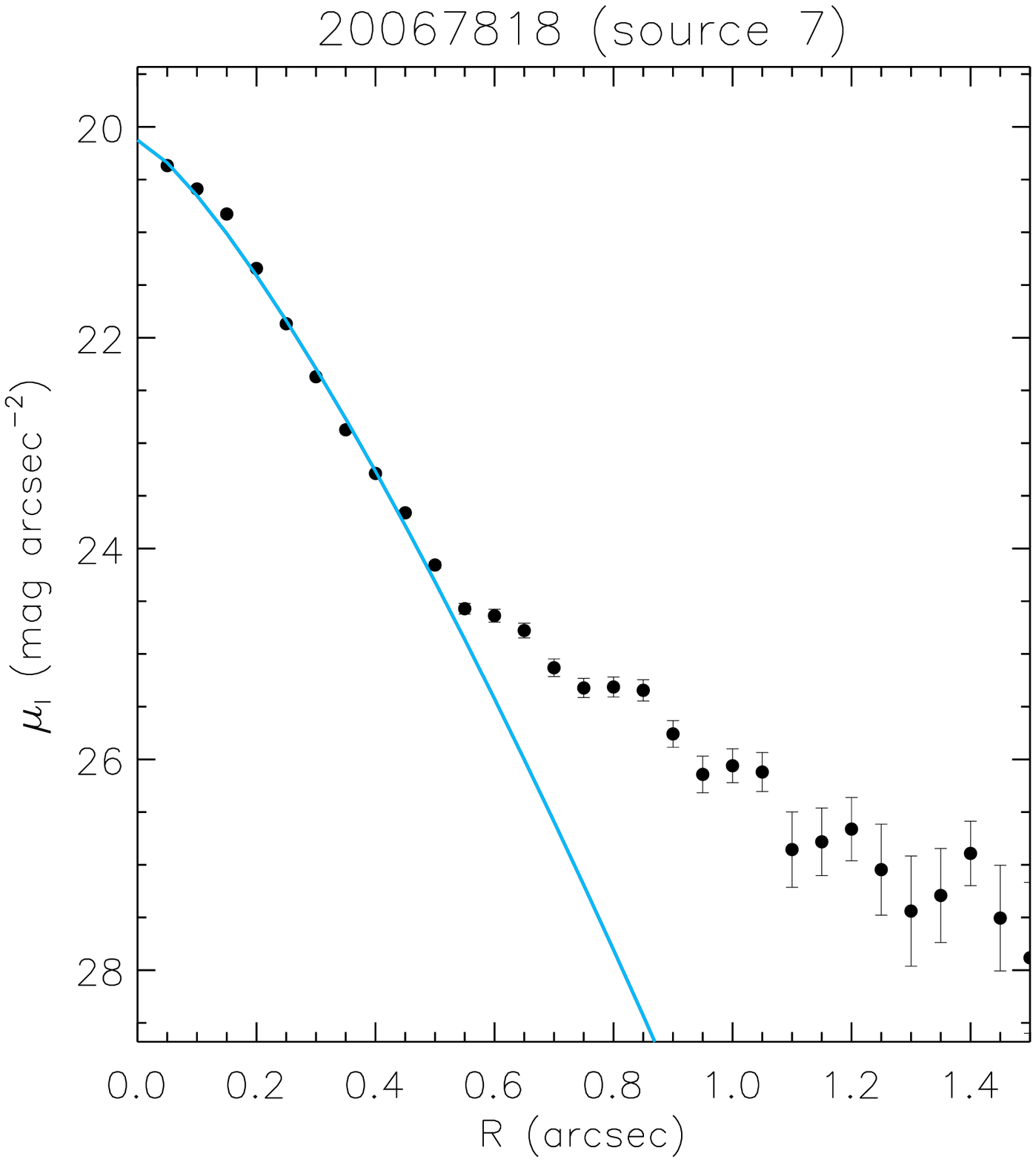}
\includegraphics[width=4cm]{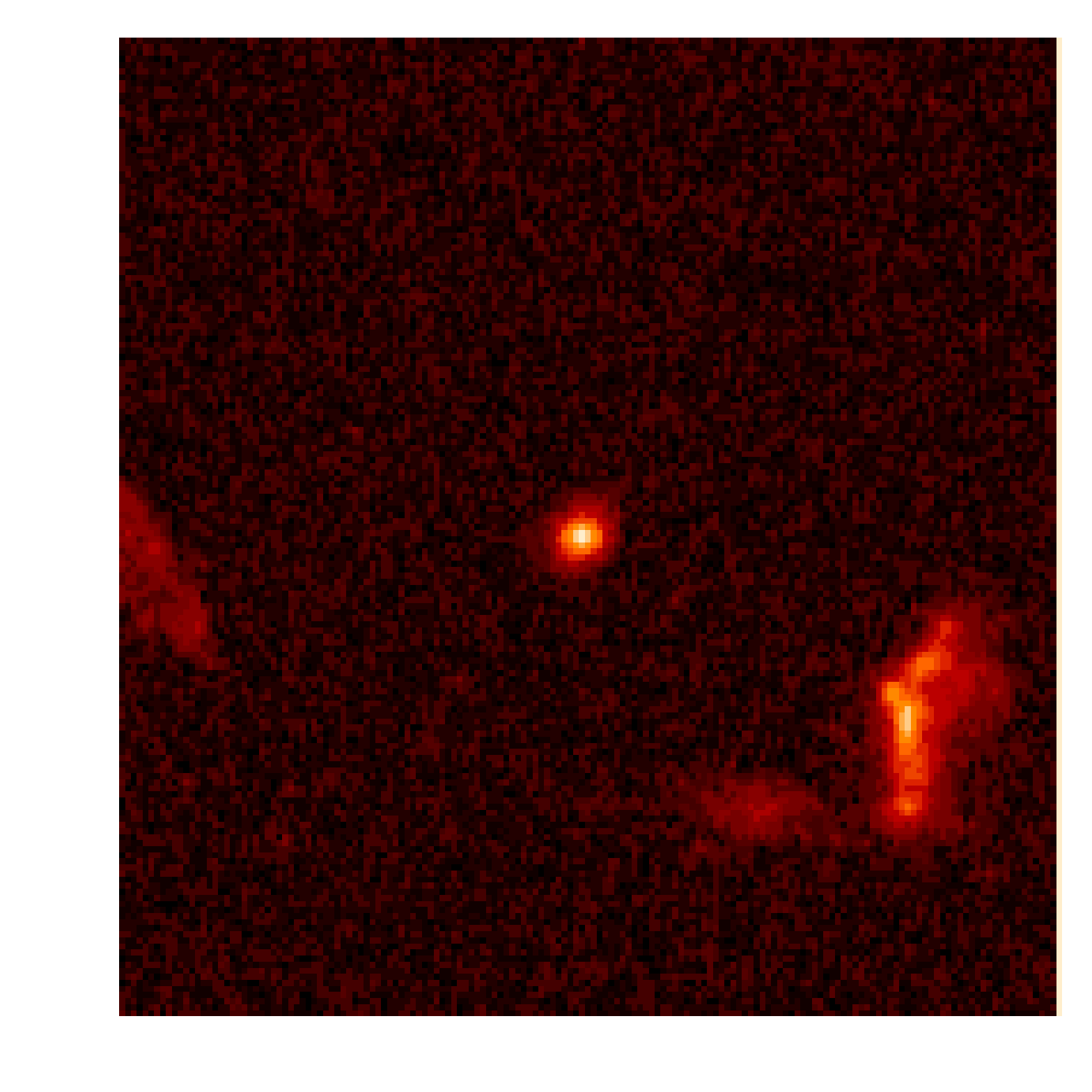}
\includegraphics[width=4cm]{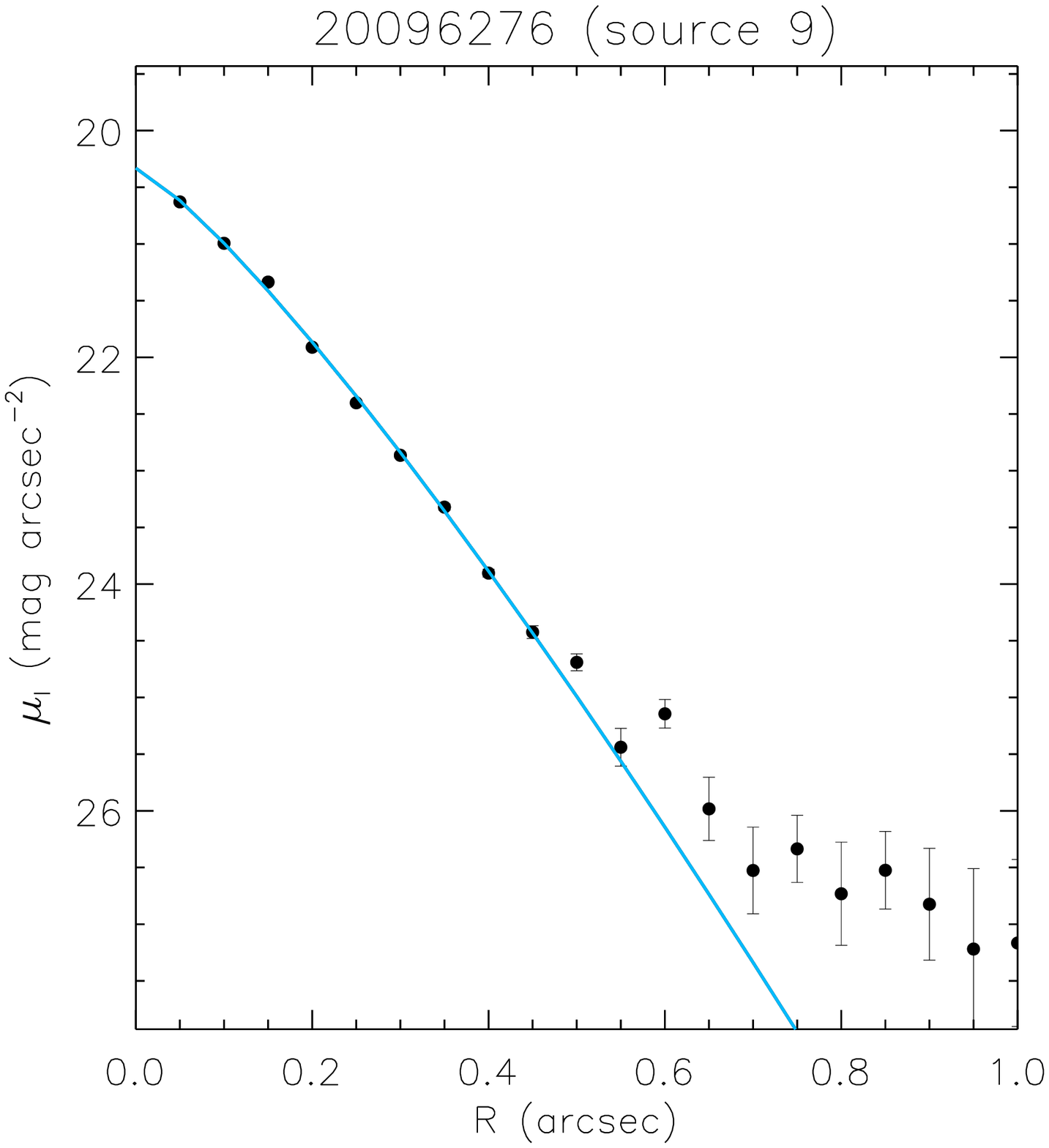}
\includegraphics[width=4cm]{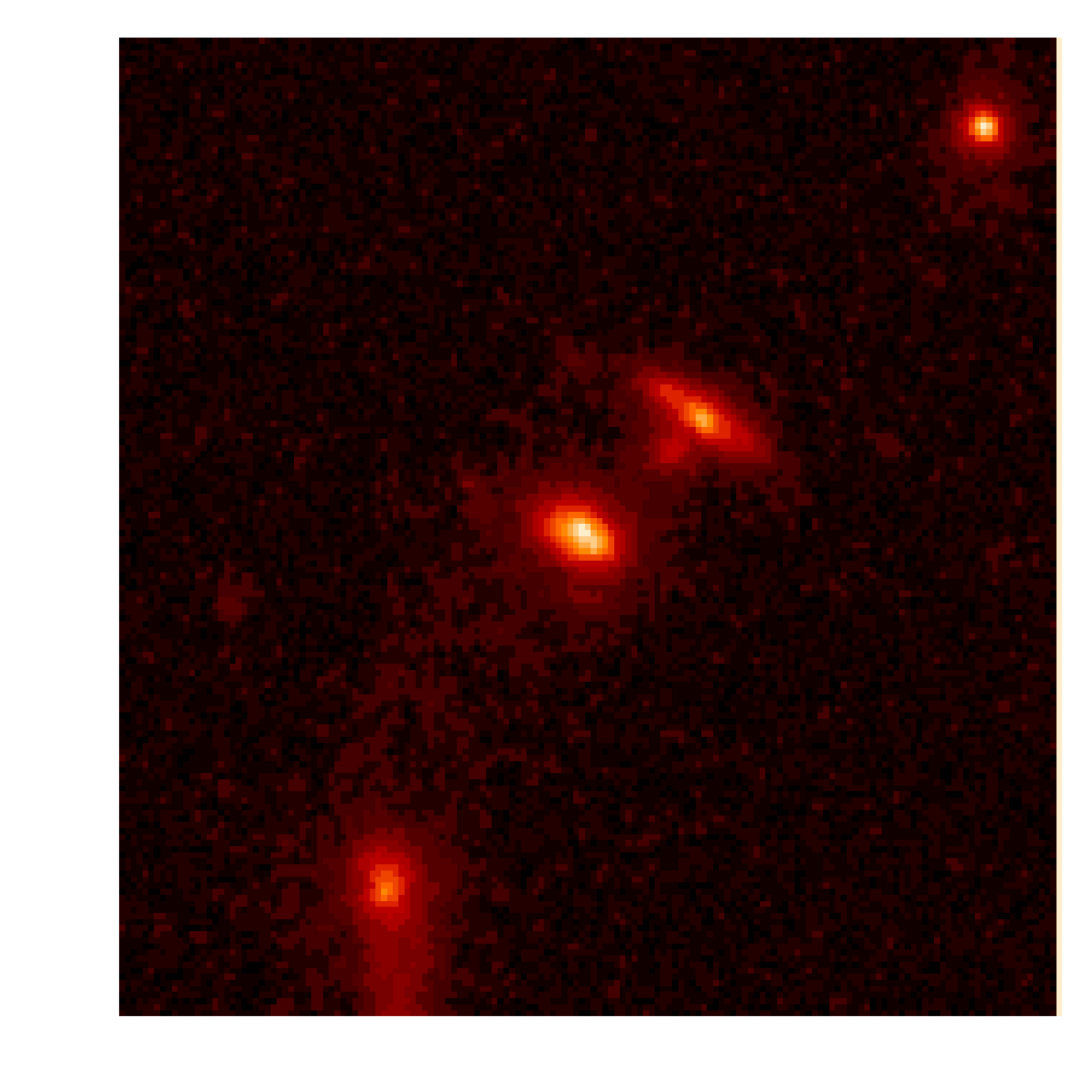}
\includegraphics[width=4cm]{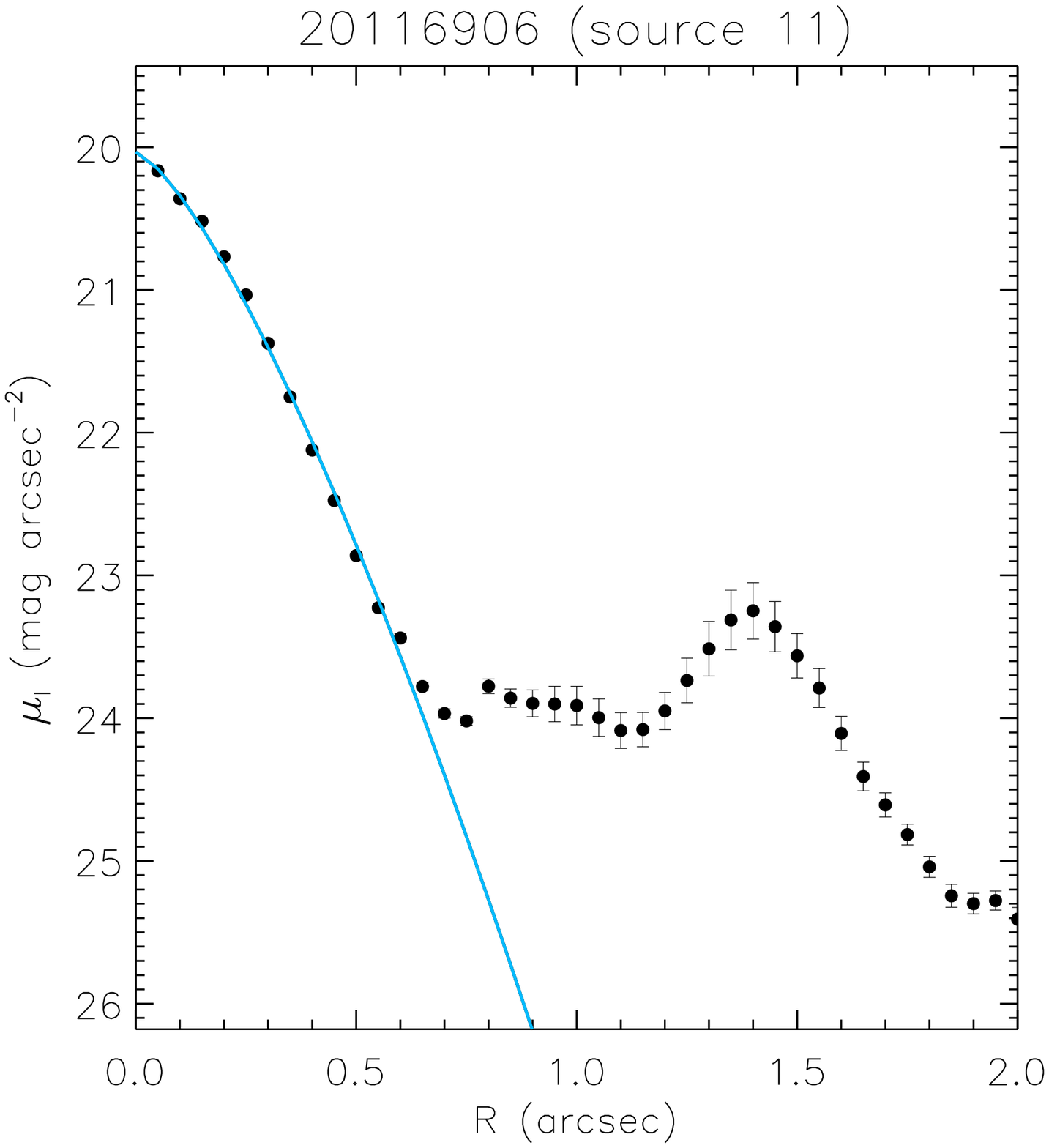}
\includegraphics[width=4cm]{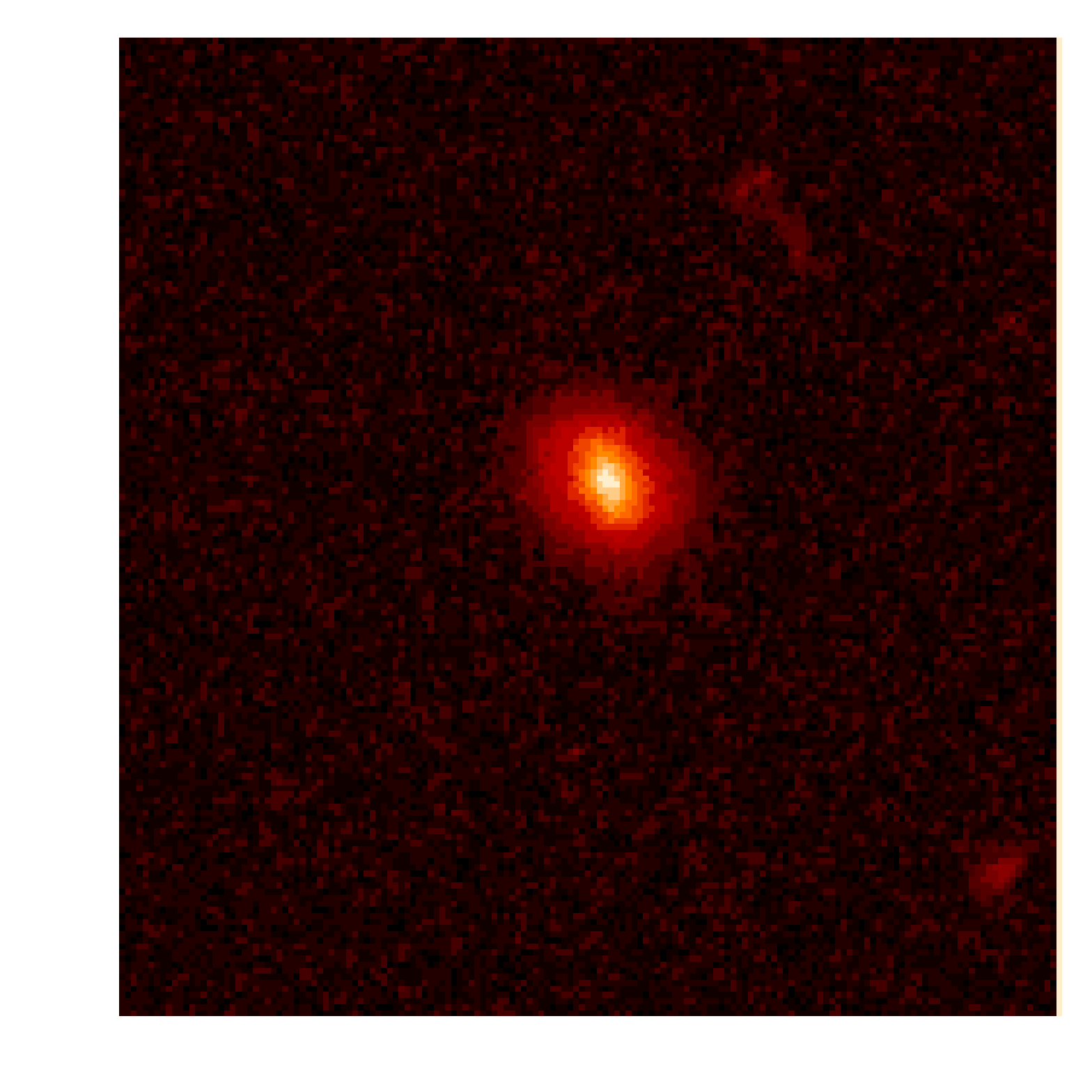}
\includegraphics[width=4cm]{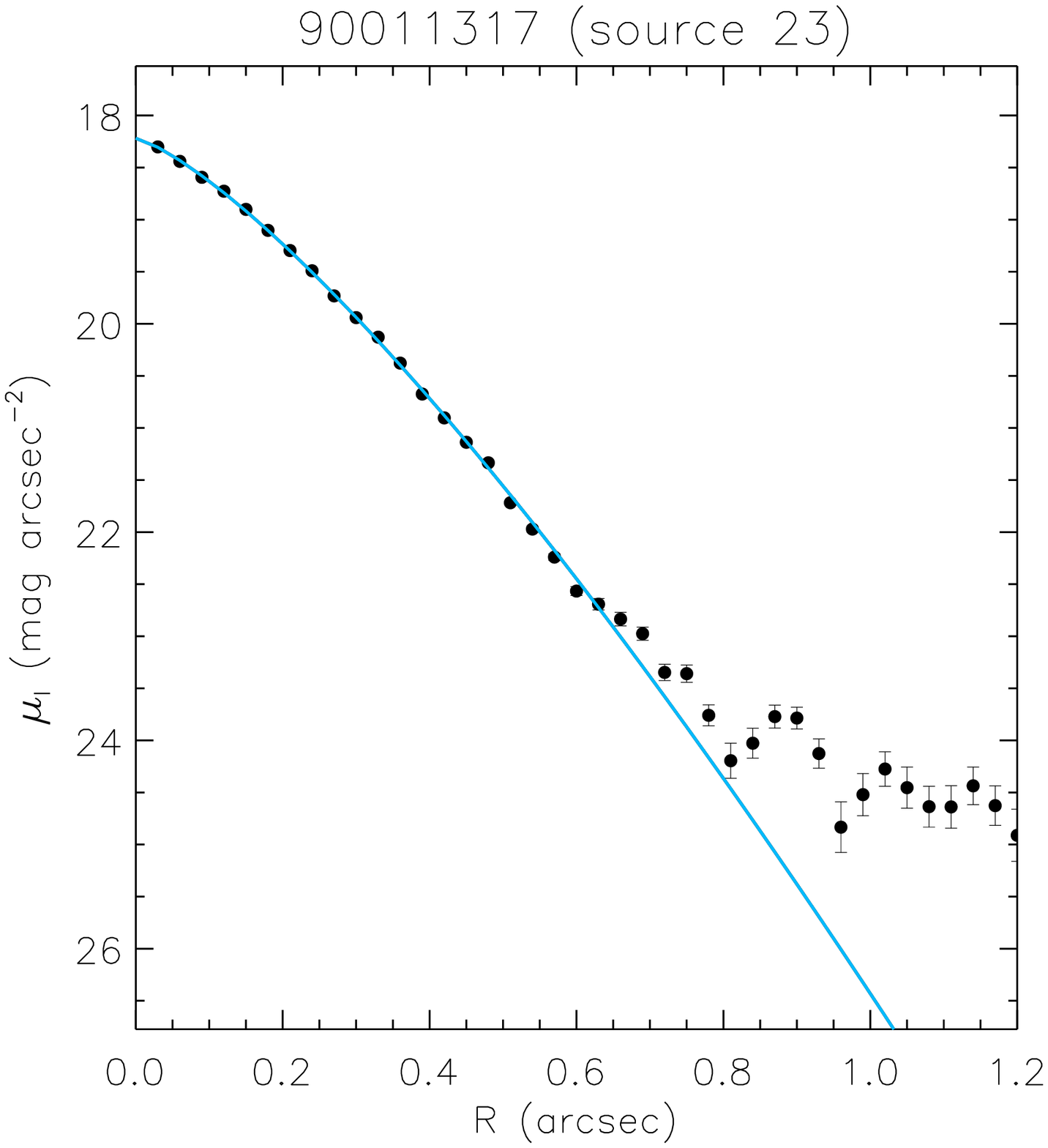}
\includegraphics[width=4cm]{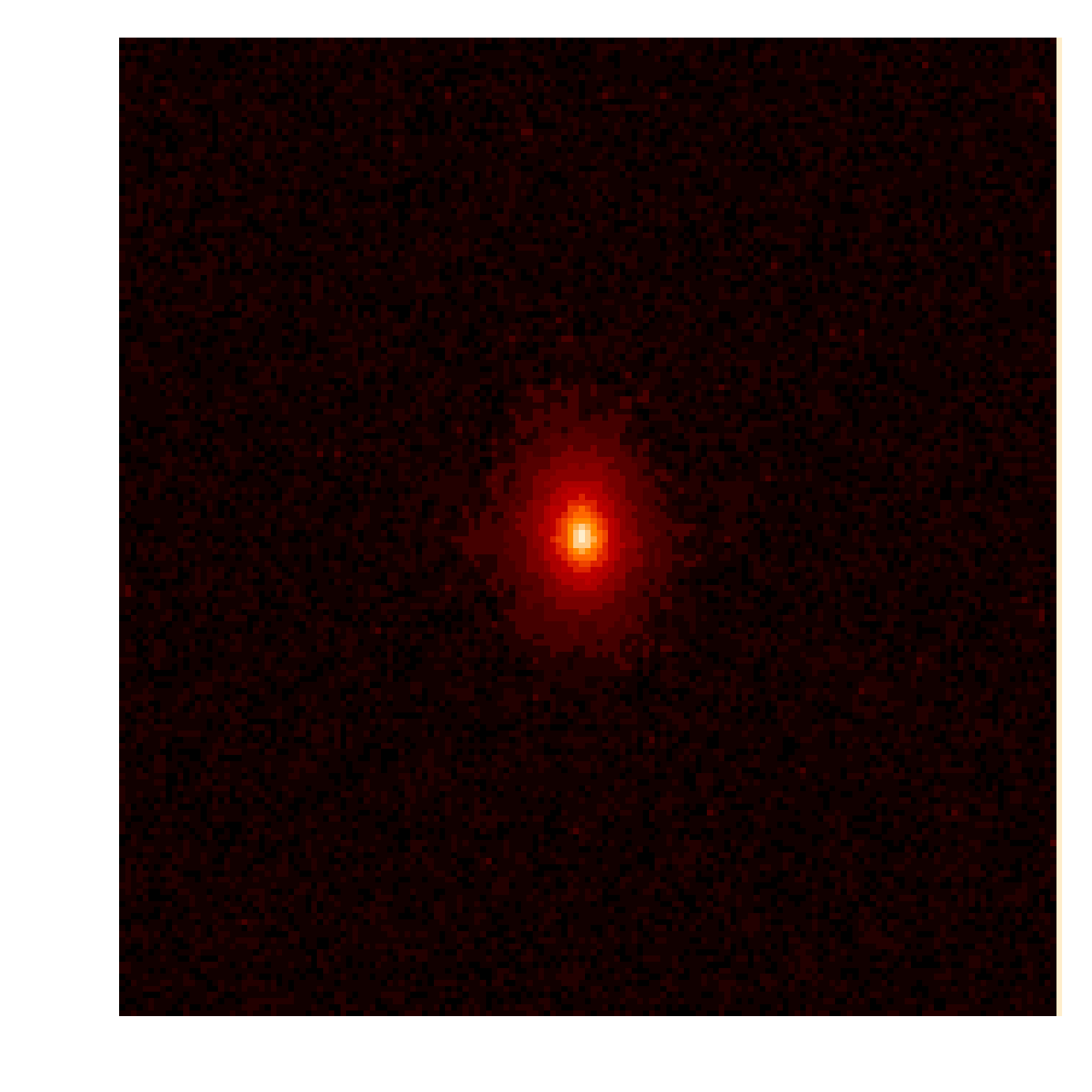}
\includegraphics[width=4cm]{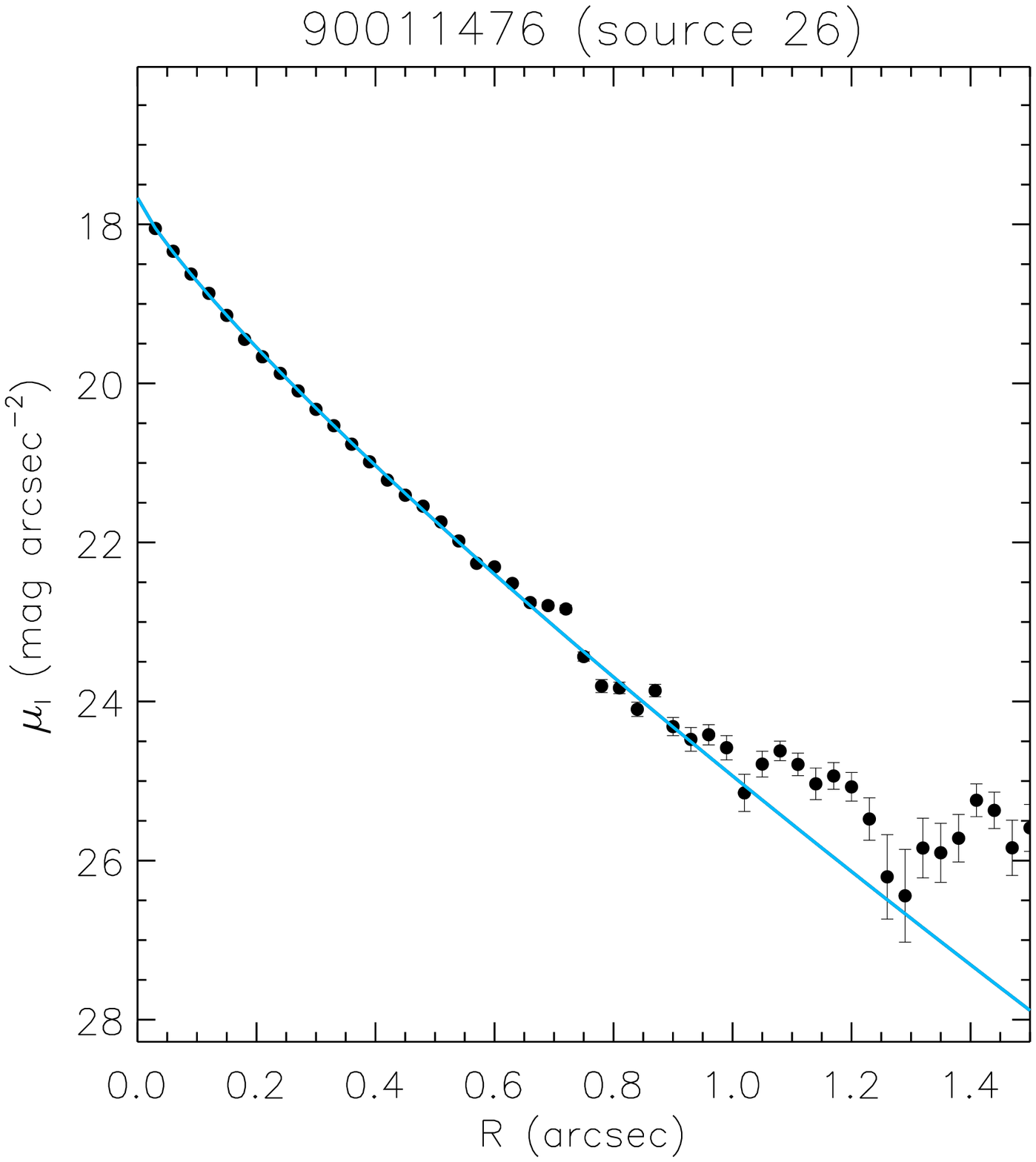}
\includegraphics[width=4cm]{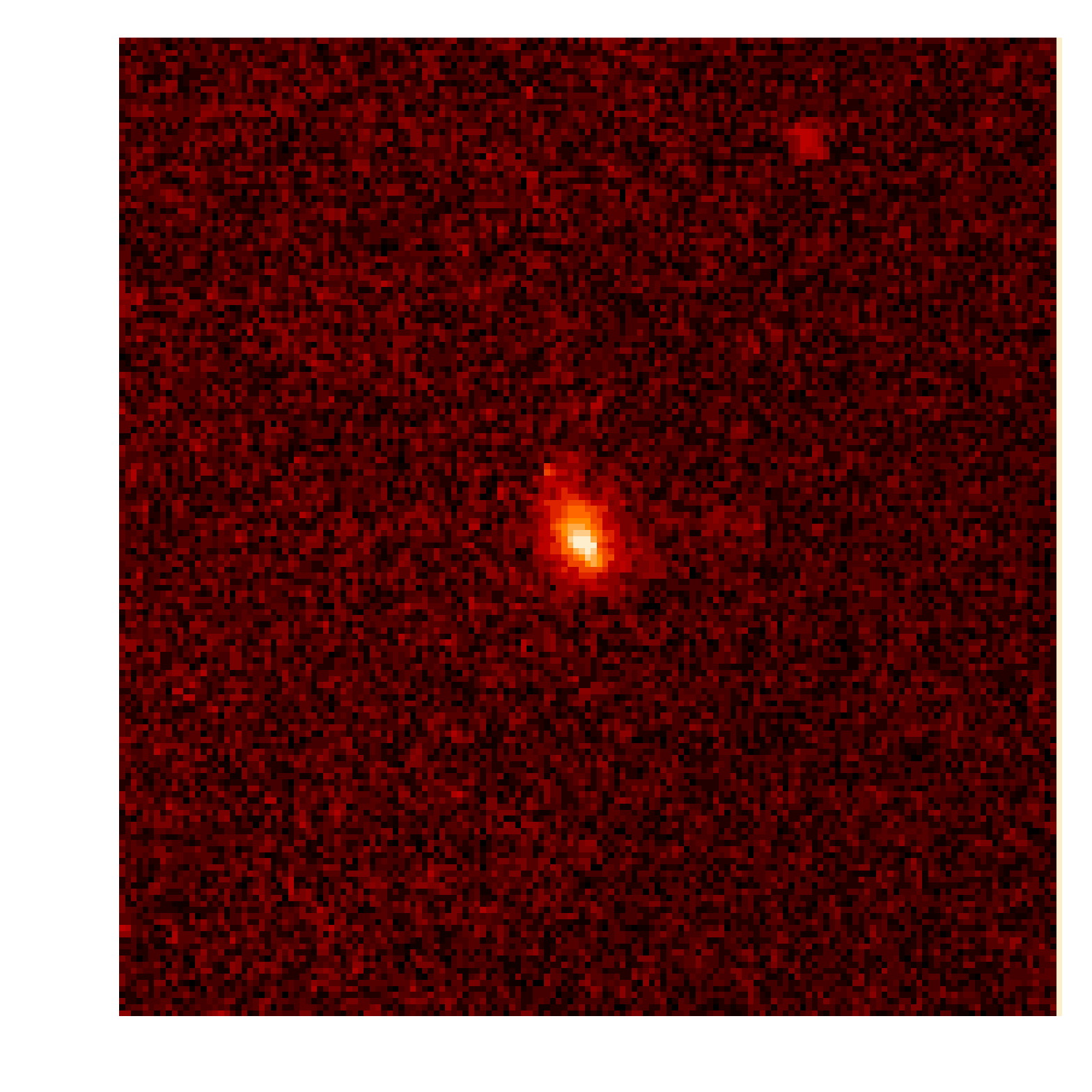}
\includegraphics[width=4cm]{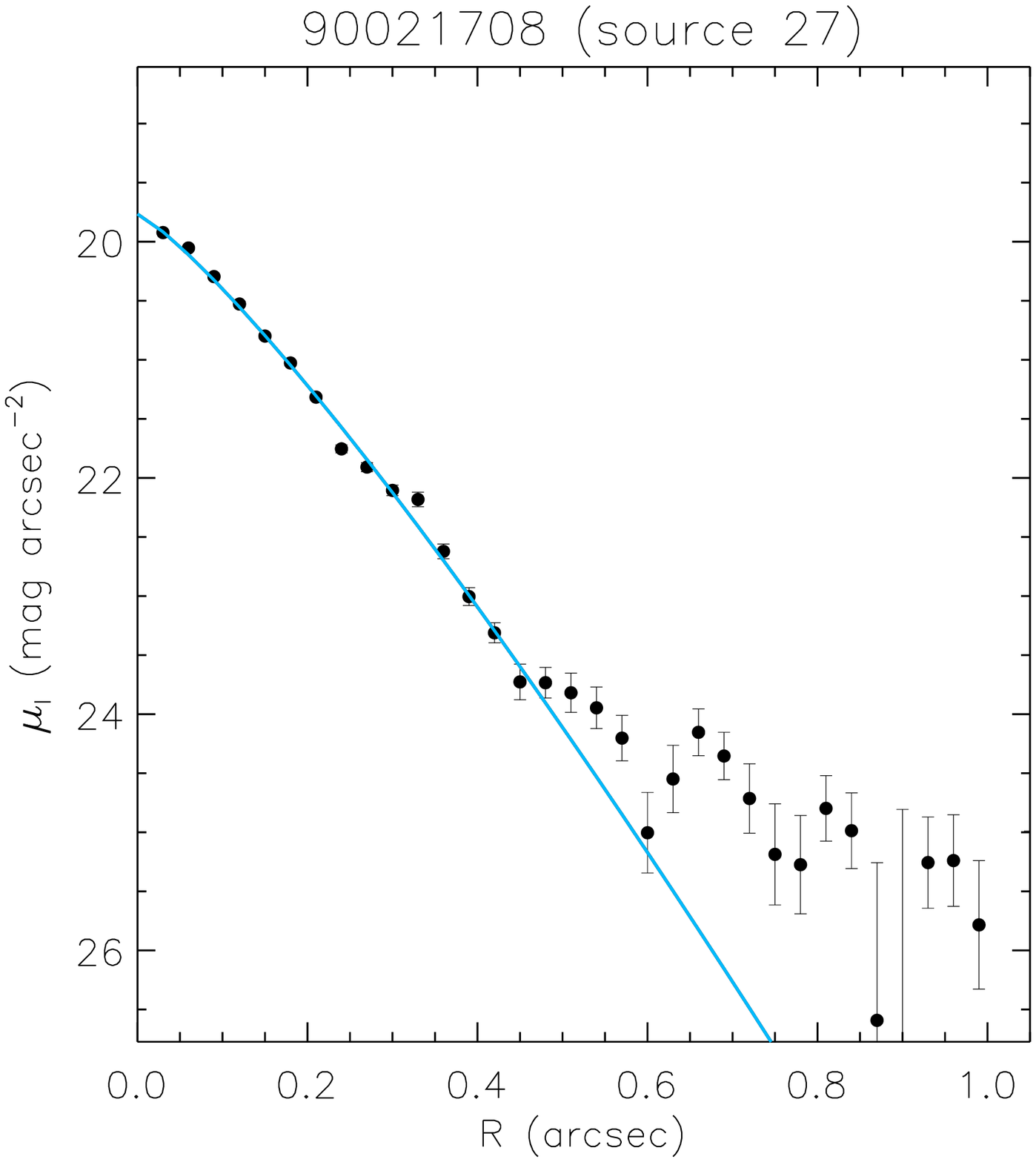}
\caption{HST/ACS images and surface brightness profile of the AGN bulgeless host galaxy 
         candidates with $n < 1.5$ and spheroidal morphology.}
\label{fig:AGN-prof-dubious}
\end{figure}
\clearpage

\begin{sidewaysfigure}
  \centering
  \includegraphics[angle=0,width=20cm]{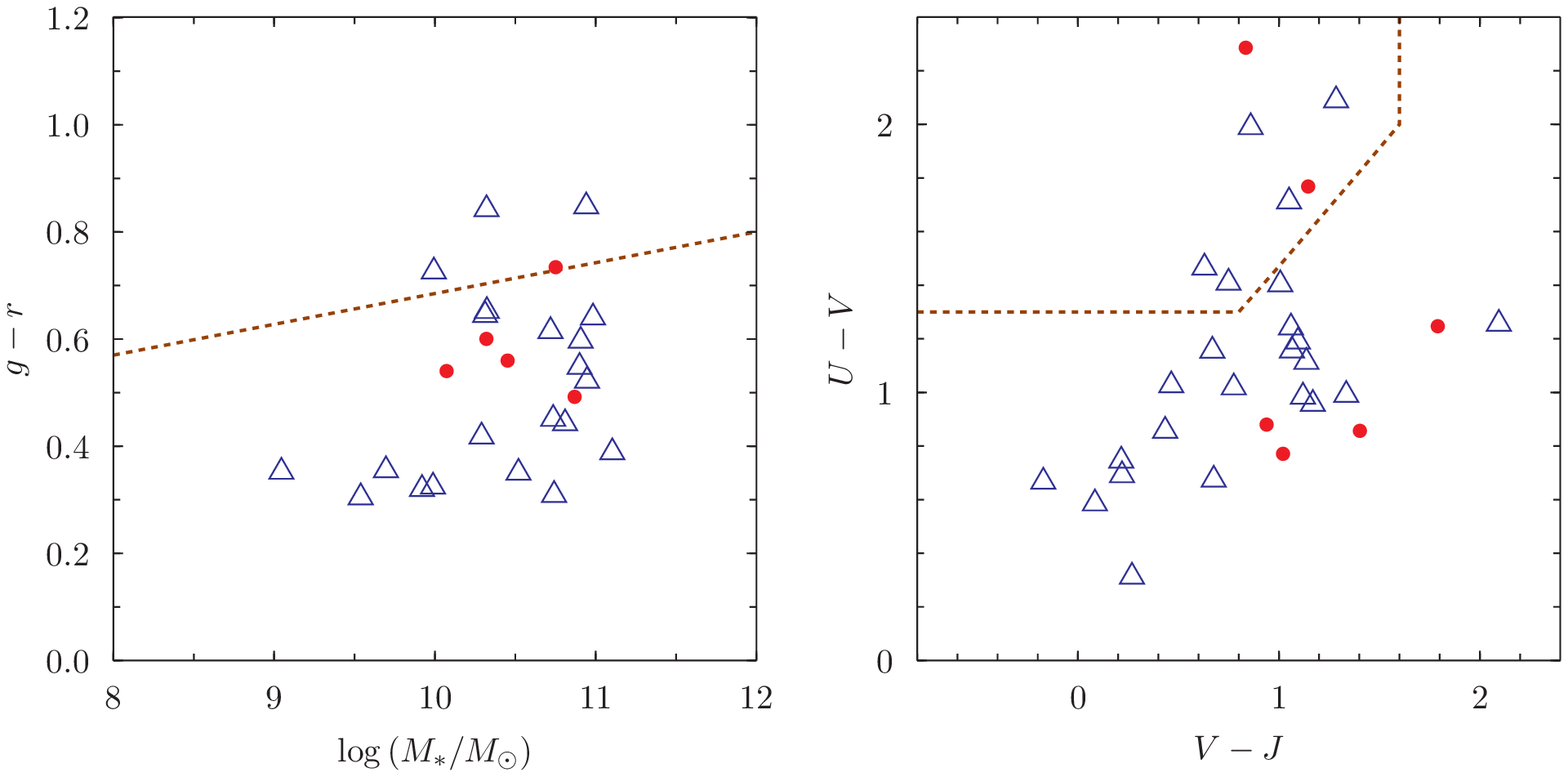}
  \caption{Left panel: $g - r$ colour--mass plot of the AGN bulgeless host galaxy candidates 
           with disc/irregular (blue triangles) and spheroidal (red dots) morphology 
           ($0.4 \leq z \leq 1.0$).
           Right panel: Rest-frame $U-V$ as a function of $V-J$ colour for the same galaxy 
           sample.
           Superimposed as dashed line are the rest-frame colours cuts defined in
           \citet{Williams-ApJ09-quiegal}.}
  \label{fig:AGN-colours}
\end{sidewaysfigure}
\clearpage

\end{document}